\documentclass[aps,prb,reprint,superscriptaddress, longbibliography]{revtex4-2}

\usepackage{blindtext}
\usepackage{graphicx}
\usepackage{amsfonts}
\usepackage{amssymb}
\usepackage{amsmath}
\usepackage{mdframed}
\usepackage{bbold}
\usepackage{physics}
\usepackage[caption=false]{subfig}
\usepackage{float} 
\usepackage{xcolor}
\usepackage[export]{adjustbox}
\usepackage{mathtools}
\usepackage{empheq}
\usepackage[most]{tcolorbox}

\definecolor{eqshade}{gray}{0.96}
\definecolor{eqborder}{gray}{0.60}

\newtcolorbox{wideeqbox}{
  enhanced,
  colback=eqshade,
  colframe=eqborder,
  boxrule=0.5pt,
  arc=1pt,
  boxsep=0pt,
  left=4pt,
  right=4pt,
  top=5pt,
  bottom=5pt,
  width=\linewidth
}

\definecolor{dodgerblue}{HTML}{1E90FF}
\definecolor{lightdodgerblue}{HTML}{4dbff7}
\usepackage[colorlinks=true,citecolor=dodgerblue,linkcolor=dodgerblue,urlcolor=dodgerblue,pdftitle={Long-lived coherences}]{hyperref}

\renewcommand{\l}{\left(}
\renewcommand{\r}{\right)}
\newcommand{\lb}{\left[}
\newcommand{\rb}{\right]}
\newcommand{\lcb}{\left\{ }
\newcommand{\rcb}{\right\} }
\newcommand{\lv}{\left|}
\newcommand{\rv}{\right|}
\newcommand{\mbf}[1]{\mathbf{\mathrel{#1}}}
\newcommand{\mc}[1]{\mathcal{\mathrel{#1}}}
\newcommand{\up}{\uparrow}
\newcommand{\down}{\downarrow}

\newcommand{\dbar}{\mathchar'26\mkern-12mu d}
\renewcommand{\tr}[1]{\, {\rm Tr} \left[ \mathrel{#1} \right] }
\newcommand{\titlemath}[1]{\texorpdfstring{$\mathrel{#1}$}{TEXT}}

\makeatletter
\def\l@subsubsection#1#2{}
\makeatother

\makeatletter
\newif\if@inappendix
\@inappendixfalse
\def\@appendixTOCsentinel{%
  \@inappendixtrue
  \vspace{8pt}%
}
\let\revtex@appendix\appendix
\renewcommand{\appendix}{%
  \revtex@appendix
  \addtocontents{toc}{\protect\@appendixTOCsentinel}%
}
\AtBeginDocument{%
  \let\orig@contentsline\contentsline
  \def\contentsline#1#2#3#4{%
    \def\@tempa{#1}%
    \if@inappendix
      \ifx\@tempa\@subsectionstring
      \else
        \nointerlineskip\vspace{-5pt}%
        \orig@contentsline{#1}{#2}{#3}{#4}%
      \fi
    \else
      \orig@contentsline{#1}{#2}{#3}{#4}%
    \fi
  }%
  \def\@subsectionstring{subsection}%
}
\makeatother

\begin{document}

\title{Long-lived local quantum coherences from hydrodynamic large deviations}

\author{Ewan McCulloch}
\affiliation{Laboratoire de Physique de l'École Normale Supérieure, CNRS, ENS \& Université PSL; 24 rue Lhomond, 75005 Paris, France}

\author{J. Alexander Jacoby}
\affiliation{Department of Physics, Princeton University, Princeton, New Jersey 08544, USA}

\author{Sarang Gopalakrishnan}
\affiliation{Department of Electrical and Computer Engineering,
Princeton University, Princeton, NJ 08544, USA}

\begin{abstract}
We develop a framework to describe how quantum coherences between distinct charge sectors evolve under generic charge-conserving dynamics. Our framework captures the nonperturbative interactions between quantum coherences and hydrodynamic large deviations---i.e., rare ``voids'' of low charge entropy. Conditional on surviving, the quantum coherence and its surrounding void form a collective polaron-like object. In one dimension, even at infinite temperature, we show that the lifetime of coherences is parametrically enhanced because they bind to voids. We use our framework to address two fundamental questions about generic quantum dynamics with a conserved charge. First, we argue that \emph{gapped} Ruelle-Pollicott resonances are absent in the weak-noise limit, even in sectors of operator space that contain no hydrodynamic slow modes: instead, the spectral gap in all sectors vanishes nonperturbatively in the noise strength. Second, we compute the spacetime asymptotics of the dynamical single-particle Green’s function, both in the weak-noise regime and in the absence of noise. In the noiseless case, we find that the void-coherence polaron undergoes subdiffusion, with an exponent we calculate. We support our general arguments with a microscopic derivation for random charge-conserving circuits, as well as numerical evidence from tensor-network simulations.
\end{abstract}

\maketitle
\tableofcontents

\section{Introduction}

It is often said \cite{BAA_06} that a many-body system ``acts as its own bath.'' Under generic chaotic dynamics, most initial states rapidly approach local thermal equilibrium: the density matrices of local subsystems are well approximated by Gibbs states, potentially with spatially varying temperatures and chemical potentials. %~\cite{mori2018thermalization}. 
On timescales beyond this rapid initial equilibration, the dynamics of local operators can be captured by hydrodynamics~\cite{PhysRevA.89.053608, Khemani_2018, wienand2023emergence, joshi2022observing, rosenberg2023dynamics, wei2022quantum, le2023observation, gross2017quantum, arute2019quantum, scholl2021microwave, rosenberg2023dynamics, 2016Sci...353.1257B, 2020PhRvL.125a0403K, heavyionlecturenotes,heavyionreview, Muller2008,Lucas2016a,Lucas2016b,Crossno2016,Narozhny2017,Lucas2018,Bal2021,McCulloch2023,SinghNavierStokes,gopalakrishnan2024non,Bauer2017,Bernard_2019,QSSEP_Bernard_2021,2026arXiv260116883A}. Hydrodynamics consists of classical stochastic equations for the dynamics of densities of conserved charges, e.g., energy and particle number~\cite{forster2018hydrodynamic,spohn2012large,SpohnNLFH}. The structure of these equations depends only on symmetries and conservation laws, and is insensitive to whether the underlying problem is classical or quantum.

Within this framework, the nontrivial early-time quantum dynamics is modeled as classical noise. Consequently, subjecting the system to \emph{extrinsic} classical noise that respects all applicable symmetries (but causes decoherence) does not change the equations of hydrodynamics. For local operators whose late-time behavior is controlled by overlap onto conserved densities and currents, the asymptotics of correlation functions is determined by hydrodynamic projection~\cite{doyon2022diffusion,2019PhRvL.122i1602C,PhysRevB.73.035113,von_Keyserlingk_2022}, and is therefore insensitive to quantum coherence.
%Since the late-time asymptotics of generic local operators or generic correlation functions can be expressed in terms of hydrodynamic projections~\cite{doyon2022diffusion,2019PhRvL.122i1602C,PhysRevB.73.035113,von_Keyserlingk_2022}, the asymptotics of these correlations is also insensitive to quantum coherence.

In systems with no conserved charges, any initial state becomes locally indistinguishable from maximally mixed, and the hydrodynamics is trivial~\cite{PhysRevX.4.041048,PhysRevB.93.155132,PhysRevE.90.012110,PhysRevX.7.031016}. Local correlation functions decay exponentially, on a characteristic $O(1)$ timescale~\cite{Prosen_2004,Mori_24,Znidaric_2024,zhang2025,SpectralGaps,Nahum_22,teretenkov_2025}. This exponential decay has a natural interpretation in terms of Heisenberg-picture operator evolution. An initially local operator becomes nonlocal under Heisenberg evolution; to contribute to a high-temperature local correlation function like $\mathrm{Tr}(A(t) A)$, the operator $A(t)$ must ``refocus'' and become local at time $t$, and this is exponentially unlikely on entropic grounds. 

Since correlation functions decay exponentially even absent noise, the addition of weak extrinsic noise again has no qualitative effect on the asymptotics of correlation functions. Indeed, the characteristic exponential decay of correlations in the weak-noise limit can be retrieved from the noisy evolution, by studying the limiting behavior of the spectral gap of the quantum channel governing the noisy evolution: this gap approaches an $O(1)$ value, the %so-called 
Ruelle-Pollicott (RP) resonance~\cite{Ruelle86,Pollicott1985}, in the appropriate weak-noise limit~\cite{Prosen_2004,Mori_24,Znidaric_2024,SpectralGaps,zhang2025}. There is considerable evidence that noisy quantum dynamics are easy to simulate on a classical computer~\cite{schuster2024polynomial,gonzalez2024pauli,Begusic2024,aharonov2023polynomial,Noh2020efficientclassical}; thus, neither hydrodynamic correlators nor RP resonances are appealing candidates for quantum advantage in quantum simulation problems.

An important class of physically relevant correlation functions does not fit into either of these categories. The most familiar example of such a correlation function is the single-particle Green's function for a system of interacting fermions, $\mathcal G(x,t) = \langle c^\dagger(x,t) c(0,0) \rangle$. The dynamics of such systems obeys both energy and charge conservation, but the fermionic field operator $c^\dagger(x,t)$ has no hydrodynamic projection, as we will discuss in more detail below: under time evolution, it has no amplitude for turning into products, derivatives, etc. of energy or charge density~\cite{von_Keyserlingk_2022}. We refer to objects like $c^\dagger(x,t)$ (i.e., products of unequally many creation and annihilation operators) as non-hydrodynamic. Naively, one might expect $\mathcal G(x,t)$ to behave as if there were no hydrodynamic modes, and to decay exponentially. %Crucially, t
This is not the case: when conservation laws are present, the local thermalization \emph{rate} depends on the local charge densities. Thermalization slows down in the low-density limit, where the dynamics consists of single quasiparticles propagating coherently above the vacuum. 

We recently observed that the existence of slowly thermalizing states infects the behavior of correlations even in typical states, at least in one dimension, through rare events: any snapshot of a finite-temperature state will contain rare low-entropy ``voids'' where coherences can persist long enough that the decay of the single-particle Green's function can be bounded by a stretched exponential~\cite{StretchedExp} (see also~\cite{Duh2026,2025arXiv251002198C}). We referred to this phenomenon as ``diffusion-limited dephasing'': the physical process that sets the lifetime of the (non-hydrodynamic) coherence is the (hydrodynamic) diffusion of charge into the void.

In Ref. \cite{StretchedExp} we established sub-exponential decay by a variational argument. In the present work we develop a systematic framework to compute the dynamics of operators like $c^\dagger(x,t)$, which create local coherences between distinct charge sectors, and their coupling to hydrodynamic large deviations. 
Our framework incorporates the dynamics of local coherences into macroscopic fluctuation theory (MFT)~\cite{Bertini_2015,Derrida_2025,PhysRevLett.129.040601,2026arXiv260102319S,10.21468/SciPostPhys.17.2.033}, the standard framework for treating these large deviations. 
Indeed, our central result is to reduce a formally quantum problem (the survival of coherences) into a classical one about hydrodynamic large deviations: we argue that this reduction holds generally, and extends beyond the cases where we are able to solve for the large-deviation properties.
The key idea behind our framework is as follows. One can expand the Heisenberg evolution of any operator as a sum of histories over a suitable operator basis. We are interested in the \emph{conditional} ensemble of histories that take a single local coherence at time $0$ into a single local coherence at time $t$. These histories are rare among all possible operator histories, but as we will see they dominate both the late-time asymptotics of local correlation functions and the leading eigen-operators of the dynamics in the weak-noise limit. We argue that these dominant histories involve a ``void-coherence polaron'': for example, the Green's function $\mathcal G(x,t)$ is dominated by trajectories where the coherence moves from $0$ to $x$ and drags the void along with it. We use standard techniques to compute the properties of this polaron. 

% \jaj{this section overclaims and should be slightly rewritten. we are not talking about one-dimesional charge conserving systems,  but those which remain diffusive at arbitrary density (a non-generic property).} \sg{The paragraph above is fine, the qualitative picture is the same in all cases.}\jaj{not necessarily talking about IP directly above. Just in general signposting on the diffusion constant remaining finite.} \sg{We do want people to read the paper so we should not advertise too aggressively that it is not applicable to any system they might care about.}\jaj{yes. tightrope walking.}
We use this framework to address two basic questions about the dynamics of one-dimensional quantum systems. First, we consider charge-conserving dynamics subject to weak noise (which for simplicity we take to be depolarizing). We compute the low-energy spectrum of the quantum channel generating this dynamics, and find that it is gapless even when restricted to non-hydrodynamic sectors. For noise strength $\gamma$ and momentum $k$, the momentum-dependent gap for non-hydrodynamic operators scales as $\sqrt{\gamma}(1 + \tilde{D}_{\mathrm{eff}} k^2)$, for some constant $\tilde{D}_{\mathrm{eff}}=\mathcal{O}(1)$ that we estimate; meanwhile, for hydrodynamic operators, this gap scales approximately as $\min(\gamma + D k^2, |k| \sqrt{D \gamma})$. Thus, at least in one dimension, there are no well-defined RP resonances in systems with a conserved charge, even if one orthogonalizes against hydrodynamic sectors. However, the gap is nonperturbative in $\gamma$, and the corresponding eigenvectors are highly collective objects at small $\gamma$. These explicit solutions rely on the further simplification that the diffusion constant remains finite at all densities (valid, e.g., for random charge-conserving circuits); however, we expect the qualitative conclusions and framework to extend to realistic Hamiltonians in which diffusion constants diverge or vanish at low density. We discuss this general case briefly in Sec.~\ref{sec:spectral-structure}.

Second, we turn to the case with no extrinsic dissipation, and explore the structure of the two-point correlation function. We recover the scaling of Ref. \cite{StretchedExp} and find the leading space- (or momentum-) dependent correction, which we argue is \emph{nonanalytic} in momentum: this nonanalyticity stems from the anomalous diffusion of the void-coherence polaron. The final result is simplest to express in real space and time: we predict that $\mathcal G(x,t) \sim \exp\big(-\sqrt{t/t_0}\big) f(x/t^{1/3})$, for some scaling function $f$. We discuss the implications of our results for near-term experiments both in real materials and synthetic quantum matter. 

The rest of this paper is organized as follows. In Sec.~\ref{sec:context} we review operator dynamics, Ruelle resonances, and the role of conserved charges. In Sec.~\ref{sec:slow-manifold} we argue %on general heuristic grounds \jaj{[again the word general seems misplaced-- if nothing else because the conditioning is not. I agree this is not the place to get into the specifics (so maybe just very minor hedging is called for).]} 
heuristically for the MFT action containing a single local coherence. In Sec.~\ref{sec:optimal_void_profiles} we derive the consequences of this action both for noisy and noiseless evolution. In Sec.~\ref{RUCsection} we provide a microscopic derivation of the effective conditioned fluid theory in a concrete random-circuit setting (as well as corrections due to the motion of the coherence). It is more technical than the rest of the paper and may be skipped on a first reading. In Sections~\ref{sec:polaron-dynamics} and \ref{sec:aging-polaron} we explore the collective ``polaron'' in more detail in the weak noise and noiseless cases respectively, deriving the $k$-dependent corrections we quoted above. In Sec.~\ref{sec:spectral-structure} we summarize our understanding of the spectral structure of generic quantum evolution in the weak-noise limit with a single conserved charge. We close in Sec.~\ref{sec:discussion} with a discussion of experimental implications and potential extensions of our results.

\section{Context}
\label{sec:context}

In this section we review some of the key background for the rest of this paper: operator growth, RP resonances, and the role of conservation laws and large deviations in determining the behavior of correlation functions. 

\subsection{Operator growth, RP resonances, correlation functions}

We first briefly review operator evolution in the Heisenberg picture for systems with no conservation laws, potentially subject to depolarizing noise. In both the unitary and dissipative cases, Heisenberg evolution is unital: %``unital'':  (why is unital in quotation marks here??)
it maps the identity operator to itself. Therefore, \emph{local} Heisenberg evolution can only grow operators inside a light cone. (This result extends to continuous-time evolution via Lieb-Robinson bounds.) The time-evolved operator $O(t)$ can be expanded in a basis of Pauli strings, $\cdots \mathbb{1} \otimes \sigma^{x} \otimes \sigma^{y} \otimes \mathbb{1} \otimes \cdots$, where all legal strings are strictly $\mathbb{1}$ outside the light cone. Each Pauli string evolves into a superposition of strings, and the overall evolution rule is linear, so one can express operator evolution as a spacetime sum over Pauli ``paths'' (analogous to a path integral). 

Under unitary dynamics, the light cone is the only constraint on operator growth: a simple local operator spreads ballistically into increasingly extended and complicated strings, so typical operator Feynman histories have sizes growing linearly in time, $S[\hat A(t)]\sim vt$~\cite{Nahum_18,von_Keyserlingk_2018}. The mechanism for the ballistic growth is entropic, since (for two-level systems/qubits) three of the four basis operators are non-identity. Heuristically, the ``endpoints'' of the operator (i.e., the leftmost and rightmost non-identity Pauli) evolve under a random walk that is biased to the left and right respectively. Bulk depolarization, by contrast, acts independently on each nontrivial site, so the total decay rate is proportional to the operator size, $\gamma S[\hat A]$. The effect of weak dissipation is therefore  to cut off this ballistic growth at a characteristic size $S_\ast\sim \gamma^{-1}$~\cite{Mori_24}. The slowest nontrivial eigenoperators of the noisy channel have size $\sim \gamma^{-1}$, and decay rate $\sim \gamma \times \gamma^{-1} = O(1)$, so the spectral gap converges to some nonzero value independent of $\gamma$ in the weak dissipation limit \cite{SpectralGaps,zhang2025}.
%
%when the, a-priori unitary, hydrodynamic spreading modes equilibrate with the large-sized cutoff (via a dirichlet boundary condition \cite{curt, tomasz...} or Airy matching \cite{jaj}), forming an $O(1)$ spectral gap. 
Their $O(1)$ decay rate in the weak-dissipation limit has been interpreted as quantum many-body Ruelle--Pollicott resonances~\cite{prosen2002ruelle,Prosen_2004,Mori_24,Znidaric_2024,SpectralGaps,zhang2025,Yoshimura_2025}.

In the absence of conservation laws, generic dynamics leads to a steady state that is (at least locally) maximally mixed. Temporal correlation functions in this state can be expressed in the form $\mathrm{Tr}(A(t) B)$, where $A, B$ are local Pauli strings, and $A(t)$ means the Heisenberg-evolved operator $A$. If one expands the evolution of $A(t)$ in terms of Pauli paths, all surviving contributions to the correlation function come from paths that start and end as the identity except at $O(1)$ sites. Such paths are exponentially unlikely, so the correlation function is exponentially suppressed. One can regard the correlation function as a sum over a \emph{conditional} ensemble of Pauli paths, which started as $A$ and ended as $B$. The paths contributing to this conditional ensemble are highly atypical throughout their spacetime history: unlike typical paths, which grow ballistically in time, these only reach sizes $\leq t^{1/2}$ \cite{Nahum_22}. These paths are much more robust against dissipation than typical ones (they are only damped at a rate $\leq \gamma \sqrt{t}$), so the leading exponential decay of correlations is robust against weak dissipation. This simple argument already illustrates a point that will be crucial for us below, namely that---if one expands the operator history as a sum over paths---the correlation function is related to large deviations of the operator-path ensemble.

\subsection{Operator growth with charge conservation}

We now turn to operator growth in the presence of a single conserved charge, or equivalently a $U(1)$ symmetry. Following \cite{Khemani_2018,Rakovszky2018}, we will consider this conserved charge to be $Q = \sum_i \sigma^{z}_{i}$, and restrict to two-qubit gates for simplicity. We will also focus on noise that ``weakly'' preserves the $U(1)$ symmetry, such as depolarizing noise: under weakly symmetric noise, the system can exchange charge with the bath, but the system and bath together remain $U(1)$ symmetric (for example, the bath is not a superfluid)~\cite{Buca_2012,PhysRevA.89.022118,PhysRevB.110.155150,ziereis2025}. Under weakly symmetric noise, the Heisenberg evolution of operators is block diagonal, in a suitably chosen basis, generated by the single-site operators $\mathbb{1}, \sigma^{z}, \sigma^+, \sigma^-$ (or $P^\uparrow, P^\downarrow, \sigma^+, \sigma^-$). We refer to the subspace spanned by $\mathbb{1}, \sigma^{z}$ (or $P^\uparrow, P^\downarrow$) as populations, and that spanned by $\sigma^+, \sigma^-$ as coherences. Many-body operators are strings of these single-site basis operators. The net ``operator charge'' $\mathcal{Q}$ of a string is the difference between the number of $\sigma^+$ and $\sigma^-$ entries that it contains. Thus, for example, $\sigma^+$ has a charge of $1$, and $\sigma^{z}$ has a charge of $0$. All hydrodynamic operators obey $\mathcal{Q} = 0$; under weakly symmetric Heisenberg evolution, $\mathcal{Q}$ is conserved, so neutral and charged operators do not mix, allowing us to define ``non-hydrodynamic'' operators in a precise way.

Charge conservation means that the operator $Q$ is invariant under the unitary part of time evolution. To ensure that this condition holds under each local gate, we require that the combination of operators $\mathbb{1} \otimes \sigma^{z} + \sigma^{z} \otimes \mathbb{1}$ is invariant under the action of the two-qubit gates. As a consequence of these constraints, long-wavelength modes like $Q_k \equiv \sum_i \sigma^{z}_{i} \exp(i k x)$ are almost stationary: $Q_k(t) = \exp(- D k^2 t) Q_k + O'$, where (under unitary dynamics) $O'$ consists of complex operators. Since unitary evolution conserves the Frobenius norm of an operator, $O'$ must exist: at each time step, the evolution ``emits'' non-conserved operators that eventually become large~\cite{Khemani_2018}. Absent dissipation, typical trajectories of these non-conserved operators grow indefinitely with the butterfly velocity; in the presence of dissipation, their growth is cut off by the same physics as RP resonances, and the size of the corresponding eigenoperators scales as $\sim D k^2/\gamma$. 

This observation applies much more broadly than just to the single-body modes $\sigma^{z}_{k}$. What matters is that any slowly varying diagonal charge profile, for example $\exp[\sum_x \mu(x) \sigma^z_x]$ for slowly varying $\mu(x)$, is only a weak source of off-diagonal operator weight. For two-local $U(1)$-conserving dynamics, one may think of operator growth as proceeding through an alternating cascade: diagonal charge inhomogeneities generate local current operators, and those current operators in turn generate new nearby diagonal inhomogeneities, which can then seed further growth. Schematically,
\[
\sigma^z _{x}\mathbb{1}_{x+1}\mathbb{1}_{x+2}
\to
(\sigma^+_{x}\sigma^-_{x+1}-\sigma^-_{x}\sigma^+_{x+1})\mathbb{1}_{x+2}
\to
\sigma^+_{x}\sigma^-_{x+1}\sigma^z_{x+2}
\]
illustrates this mechanism. The first step shows how an inhomogeneous diagonal charge profile couples to the current operator $j_{x,x+1}\propto i(\sigma_x^-\sigma_{x+1}^+ - \sigma_x^+\sigma_{x+1}^-)$, while the second shows how those locally off-diagonal operators seed new diagonal charge inhomogeneity. By contrast, a locally flat combination such as $\mathbb{1}_{x}\sigma^z_{x+1}+\sigma^z_{x}\mathbb{1}_{x+1}$ is just the conserved charge on a bond, so that there is no coupling to the current. The important point is that the leakage of operator weight from slowly varying charge profiles, which need not be exact microscopic eigenoperators, is gradient-controlled rather than $O(1)$.

\subsection{Non-hydrodynamic operators}

Having discussed operator growth in the neutral $\mathcal{Q}=0$ sector, we next ask how operator growth proceeds for locally off-diagonal operators such as $\sigma^{+}$. Here the naive expectation is different. A local coherence is not itself a smooth diagonal charge profile, and so does not obviously inherit the hydrodynamic bottleneck that makes such profiles slow. In a typical finite-entropy background it has available $O(1)$ local processes that generate additional nearby nontrivial operator content, which can then seed further growth. Repeated application of this mechanism suggests rapid spreading into increasingly complicated strings. This is precisely the operator-growth reasoning that would suggest simple exponential decay of correlators such as $\langle \sigma_x^+(t)\sigma_0^-(0)\rangle$.

This expectation fails because the above growth mechanism assumes a typical local background. However, as we already saw, the histories that contribute to correlation functions are always atypical ones~\cite{Nahum_22,SpectralGaps}. In this case, the dominant atypical histories are those in which the operator $\sigma^+$ is surrounded by a ``void,'' i.e., a region of charge density that is either close to zero or close to maximal. For concreteness we consider a void that is represented as a string of $P^\downarrow$ (or $P^\up$) projection operators, with a single $\sigma^+$ inside it. Inside a void, $\sigma^+$ creates a single particle, which remains a single particle under time evolution (since one cannot create a particle-hole excitation in the void). In terms of operator growth, this observation corresponds to the constraint that $P^\downarrow \otimes \sigma^+$ can only evolve into $P^\downarrow \otimes \sigma^+$ and $\sigma^+ \otimes P^\downarrow$, while strings of contiguous $P^\downarrow$ are invariant. Thus, a $\sigma^+$ surrounded by a void simply diffuses inside the void. 

\begin{figure}
    \centering
    \includegraphics[width=1.\linewidth]{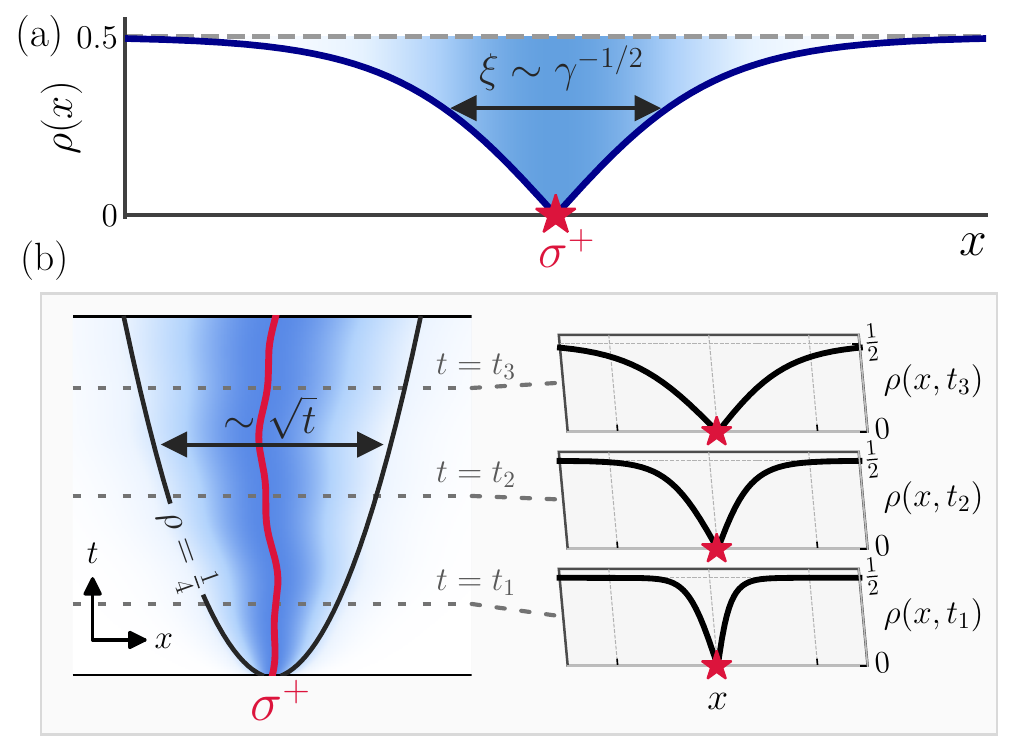}
    \caption{
\textbf{Schematic of low-entropy voids.} Voids support coherent single-particle/magnon evolution of $\sigma^+$.
(a) In the weak-noise regime, conditional on the survival of a local coherence (shown here as \(\sigma^+\)), the surrounding charge profile is depleted over a finite scale \(\xi\sim \gamma^{-1/2}\).
(b) In the noiseless case, the same diffusion-limited dephasing mechanism produces an aging void whose width grows diffusively in time, \(\xi(t)\sim \sqrt{t}\).
}
    \label{fig:slow-mode-cartoon}
\end{figure}

We now turn to the decay of a local correlation function like $\langle \sigma^+_0(t) \sigma^-_0(0) \rangle$ under unitary dynamics, following Ref. \cite{StretchedExp}.
One can write $\sigma^+(0) = P_l \sigma^+(0) P_r + O'$, where $P_l$ and $P_r$ project onto the $\downarrow$ state on regions of size $\ell \geq C \sqrt{D t}$ surrounding the coherence, with $C$ some large $O(1)$ number, and $O'$ is the rest of the operator.
The time evolution of $P_l \sigma^+(0) P_r$ is simple. By time $t$ a region of size $\sqrt{D t}$ from the center of the void is essentially fully polarized, so the initial coherence only sees polarized spins and wanders as a magnon without undergoing operator growth. Near the edge of the void, a region of width $\sqrt{D t}$ evolves nontrivially under hydrodynamics. Putting these things together, one finds that the piece of the operator $\langle U^\dagger_t P_l \sigma^+(0) P_r U_t \sigma^-(0)\rangle \sim \exp\big(-C\sqrt{t/t_0}\big)$. 
This is parametrically slower than the behavior expected in typical configurations, where the operator should decay exponentially. 
%We expand the initial operator in the basis $P_\uparrow, P_\downarrow, \sigma^+, \sigma^-$ inside the light cone. In a typical trajectory, the initial coherence is surrounded by a random background configuration of $P_\uparrow, P_\downarrow$. The operator growth in these trajectories is not constrained by the conservation law, so their weight is exponentially suppressed in time in the conditional ensemble. On the other hand, in trajectories where the coherence is initially surrounded by a void of size $\ell(t) \sim \sqrt{t}$, the coherence does not undergo operator growth until a time $\sim t$, when the void fills in from the outside. These trajectories are only suppressed by a probability cost $\sim \exp(-\sqrt{t/t_0})$ in the conditional state, and therefore dominate over typical trajectories. 
These considerations suggest that the long-time behavior of the correlation function is dominated by increasingly large voids (Fig.~\ref{fig:slow-mode-cartoon}).

%These considerations suggest 
%a distinct slow manifold in the $\mathcal Q=1$ sector. Shallow gradients make locally diagonal operators slow, while a well-polarized region suppresses the rapid operator growth of a local charge coherence $\sigma^+$. Combining the two yields a mechanism by which an off-diagonal operator can become parametrically slower than the naive expectation would suggest. 
%that the natural long-lived object is therefore not a bare $\sigma^+$, but a $\sigma^+$ dressed by a smooth polarized profile. The two cases relevant below --- a quasi-stationary screened void in the weak-noise regime and a diffusively growing void in the isolated limit --- are sketched in Fig.~\ref{fig:slow-mode-cartoon}.

\section{Macroscopic fluctuations with a single coherence}
\label{sec:slow-manifold}

In this section we provide a general heuristic argument relating the survival of local coherences to hydrodynamic large deviations, which we will then describe using macroscopic fluctuation theory (MFT)~\cite{Bertini_2015,Derrida_2025,PhysRevLett.129.040601,2026arXiv260102319S,10.21468/SciPostPhys.17.2.033}. This argument leads to the following simple conclusion: the survival probability of a local coherence that propagates along a trajectory $X(t)$ can be directly related to the probability of the optimal fluid fluctuation that supports a spacetime void along that trajectory. Thus, the nominally quantum-mechanical problem of the survival of local coherences can be fully re-expressed in the language of hydrodynamic large deviations. 

\subsection{Source manifold}

The previous two observations---that operator growth from a smooth density profile is gradient-limited, and that operator growth inside a locally void-like region is entropically suppressed so that \(\sigma^+\) evolves essentially as a single particle---suggest a natural slow manifold of charged operators. Concretely, one is led to a local coherence \(\sigma^+\) embedded in a smooth density profile that is void-like near the coherence. Such an operator is nowhere able to undergo rapid operator growth.

This construction naturally extends the familiar notion of a ``source'' manifold from the hydrodynamic sector ($\mathcal Q =0$) to charged operator sectors ($\mathcal Q \neq 0$). By a source we mean an operator, or more generally a subspace of operators, whose decay into more complicated operators is parametrically slower than typical operator growth. In the hydrodynamic sector, slowly varying density profiles provide the standard example: their leakage into generic operator space is gradient-limited, with rate \(\mathcal{O}(Dk^2)\) rather than \(O(1)\). The corresponding source manifold is just the familiar hydrodynamic slow modes controlling long-time density observables. When such a parametric separation of timescales exists, the remaining operator space may be regarded as an effectively Markovian reservoir for emitted operator weight, and the source manifold becomes a well-defined slow sector of the dynamics. For simple local observables, such as autocorrelation functions, the leading contribution then comes from operator histories that remain within or near this source manifold, since once operator weight has escaped into generic, highly extended operators it is unlikely to refocus onto a simple local operator~\cite{Khemani_2018,von_Keyserlingk_2022}.

Our key ansatz is that such sources for the evolution of operators in the $\mathcal{Q} = 1$ sector (i.e., single coherences) are made up of operators containing a single local coherence dressed by a smooth hydrodynamic density profile. Concretely, we consider operators of the form
\begin{equation}
\sigma_x^+ \exp\!\Big(-\sum_y \mu(y)\sigma_y^z\Big),
\end{equation}
with \(\mu(y)\) smooth on hydrodynamic scales and vanishing at large \(|x-y|\), so that the operator remains local at any finite time. The corresponding source manifold has two symmetry-related branches, according to whether the local density deformation is predominantly down- or up-polarized. For concreteness we work with the down-polarized branch. For simplicity, in this section we will first consider a semi-infinite chain on $[0, \infty)$ in which the coherence is artificially pinned to the leftmost site $x = 0$. The task is to find profiles $\mu(y,t)$ that minimize the probability that time evolution will take the system out of the source manifold.

Far from the coherence, operators in the source manifold evolve by diffusion of the hydrodynamic dressing profile. The nontrivial dynamics is local to the coherence itself, which sets the boundary condition at \(y=0\). Near that boundary, the slow operator is approximately a local coherence embedded in a predominantly down-polarized void, with only a low density of \(P^\uparrow\) excitations. In this regime it is more natural to think in quasiparticle terms than in terms of generic operator growth. For example, the correlator \(\langle \sigma^-_0(t)\sigma^+_0(0)\rangle\) may be viewed as an overlap between a ``bra'' and a ``ket'' history, one of which contains an extra particle pinned at the origin. Each time the particle at the origin collides with another particle, it picks up a random phase. These phases cancel destructively when one sums over collision histories. On the other hand, when the coherence at the origin is undergoing free evolution, it picks up a deterministic phase, which is the same across all histories of the rest of the fluid. In the full operator evolution, this loss of weight from the source manifold due to dephasing must be accompanied by emission into the larger space of generic, more complicated operators. (The autocorrelation function isolates only the survival of weight within the source manifold.) The effective dynamics within this manifold are therefore \emph{conditioned}: one retains precisely those histories in which the local coherence avoids collisions with any $P^\up$ excitation.

This conditioning has a direct consequence for the surrounding fluid. In the down-polarized branch, remaining within the source manifold requires suppressing precisely those histories in which a \(P^\uparrow\) reaches the coherence. Microscopically, this is equivalent to a local projection onto the branch-compatible state \(P^\downarrow\) on the two sites neighboring the coherence $\sigma^+$,
\begin{equation}
    A\otimes\sigma^+\otimes B \to (P^\down A P^\down) \otimes \sigma^+ \otimes (P^\down B P^\down).
\label{eq:branch-projection}
\end{equation}
The projection in Eq.~\eqref{eq:branch-projection} assumes that a single collision with a \(P^\uparrow\) completely dephases the local coherence. This assumption is realized in random unitary circuits and Brownian Hamiltonian dynamics, where different collision histories acquire independent random phases; in Sec.~\ref{RUCsection}, we derive this conditioned fluid dynamics explicitly for \(U(1)\)-conserving random circuits. For a fixed deterministic Hamiltonian or Floquet evolution, by contrast, there is no spacetime-random phase attached to each collision. Dephasing is then expected to build up gradually through interference among many collision histories. The corresponding conditioning is therefore better viewed as a soft local penalty on the fluid density near the coherence~\cite{StretchedExp}. (We return to this distinction in Sec.~\ref{sec:spectral-structure} when discussing translation-invariant systems.)

Although charge is conserved on every microscopic history, this postselection changes the expected charge of the surviving ensemble: pruning configurations with \(P^\uparrow\) adjacent to the coherence makes the coherence appear as an effective charge sink. For example, locally filtering a charge-uncertain state, schematically \(\mathbb{1}\to P^\downarrow\), changes the expected charge. If the evolution is subject to weak depolarizing noise, the absorption of expected charge at the coherence is balanced by injection from the noise. The conditioned fluid can then settle into a quasi-stationary void, and one is led to a long-lived mode of the dissipative dynamics built from a source operator of fixed shape together with an emitted tail in the larger space of generic operators. In the absence of noise, by contrast, there is no compensating injection: the coherence is instead surrounded by an increasingly large void at late times.

In both cases, leakage out of the source manifold is parametrically slow. Without noise, the void broadens diffusively, and the rate at which histories are pruned at the coherence decays as \(t^{-1/2}\) at large times; equivalently, the emission rate out of the source manifold is asymptotically \(\mathcal{O}(t^{-1/2})\). This gives the aforementioned stretched-exponential decay of the charged autocorrelator, \(\mathcal G(0,t)\equiv \langle \sigma_0^+(t)\sigma_0^-(0)\rangle \sim \exp[-O(\sqrt{t})]\). With weak depolarizing noise, by contrast, the void growth is cut off at \(t\sim \gamma^{-1}\), leaving a stationary void of width \(\xi\sim \gamma^{-1/2}\). The corresponding pruning rate, and equivalently the emission rate out of the source manifold, is then of order \(\sqrt{\gamma}\). Thus, both in the weak-noise limit \(\gamma\ll 1\), and in the noiseless case at late times \(t\gg 1\), the leakage out of the source manifold is parametrically small. The conditioned fluid profile in both the noiseless and weak noise cases is schematically shown in Fig.~\ref{fig:slow-mode-cartoon}.

In the weak-noise problem, this source manifold therefore supports genuine long-lived eigenoperators of the dissipative dynamics, by the same general logic as in the hydrodynamic sector. The corresponding eigenoperator consists of a slow source together with an emitted tail in the larger space of generic, more complicated operators~\cite{Khemani_2018}. Weak depolarizing noise dissipates this large-operator tail, while the source manifold feeds weight into it. The resulting balance produces an eigenmode whose decay rate is set not by an \(O(1)\) microscopic timescale, but by the parametrically slow emission out of the source manifold. In this sense, the source manifold also controls the associated Ruelle--Pollicott resonance structure.

\subsection{Effective continuum description}
\label{subsec:effective_continuum_description}

As described above, in the down-polarized branch the effective source-manifold dynamics is conditioned by pruning precisely those histories in which a \(P^\uparrow\) reaches the coherence, equivalently, by the local projection in Eq.~\eqref{eq:branch-projection} acting on the sites adjacent to the coherence. Implementing this projection directly is too singular for a continuum description. A more convenient but equivalent formulation acts one update earlier, on the precursor configurations that would be removed at the next step.

Concretely, on either side of the coherence, a \(P^\uparrow\) on the next site outward is such a precursor: at the next bond update it has a fixed probability to hop inward, only then to be pruned by Eq.~\eqref{eq:branch-projection}. Rather than discarding histories only after this happens, one may equivalently assign a cost directly to the precursor occupation itself. This reformulation is better suited to coarse-graining, because in the continuum the precursor occupation becomes the outward density gradient at the coherence, as we now show.

At hydrodynamic scales, the diagonal dressing around the coherence is described by a density field \(\rho(x,t)\) measuring excitations above the chosen polarized background. For the down-polarized branch we identify
\begin{equation}
\rho(x,t)\equiv \langle P^\uparrow_x(t)\rangle .
\end{equation}
Away from the coherence, \(\rho\) evolves according to diffusive fluctuating hydrodynamics,
\begin{equation}
\partial_t\rho=-\partial_x j,
\qquad
j=-D(\rho)\partial_x\rho+\sqrt{\sigma(\rho)}\,\eta,
\label{eq:MFT_current}
\end{equation}
where \(\eta\) is unit white noise. In the continuum, the projection in Eq.~\eqref{eq:branch-projection} on the site adjacent to the coherence becomes a point Dirichlet condition at the coherence position \(X(t)\),
\begin{equation}
\rho(X(t),t)=0,
\label{eq:rho-Dirichlet-1rep-full}
\end{equation}
up to an irrelevant lattice-scale offset.

With this boundary condition, a gradient expansion identifies the precursor occupations on either side with the outward density gradients at the coherence. Using Eq.~\eqref{eq:rho-Dirichlet-1rep-full}, we have
\begin{equation}
\rho(X(t)^\pm \pm a,t)\simeq \pm \,a\,\partial_x\rho(X(t)^\pm,t),
\end{equation}
with \(a\) the lattice spacing (we now assume $a=1$). Thus the microscopic penalty on precursor-site occupancy becomes a bias on the outward density gradients at the defect. For a prescribed trajectory $X(t)$ this is encoded by the tilted fluid measure
\begin{center}
\refstepcounter{equation}\label{eq:gradient_tilt}
\begin{wideeqbox}
\centering
\(
\displaystyle
\mathcal Z_T[X]\equiv \Big\langle e^{-S_{\rm bdy}[\rho,X]} \Big\rangle,
\)

\vspace{6pt}

\(
\displaystyle
S_{\rm bdy}[\rho,X]
=
s\int_0^T \!dt
\Big(
\partial_x\rho(X(t)^+\!,t)-\partial_x\rho(X(t)^-\!,t)
\Big).
\)
\end{wideeqbox}

\makebox[\linewidth][r]{(\theequation)}
\end{center}
and where $\langle\cdots\rangle$ denotes the untilted fluid measure. Tilted measures are a standard device in large-deviation theory~\cite{Doob1957,Doob1984,PhysRevLett.98.195702,Garrahan_2009,PhysRevLett.111.120601,PhysRevLett.95.010601,Chetrite2015,2010PThPS.184..304J,PhysRevE.81.011111,PhysRevE.81.011110,Touchette2013,Gopalakrishnan2026,DerridaI2019,DerridaII2019}: the tilted measure reweights each spacetime history by the value of the boundary action $S_{\rm bdy}$ along that history, which in this case has the effect of suppressing trajectories that underwent operator growth and left the source. The full counting statistics of current at the origin can be retrieved by taking derivatives with respect to the tilting field $s$ at $s = 0$. In our model, we are interested in specific large deviations, so $s$ has a finite value. For the $U(1)$-conserving random circuit models that we benchmark below, the precursor occupation on either side of the coherence is removed with probability $1/2$ when the coherence bond is updated. In our time normalization this fixes $s=1/2$, as discussed in more detail in Sec.~\ref{RUCsection}.

Thus the local quantum coherence enters the continuum fluid description only through boundary data at $X(t)$: branch projection at the coherence imposes the point condition \eqref{eq:rho-Dirichlet-1rep-full}, while precursor occupations are suppressed by a bias on the outward diffusive fluxes that attempt to refill it. The resulting effective dynamics are therefore those of a diffusive fluid conditioned at a moving charged defect.

Within the slow manifold, the coherence itself undergoes only single-particle propagation through the polarized background. For a fixed microscopic realization of the dynamics, the corresponding propagator may be written as a sum over single particle trajectories $X(t)$,
\begin{equation}
K_T(x_f,x_i)
=
\sum_{X:\,x_i\to x_f}
\mathcal A[X]
=
\sum_{X:\,x_i\to x_f}
\sqrt{\mathbb P[X]}\,e^{i\Phi[X]}.
\label{eq:coh_pathsum_general}
\end{equation}
Here $\mathbb P[X] \propto \exp(-S_X[X])$ is the bare propagation weight assigned to the path $X$ within the slow manifold, while $\Phi[X]$ is the phase accumulated along the path. In deterministic settings $\Phi[X]$ is an ordinary path phase functional, while in random circuits it becomes random and realization-dependent. %This form therefore accommodates both deterministic and random coherent propagation.

In the next subsection we encode these two ingredients---the fluid point conditioning at $X(t)$ and the single-particle evolution of $\sigma^+$---in a hydrodynamic large-deviation functional for the coupled fluid--coherence dynamics.

\subsection{Macroscopic fluctuation theory with a local quantum coherence}
\label{subsec:hydro_ld_local_coherence}
We now represent the tilted fluid measure Eq.~\eqref{eq:gradient_tilt} in the Hamiltonian formulation of macroscopic fluctuation theory. The auxiliary field \(\pi(x,t)\) is the MFT response field conjugate to the density \(\rho(x,t)\). Equivalently, it is the counting field that biases the local current and reaction events in a density history, and thereby encodes fluid large deviations. For a given trajectory $X(t)$, the fluid weight is
\vspace{4pt}
\refstepcounter{equation}\label{eq:Z1rep_MFT_prevar}
\begin{wideeqbox}
\vspace{4pt}
\noindent
\makebox[\linewidth][l]{%
  \makebox[3.5em][l]{}%
  \makebox[\dimexpr\linewidth-8em\relax][c]{%
    $\mathcal Z_T[X]
=
\int_{\rho(X(t),t)=0}
\mathcal D\rho\,\mathcal D\pi\;
e^{-S_{\rm bulk}[\rho,\pi]-S_{\rm bdy}[\rho;X]},
    $%
  }%
  \makebox[4.5em][r]{(\theequation)}%
\vspace{4pt}
}
\end{wideeqbox}
\vspace{4pt}
\noindent where the point condition enforces the branch projection at the coherence position. For weak-depolarizing noise strength $\gamma$, the conservative dynamics are supplemented by a weak Poisson bath. At the level of fluctuating hydrodynamics, this promotes the continuity equation to a conservative-plus-reactive Langevin evolution. In the Hamiltonian formulation of macroscopic fluctuation theory~\cite{Bertini_2015,Derrida_2025,PhysRevLett.129.040601,2026arXiv260102319S,10.21468/SciPostPhys.17.2.033}, the corresponding bulk fluid action is
\begin{align}
S_{\rm bulk}[\rho,\pi]
=
\int_0^T &dt \int dx\,
\Big[
\pi\,\partial_t\rho
+ D(\rho)\,\partial_x\pi\,\partial_x\rho \qquad \nonumber\\
&\ -\frac{\sigma(\rho)}{2}(\partial_x\pi)^2
-\gamma\,\mathcal H_{\rm bath}(\rho,\pi)
\Big],
\label{eq:Sbulk_1rep_general}
\end{align}
with
\begin{equation}
\mathcal H_{\rm bath}(\rho,\pi)
=
\frac12\Big[
\rho\,(e^{-\pi}-1)
+
(1-\rho)\,(e^{\pi}-1)
\Big],
\label{eq:Hbath_1rep_general}
\end{equation}
while the coherence-induced boundary tilt is as given in Eq.~\eqref{eq:gradient_tilt}. For $\gamma=0$ the bath term is absent. Throughout, the spatial integral in Eq.~\eqref{eq:Sbulk_1rep_general} is understood as the sum of the two half-line integrals on either side of the coherence position $X(t)$.

The full path sum representation of the charged raising-operator correlator
\(
\langle \sigma^+_{x_t}(t)\sigma^-_{x_0}(0)\rangle
\) is then
\vspace{2pt}
\refstepcounter{equation}\label{eq:G1rep_full_general}
\begin{wideeqbox}
\vspace{4pt}
\noindent
\makebox[\linewidth][l]{%
  \makebox[3em][l]{}%
  \makebox[\dimexpr\linewidth-8em\relax][c]{%
    $\displaystyle
    \mathcal G(x_t,x_0;t)
    =
    \sum_{X:\,x_0\to x_t}
    \sqrt{\mathbb P[X]}\,e^{i\Phi[X]}\,
    \mathcal Z_t[X].
    $%
  }%
  \makebox[5em][r]{(\theequation)}%
}
\end{wideeqbox}
\noindent where the amplitude $\sqrt{\mathbb P[X]}$ and phase $\Phi[X]$ are defined by a model-specific single-particle propagator (Eq.~\eqref{eq:coh_pathsum_general}).

Equation~\eqref{eq:G1rep_full_general} makes the structure of the effective theory particularly transparent. It is intrinsically hybrid. The coherence itself propagates quantum mechanically, through the single-body path sum over $X(t)$ and its associated phase functional $\Phi[X]$. The low-entropy void that supports this coherent quantum propagation is itself a rare fluctuation of a classical fluid. In this sense, long-lived charged dynamics are built by stitching together two ingredients that would ordinarily be discussed separately: coherent quantum evolution in a low-entropy region, and a classical hydrodynamic large deviation that creates and sustains that region inside a finite-entropy bulk.

A further simplification occurs at the level of the macroscopic variational problem. The boundary tilt $S_{\rm bdy}$ in Eq.~\eqref{eq:gradient_tilt} may be traded for a Dirichlet condition on the conjugate field. To see this, one splits the bulk action into the two half-lines adjacent to $X(t)$ and varies the boundary gradients $\partial_x\rho(X(t)^\pm,t)$. The mixed bulk term then produces the boundary contribution
\begin{align}
\int_0^T dt\,
\Big[&
\pi(X(t)^+,t)\,\partial_x\rho(X(t)^+,t)\nonumber\\
& -
\pi(X(t)^-,t)\,\partial_x\rho(X(t)^-,t)
\Big].
\label{eq:boundary_term_general}
\end{align}
Combining this with Eq.~\eqref{eq:gradient_tilt}, stationarity with respect to the boundary gradients yields
\begin{equation}
\pi(X(t)^+,t)=\pi(X(t)^-,t)=-s.
\label{eq:pi_Dirichlet_1rep_general}
\end{equation}
Thus, in the MFT saddle-point problem, the fluid partition sum may be written in the simpler equivalent form
\begin{equation}
\mathcal Z_T[X]
=
\int_{\substack{\rho(X(t),t)=0\\[2pt]\pi(X(t),t)=-s}}
\mathcal D\rho\,\mathcal D\pi\;
e^{-S_{\rm bulk}[\rho,\pi]}.
\label{eq:Z1rep_MFT_dirichlet}
\end{equation}

Equation~\eqref{eq:Z1rep_MFT_dirichlet} is the central continuum description that we will use below. The local coherence enters only through boundary data at the defect: $\rho(X(t),t)=0$ enforces the empty void core, while $\pi(X(t)^\pm,t)=-s$ encodes the conditioning cost associated with pruning histories in which charge reaches the coherence. The distinction between the weak-noise and noiseless problems is only in the bulk dynamics (apart from the trivial exponential decay at rate $\gamma$ from depolarization events directly at the coherence position). For $\gamma>0$, the depolarizing bath breaks the conservative gauge freedom of $\pi$ and admits a stationary saddle solution. For $\gamma=0$, the bulk is purely conservative, and the nontrivial saddle is instead enforced by the terminal condition provided by the charged correlator $\langle \sigma^+_x(t)\sigma^-_0(0)\rangle$.

%\subsection{Roadmap}

%\ewan{Need to explain that we show the optimal voids first at stationary coherence $X(t)=0$, and that we later show that the motion of the coherence is a subleading effect. Also need to explain that a curious reader should read the microscopic derivation of the hybrid action above, but that the main results for the optimal coarse-grained fluid profile follow from the above action obtained on general grounds.}
%\jaj{I believe this comment is quite out of date.}
%\ewan{It is.}

\section{Optimal density fluctuations from hydrodynamic large deviations}
\label{sec:optimal_void_profiles}

We now use the continuum description developed in Sec.~\ref{sec:slow-manifold} to determine the optimal void profile in the two regimes of interest: weak noise $\gamma\ll 1$ and the strictly zero noise limit $\gamma=0$. %\sg{For the purposes of this section we will ignore the kinetic energy of the coherence inside the void; we will reinstate this effect later and check that it does not affect the conclusions in this section}. 
In both cases we first hold the coherence fixed and solve for the optimal conditioned fluid configuration. This isolates the leading hydrodynamic structure of the void itself; the motion of the coherence will be restored later and will be shown to be a subleading effect. The local coherence then enters throughout only through the defect boundary data in Eq.~\eqref{eq:Z1rep_MFT_dirichlet}. What distinguishes the two regimes is how the bulk and far-field conditions select the saddle: with weak depolarizing noise the optimal void is stationary, whereas in the noiseless problem it is aging.

\subsection{Weak-noise stationary saddle}
\label{subsec:weak_noise_stationary_saddle}

We first consider weak depolarizing noise, $\gamma>0$, for which the optimal void is stationary. We hold the coherence fixed at the origin, $X(t)=0$. The full-line problem then decouples into two identical half-line problems, so it is enough to work on $x\ge 0$ and reconstruct the full profile by reflection. We further specialize the action Eq.~\eqref{eq:Z1rep_MFT_dirichlet} to diffusive charge transport with the symmetric simple exclusion process (SSEP) diffusivity and mobility~\cite{Bertini_2015,PhysRevLett.129.040601},
\begin{equation}
D(\rho)=D,
\qquad
\sigma(\rho)=2D\rho(1-\rho).
\label{eq:weak_noise_SSEP_data}
\end{equation}
%%%
This choice is minimal among transport models with bounded and nonzero diffusion constants across the full range of densities and for spin-$\tfrac12$/hard-core exclusion systems, in which the local occupancy is Bernoulli in equilibrium. %(so the static susceptibility vanishes at full and empty polarization and takes the form $\sigma(\rho)=2\rho(1-\rho)$). 
Additional nonlinear density dependence in $\sigma $ and $D$ can then modify the scaling functions in the MFT saddle equations, but not the underlying scaling variables. Therefore, we expect the saddle point scaling exhibited by this model will remain robust to the addition of nonlinearities.
%%%
The bulk action is as given in Eq.~\eqref{eq:Sbulk_1rep_general} (restricted to the half-line), while the coherence imposes the boundary conditions 
\begin{equation}
    \rho(0,t)=0,\quad \pi(0,t)=-s.
\label{eq:defect-bdy-conditions}
\end{equation}

In the far field we require the unforced bulk fixed point. In a purely conservative problem this would amount only to $\partial_x\pi\to 0$, reflecting the gauge freedom $\pi\to \pi+c$. The weak depolarizing bath breaks this symmetry: the bulk Hamiltonian depends explicitly on $\pi$, so any nonzero constant $\pi$ biases the local reset dynamics and therefore forces the density field. The correct far-field condition is thus
\begin{equation}
\rho(\infty,t)=\frac12,
\qquad
\pi(\infty,t)=0.
\label{eq:weak_noise_bc_farfield}
\end{equation}

In the quasi-stationary regime we seek a time-translation invariant saddle,
\begin{equation}
\rho(x,t)\equiv \rho_\star(x),
\qquad
\pi(x,t)\equiv \pi_\star(x).
\label{eq:weak_noise_stationary_ansatz}
\end{equation}
Extremizing Eq.~\eqref{eq:Sbulk_1rep_general} gives the stationary saddle equations
\begin{align}
0&=
D\,\partial_x^2\rho_\star
-\partial_x\big(\sigma(\rho_\star)\partial_x\pi_\star\big)
+\gamma\,\partial_\pi \mathcal H_{\rm bath}(\rho_\star,\pi_\star),
\label{eq:weak_noise_stat_rho_phys}\\
0&=
-D\partial_x^2\pi_\star
-\frac{\sigma'(\rho_\star)}{2}(\partial_x\pi_\star)^2
-\gamma\partial_\rho \mathcal H_{\rm bath}(\rho_\star,\pi_\star).
\label{eq:weak_noise_stat_pi_phys}
\end{align}
The weak-noise scaling follows by introducing the rescaled coordinate $z\equiv \sqrt{\gamma/{D}}x$ and writing $\hat\rho(z)\equiv \rho_\star(x)$, $\hat\pi(z)\equiv \pi_\star(x)$. The stationary saddle then obeys a dimensionless boundary-value problem
\begin{align}
\hat\rho'' &=
\big(2\hat\rho(1-\hat\rho)\hat\pi'\big)'
-\frac12\Big[(1-\hat\rho)e^{\hat\pi}-\hat\rho e^{-\hat\pi}\Big],
\label{eq:weak_noise_dimless_rho}\\
\hat\pi'' &=
\sinh(\hat\pi)
-\frac12\partial_{\hat\rho}\!\big[2\hat\rho(1-\hat\rho)\big](\hat\pi')^2,
\label{eq:weak_noise_dimless_pi}
\end{align}
with
\begin{equation}
\hat\rho(0)=0,
\ \
\hat\pi(0)=-s,
\ \
\hat\rho(\infty)=\tfrac12,
\ \
\hat\pi(\infty)=0.
\label{eq:weak_noise_dimless_bc}
\end{equation}
Thus the weak-noise problem reduces to a dimensionless boundary-value problem in the scaled coordinate $z=x/\xi$, with
\begin{equation}
\xi\sim \sqrt{D/{\gamma}}.
\label{eq:weak_noise_xi_scaling}
\end{equation}

The far-field structure is fixed by linearizing Eqs.~\eqref{eq:weak_noise_dimless_rho}--\eqref{eq:weak_noise_dimless_pi} about the unforced fixed point $(\hat\rho,\hat\pi)=(\tfrac12,0)$. One finds $\delta\hat\rho(z)\equiv \hat\rho(z)-\tfrac12 \sim e^{-z}$, and $\hat\pi(z)\sim e^{-z}$, as $z\to\infty$. In physical units, the optimal void therefore approaches equilibrium exponentially, $|\rho_\star(x)-\tfrac12| \sim e^{-x/\xi}$.

For large $T$, the corresponding conditioned fluid functional has the quasi-stationary large-deviation form
\begin{equation}
\mathcal Z_T(s)\asymp e^{-T\Lambda(s,\gamma)},
\label{eq:weak_noise_partition_sum}
\end{equation}
where $\Lambda(s,\gamma)$ is the principal eigenvalue of the tilted fluid evolution operator~\cite{PhysRevLett.95.010601,Lebowitz1999}. Evaluating the fluid action $S_{\rm bulk}$ on the stationary saddle gives
\begin{equation}
\Lambda(s,\gamma)
=
\sqrt{D\gamma}\,\hat\Lambda(s),
\label{eq:weak_noise_Lambda_scaling}
\end{equation}
with dimensionless rate function
\begin{equation}
\hat\Lambda(s)
=
\int_0^\infty dz\,
\Big[
\hat\pi'\hat\rho'
-\hat\rho(1-\hat\rho)(\hat\pi')^2
-\mathcal H_{\rm bath}(\hat\rho,\hat\pi)
\Big].
\label{eq:weak_noise_hatLambda}
\end{equation}
Thus weak noise selects a stationary optimal void of width $\xi\sim \sqrt{D/\gamma}$, with an extensive-in-time conditioning cost set by the same scale, $\Lambda(s,\gamma)\sim \xi^{-1}\sim \sqrt{{\gamma}/{D}}$.

\subsection{Noiseless aging saddle}
\label{subsec:noiseless_aging_saddle}

We now turn to the strictly noiseless limit, $\gamma=0$, and to the charged correlator itself, $\langle \sigma_x^+(t)\sigma_0^-(0)\rangle$. In the continuum description of Sec.~\ref{sec:slow-manifold}, this correlator is given by the partition sum Eq.~\eqref{eq:G1rep_full_general} over coherence trajectories $X(t)$ and conditioned fluid histories. As in the weak-noise problem, however, the leading hydrodynamic large-deviation structure is controlled by the void rather than by the motion of the coherence inside it. The optimal fluid profile and leading decay are therefore determined by the pinned-coherence problem, so we set $X(t)=0$ for the remainder of this subsection and return to the mobile-coherence problem in Sec.~\ref{sec:polaron-dynamics}. 

In the absence of bulk depolarization there is no intrinsic screening length and no quasi-stationary balance: the optimal void broadens in time rather than saturating, so the saddle is aging rather than stationary. As in the weak-noise case, we work on $x\ge 0$. The conservative SSEP bulk saddle equations are~\cite{PhysRevLett.129.040601}
\begin{align}
\partial_t \rho
&=
\partial_x\Big(D\,\partial_x \rho-\sigma(\rho)\,\partial_x \pi\Big),
\label{eq:noiseless_ham_rho}\\
\partial_t \pi
&=
- D\,\partial_x^2 \pi-\frac{1}{2}\sigma'(\rho)\,(\partial_x \pi)^2,
\label{eq:noiseless_ham_pi}
\end{align}
with diffusivity and mobility as given in Eq.~\eqref{eq:weak_noise_SSEP_data}, and the same defect boundary conditions $\rho(0,t)=0$ and $\pi(0,t)=-s$ as in the weak noise case.

Unlike the weak-noise case, the far field no longer fixes $\pi(\infty,t)$ itself: in the absence of a bath the conservative gauge freedom $\pi\to\pi+c$ is restored, so only gradients of $\pi$ enter the bulk dynamics. The far-field conditions are therefore $\rho(\infty,t)=\frac12$ and $\partial_x\pi(\infty,t)=0$. The nontrivial saddle is then selected by the two-time conditions appropriate to the charged correlator. The initial state is the infinite-temperature ensemble, while the final-time contraction fixes the coherence position but sums uniformly over the surrounding fluid configurations. In Hamiltonian MFT, these become $\rho(x,0)=\frac12$ and $\pi(x,T)=0$.

For present purposes it is useful to revert briefly to the equivalent gradient-tilt formulation of Sec.~\ref{subsec:effective_continuum_description}, since this makes contact with known current large-deviation results transparent. The optimal current is
\begin{equation}
j(x,t)= -D\,\partial_x\rho(x,t)+\sigma(\rho(x,t))\,\partial_x\pi(x,t).
\label{eq:noiseless_current}
\end{equation}
At the coherence, however, $\rho(0,t)=0$ forces $\sigma(\rho(0,t))=0$, so the boundary current reduces to
\begin{equation}
j(0,t)=-D\,\partial_x\rho(0^+,t).
\label{eq:noiseless_boundary_current}
\end{equation}
Thus, at the level of the conditioned ensemble, the boundary-gradient bias in Eq.~\eqref{eq:gradient_tilt} is equivalent to a bias on the integrated current into a zero-density reservoir at the defect.

\begin{figure*}
    \centering
    \includegraphics[width=\textwidth]{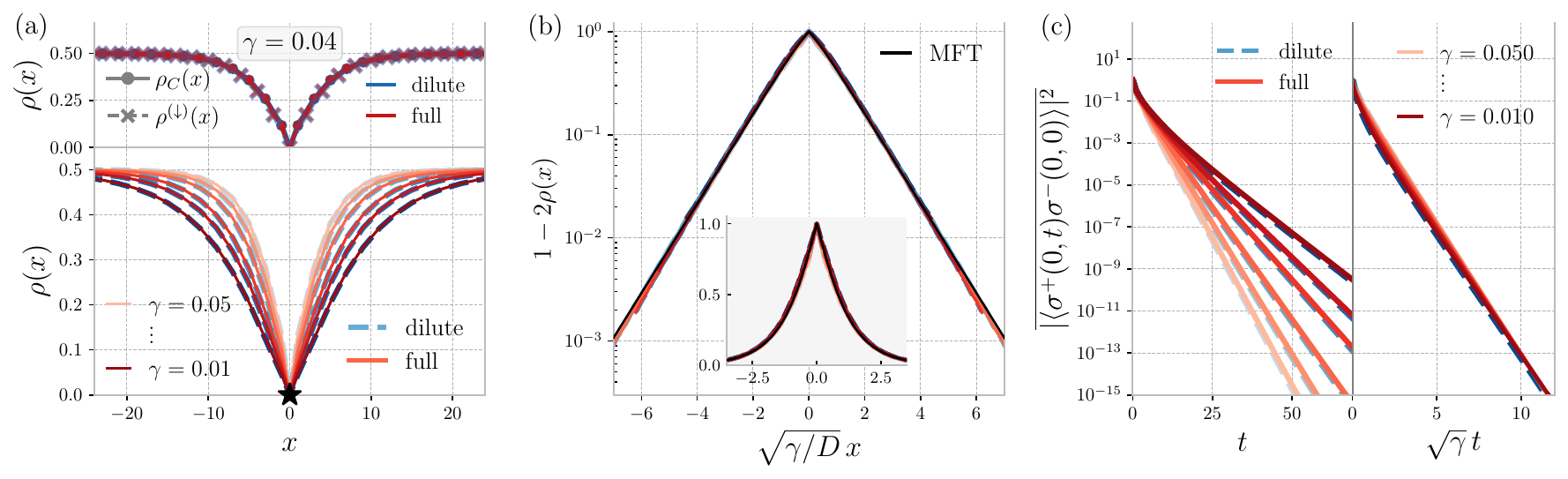}
    \caption{
\textbf{Weak-noise stationary void}. 
(a) Top: branch-resolved density profile in the down-polarized branch, \(\rho^{(\downarrow)}(x)\), compared with the profile \(\rho_C(x)\) reconstructed from the particle-hole-symmetric diagnostic \(C(x,t)\approx (1-2\rho(x,t))^2\) (Eq.~\eqref{eq:conditional-profile}), shown here for \(\gamma=0.04\). Their agreement confirms that \(C(x,t)\) faithfully captures the branch-resolved stationary void profile. Bottom: stationary density profiles for \(\gamma\in\{0.01,\cdots,0.05\}\). The close agreement between the full two-replica statistical mechanics model (``full'') and the dilute-coherence approximation (``dilute'') shows that the latter reproduces the same conditioned hydrodynamic voids.
(b) Collapse of the stationary void profiles under the rescaling \(x\to x/\xi\), with \(\xi\sim \sqrt{D/\gamma}\), together with the numerical solution of the stationary MFT saddle equations~\eqref{eq:weak_noise_dimless_rho}-\eqref{eq:weak_noise_dimless_bc} (with \(s=\tfrac12\) for \(U(1)\)-conserving RUCs, see Sec.~\ref{subsec:selection-reset-current}).
(c) Decay of the phase-averaged norm-squared coherence Green's function \(\overline{|\mathcal G(0,t)|^2}\). After transients, the decay becomes purely exponential, \(\sim e^{-\Lambda t}\), demonstrating entrance into the quasi-stationary regime governed by a single dominant eigenvalue, with \(\Lambda\propto \sqrt{\gamma}\). Numerical data are obtained from TEBD transfer matrix evolution~\cite{tenpy2024} of the random-circuit two-replica statistical mechanics model (Sec.~\ref{RUCsection}). These simulations used an open chain of length \(L=60\) with bond dimension \(\chi=500\), initialized as \( |(\sigma^+\!\otimes \sigma^-)_{0}\rangle \), and evolved up to times \(t\le 120\) for \(\gamma\in\{0.01,\,0.015,\,0.02,\,0.03,\,0.04,\,0.05\}\); the quasi-stationary profiles shown in (a,b) are evaluated at \(t=120\), chosen to lie in the quasi-stationary regime (the onset of which is indicated in panel (c) by the crossover to a single-exponential decay).
}
    \label{fig:screened-void-profile}
\end{figure*}

The finite-time noiseless saddle problem therefore coincides with the current large-deviation problem studied by Grabsch \textit{et al.}~\cite{Grabsch} for a semi-infinite SSEP with reservoir upon choosing reservoir density $\rho_L=0$ and bulk density $\bar\rho=\tfrac12$~\footnote{Microscopically, the coherence does not literally absorb charge: each individual history still conserves charge. The effective sink is an ensemble-level effect: starting from an infinite-temperature initial ensemble, conditioning reweights fluid histories and can therefore change the ensemble-averaged charge without violating charge conservation on any history.}; see also Ref.~\cite{2024arXiv240500654S}. The large-time cumulant generating function is therefore
\begin{equation}
\Big\langle e^{-sN(T)}\Big\rangle
\asymp
e^{-\sqrt{T}\hat\Lambda_{0}(s)},
\ \
\hat\Lambda_{0}(s)
=
\frac{1}{2\sqrt{\pi}}\,
{\rm Li}_{3/2}\!\left(1-e^{-2s}\right),
\label{eq:noiseless_grabsch_cgf}
\end{equation}
where $N(T)\equiv\int_0^T dt\, \partial_x\rho(X(t)^+,t)$ is the integrated number of boundary-filtering events on the half-line. The optimal final-time density profile has the diffusive similarity form
\begin{equation}
\rho(x,T)\approx \Phi(u),
\qquad
u={x}/{\sqrt{DT}},
\label{eq:noiseless_similarity_form}
\end{equation}
where $\Phi$ is reconstructed from the solution of the integral equation~\cite{Grabsch}
\begin{gather}
\Omega(u)+\int_0^\infty dz\,\Omega(z)\Omega(u+z)
=
(e^{-2s}-1)\frac{e^{-u^2/4}}{\sqrt{4\pi}},
\label{eq:noiseless_Omega_integral_eq}\\
\Phi(u)
=
\frac12\,
\frac{\int_0^u dz\,\Omega(z)}{\int_0^\infty dz\,\Omega(z)}.
\label{eq:noiseless_Phi_from_Omega}
\end{gather}
Thus the optimal void broadens diffusively on the scale
\begin{equation}
\xi_{\gamma=0}(T)\sim \sqrt{DT},
\label{eq:noiseless_void_broadening}
\end{equation}
and the fluid partition sum obeys $\mathcal Z_T(s)\asymp e^{-\sqrt{T}\hat\Lambda_0(s)}$. Consequently, the charged correlator has the leading asymptotic form
\begin{equation}
\langle \sigma_x^+(T)\sigma_0^-(0)\rangle
\sim
e^{-\sqrt{T}\hat\Lambda_0(s)},
\label{eq:noiseless_correlator_scaling}
\end{equation}
which is the hydrodynamic large-deviation origin of the stretched-exponential decay found previously in Ref.~\cite{StretchedExp}.

\subsection{Numerical evidence}
\label{subsec:optimal_void_numerics}

To test these predicted optimal fluid density profiles we use $U(1)$-conserving random unitary circuits, which provide a convenient setting in which to study charged operator slow modes. Their coarse-grained charge dynamics are known to be diffusive, with the circuit-averaged hydrodynamics described by SSEP, and charge-transport fluctuations in typical circuits approach the corresponding SSEP large-deviation form at long times~\cite{Rakovszky2018,McCulloch2023}. We therefore assume that the conditioned void profile is self-averaging on hydrodynamic scales (up to the choice of void polarization).

Because up- and down-polarized voids protect the coherence equally well, the long-lived conditioned state is particle-hole symmetric, so any probe linear in the half-filling deviation $\rho-\tfrac12$ averages to zero. We therefore diagnose the void using an observable that is even under the particle-hole symmetry, namely the circuit-averaged, coherence-conditioned second moment of the local half-filling deviation:
\begin{equation}
C(x,t)
=
\frac{
\mathbb E_U\!\left[
\left|\langle \sigma_0^+(t)\,\sigma^z_x\,\sigma_0^-(0)\rangle_U\right|^2
\right]
}{
\mathbb E_U\!\left[
\left|\langle \sigma_0^+(t)\,\sigma_0^-(0)\rangle_U\right|^2
\right]
}.
\label{eq:conditional-profile}
\end{equation}
The conditioned local polarization profile is the same (on hydrodynamic scales) for typical circuits up to the random choice of void branch. Thus for a typical circuit $U$ one has
$\left|\langle \sigma_0^+(t) Z_x \sigma_0^-(0)\rangle_U\right|
\approx
\left|\langle \sigma_0^+(t)\sigma_0^-(0)\rangle_U\right|\,|m_{\rm typ}(x,t)|$, so the circuit-dependent coherence-survival factor cancels in Eq.~\eqref{eq:conditional-profile}, yielding
\begin{equation}
C(x,t)\approx |m_{\rm typ}(x,t)|^2 \approx (1-2\rho(x,t))^2.
\label{eq:rhoC_def}
\end{equation}

Numerically, $C(x,t)$ is evaluated efficiently by time-evolving block decimation (TEBD)~\cite{PhysRevLett.93.040502,PhysRevLett.93.207204,PhysRevLett.93.076401,tenpy2024} evolution of the circuit-averaged doubled transfer matrix $\mathbb T$ acting on the two-copy operator state $|\Psi(0)\rangle \equiv |(\sigma^+\!\otimes \sigma^-)_{0}\rangle$. Equivalently,
\begin{equation}
C(x,t)
=
\frac{\langle (\sigma^+\!\otimes \sigma^-)_{0}(Z\!\otimes\! Z)_x \,|\, \Psi(t)\rangle}
{\langle (\sigma^+\!\otimes \sigma^-)_{0} \,|\, \Psi(t)\rangle},
\label{eq:C_doubled_overlap}
\end{equation}
with $|\Psi(t)\rangle=\mathbb T^{\,t}|\Psi(0)\rangle$. We defer the explicit microscopic construction of $\mathbb T$ to Sec.~\ref{RUCsection}.

The same two-replica evolution also gives access to a direct branch-resolved probe. To select the down-polarized branch, we project one replica onto $P^\downarrow$ on every site away from the coherence at site $0$ and measure the charge density in the other at each $x$. Cross terms between the up- and down-void branches vanish under circuit averaging, since the two branches acquire independent random phases. The resulting profile, denoted $\rho^{(\downarrow)}(x,t)$, therefore directly probes the down-branch void profile.

\textbf{Weak noise.}
We test the weak-noise stationary saddle by simulating $C(x,t)$ using TEBD with bond dimension $\chi=500$ on an open chain of length $L=60$~\cite{tenpy2024}. We initialize the coherence at the center of the chain, $|\Psi(0)\rangle\equiv |(\sigma^+\!\otimes \sigma^-)_{0}\rangle$, with the remaining sites at infinite temperature, and evolve up to times $t\le 120$ for various noise strengths.

Figure~\ref{fig:screened-void-profile}(a) compares the direct branch-selected profile $\rho^{(\downarrow)}(x)$ with the reconstructed profile $\rho_C(x)$ obtained from Eq.~\eqref{eq:rhoC_def} in the stationary regime. The two agree very well, confirming that $C(x,t)$ indeed captures the branch-resolved void profile. In all cases the density is strongly depleted near the coherence and approaches zero at the coherence itself. Figure~\ref{fig:screened-void-profile}(b) then shows a collapse of the profiles under the rescaling $x\to x/\xi$, with $\xi\sim (D/\gamma)^{1/2}$, together with the numerical solution of the stationary weak-noise saddle equations \eqref{eq:weak_noise_dimless_rho}-\eqref{eq:weak_noise_dimless_bc} (with $s=\tfrac12$ for $U(1)$-conserving random unitary circuits), showing excellent agreement. Finally, Fig.~\ref{fig:screened-void-profile}(c) shows the decay of the norm-squared coherence autocorrelation function $|\mathcal G(0,t)|^2 = |\langle \sigma^+_0(t)\sigma^-_0(0)\rangle|^2$. Once transients have died out, this becomes purely exponential, $|\mathcal G(0,t)|^2\sim e^{-\Lambda t}$, demonstrating entrance into the quasi-stationary regime governed by a single dominant eigenvalue and consistent with the scaling $\Lambda\propto \sqrt{\gamma}$.

\begin{figure*}
    \centering
    \includegraphics[width=0.8\linewidth]{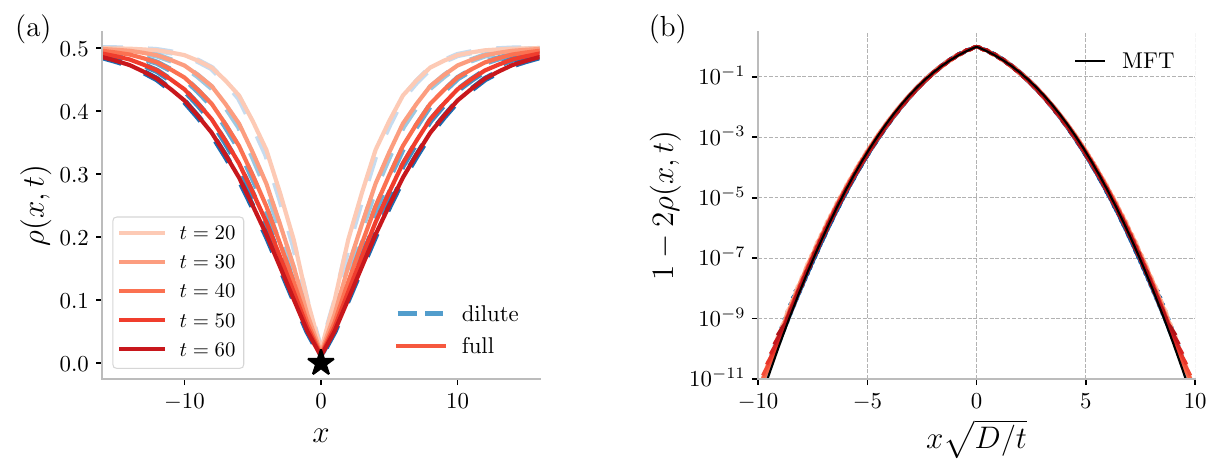}
    \caption{
\textbf{Noiseless aging void}. 
(a) Conditioned density profiles \(\rho(x,t)\) at times \(t=20,30,40,50,60\) (reconstructed from the overlap \(C(x,t)\) defined in Eq.~\eqref{eq:C_doubled_overlap}). The density is strongly depleted near the coherence and relaxes back toward half filling over a scale that broadens with time, consistent with an aging void. (b) The same profiles plotted against the similarity variable \(x/\sqrt{Dt}\) on a logarithmic scale, demonstrating diffusive collapse. The solid curve is the MFT prediction obtained by numerically solving Eqs.~\eqref{eq:noiseless_Omega_integral_eq}--\eqref{eq:noiseless_Phi_from_Omega} for the similarity profile \(\Phi(u)\), again in excellent agreement with the numerical data. Numerical data are obtained from TEBD transfer matrix evolution~\cite{tenpy2024} of the two-replica statistical mechanics model (``full'') describing \(U(1)\)-symmetric random unitary circuits (see Sec.~\ref{RUCsection}), and the dilute-coherence approximation (``dilute''), showing excellent agreement. We use an open chain of length \(L=100\) with bond dimension \(\chi=500\).%, initialized with the coherence at the system center.
}
    \label{fig:aging_void_profile}
\end{figure*}

\textbf{Noiseless case.}
At $\gamma=0$ we again use TEBD simulations with $\chi=500$, now on a chain of length $L=100$ initialized in the same way as with weak noise. We use the same diagnostic $C(x,t)$ for the void profile. Figure~\ref{fig:aging_void_profile}(a) shows that the density is strongly depleted near the coherence and relaxes back toward half filling over a scale that grows with time. Figure~\ref{fig:aging_void_profile}(b) shows the profile on a logarithmic scale in terms of the similarity variable $x/\sqrt{Dt}$, demonstrating a clear diffusive collapse. We also solve Eqs.~\eqref{eq:noiseless_Omega_integral_eq}--\eqref{eq:noiseless_Phi_from_Omega} numerically for the similarity profile $\Phi(u)$ and overlay this prediction, again finding excellent agreement. 

Together, the weak-noise and noiseless data confirm the two hydrodynamic saddles derived above: a stationary weak-noise void of width $\xi\sim \gamma^{-1/2}$ and a noiseless aging void with $\xi_{\gamma=0}(t)\sim \sqrt{t}$. In the next section we derive the point conditioning of the fluid at the coherence in a concrete $U(1)$-conserving random unitary circuit model. In doing so, we will also present the explicit two-replica transfer-matrix evolution used in the numerical simulations.

\section{$U(1)$-conserving random circuits}\label{RUCsection}

We now turn to the microscopic origin of the conditioned fluid--coherence picture developed above, using a $U(1)$-conserving random unitary circuit with qubit degrees of freedom as a concrete testbed. The preceding sections introduced the single-replica hydrodynamic description heuristically and showed that its optimal saddles accurately capture the numerical data. Our goal here is to justify that description microscopically. In particular, we now derive the local point conditioning of the fluid at the coherence. Readers primarily interested in the coherence--void polaron dynamics may skip this section on a first reading and proceed directly to Secs.~\ref{sec:polaron-dynamics} and ~\ref{sec:aging-polaron}. 

We consider a one-dimensional spin-$1/2$ chain evolved by a brickwork circuit of two-site Haar-random $U(1)$-conserving unitaries. The conserved charge is total magnetization $Q=\sum_x \sigma^z_x$, so each two-site gate $U_{x,t}$ is block diagonal in the local $z$-basis, and thus conserves $\sigma^z_x+\sigma^z_{x+1}$. In the noisy case, we apply an on-site depolarizing channel between the even and odd bond layers,
\begin{equation}
  \mathcal D_\gamma(O)
  =
  (1-\gamma)O+\gamma\frac{\rm Tr(O)}{2}\mathbb 1,
  \label{eq:depol_superop}
\end{equation}
so that a single time step is described by the channel
\begin{equation}
  \mathcal C_t
  =
  \prod_{x\ \mathrm{odd}}\mathcal U_{x,t}\,
  \Big(\bigotimes_x \mathcal D_\gamma\Big)\,
  \prod_{x\ \mathrm{even}}\mathcal U_{x,t}\,
  \Big(\bigotimes_x \mathcal D_\gamma\Big),
  \label{eq:single_step_channel}
\end{equation}
where $\mathcal U_{x,t}(O)\equiv U_{x,t}^\dagger O U_{x,t}$.

\subsection{Two-replica transfer matrix rules}
\label{subsec:stationary_ensemble}

To understand the need for replicated dynamics we return to our most basic physical object: the coherence Green's function, which we seek to average over circuit realizations. This can be written simply in the path sum formalism (\autoref{eq:G1rep_full_general}) as
%%% write the path integral with weight + phase, do average, comment it vanishes and then do the square. notation is suboptimal for this particular application.
\begin{equation}
    \mathbb{E}\lb \mathcal G(x_t,x_0;t)\rb = \sum_{X:\,x_0\to x_t} \mathbb{E}\lb \sqrt{\mathbb P[X]} \,  e^{i\Phi[X]}\rb\, \mathcal Z_t[X].
    \label{eq:1rep_trivial_RUC}
\end{equation}
We can then adopt the assumption that the single-body weights, $\sqrt{\mathbb{P}\lb  X \rb}$, for typical spacetime circuit realizations on sufficiently large time and length scales, become self-averaging.
However, $\mathbb{E}\lb e^{i\Phi[X]}\rb=0$ and the direct circuit average of the correlation function vanishes (identically, due to the invariance of the Haar measure under phases). The first non-trivial moment of the coherence Green's function in the random circuit setting is therefore the circuit-averaged \textit{variance}, $C^{(2)}\l x_{t}, x_{0}; t\r = \mathbb{E}\big[ \lv \mc{G}(x_t,x_0; t)\rv^{2} \big]$. In the path sum, this is
\begin{align}
    C^{(2)}(x_{t}, x_{0}; t) =& \sum_{X^{(1,2)}:x_{0} \to x_{t} } \mathbb{E}\lb e^{i\l \Phi_{1} - \Phi_{2} \r}\rb \nonumber \\
    &\  \times \prod_{\alpha = 1,2} \sqrt{\mathbb{P}\left[ X^{(\alpha)}\right]} \mc{Z}_{t}\left[ X^{(\alpha)}\right],
\label{eq:2rep_variance_PS}
\end{align}
where $\Phi_a \equiv \Phi[X^{(a)}]$. Only the diagonal contribution \(X^{(1)}=X^{(2)}\) survives circuit averaging. This replicated average can then be formalized from microscopics in a two-replica statistical mechanics model, which emerges after circuit averaging.

Using the %Choi/vectorization 
map $O\mapsto |O\rangle$, with Hilbert--Schmidt inner product $\langle O|O'\rangle={\rm Tr}(O^\dagger O')/{\rm Tr}(\mathbb{1})$, the averaged doubled evolution is governed by a time-independent one-step transfer matrix,
\begin{equation}
  |A(t)\boxtimes B(t)\rangle
  =
  \mathbb T^{\,t}\,
  |A(0)\boxtimes B(0)\rangle,
  \label{eq:doubled_transfer}
\end{equation}
with $\mathbb T = \mathbb E_{U_{x,t}} \left[\mathcal C_t\otimes \mathcal C_t\right]$. Here we use $\boxtimes$ to refer to a tensor product between the two replicas. 
Because the circuit is local and gates are drawn independently on each bond and time step, $\mathbb T$ factorizes into local bond and on-site pieces. Writing
\begin{equation}
  \mathbb T_x
  \equiv
  \mathbb E_{U_{x,t}}
  \Big[\mathcal U_{x,t}\boxtimes \mathcal U_{x,t}\Big],
  \label{eq:Tx_def}
\end{equation}
we obtain the explicit brickwork form
\begin{equation}
  \mathbb T
  =
  \l\, \mathop{\textstyle\prod}_{x\ \mathrm{odd}} \mathbb T_x\;
  \mathop{\textstyle\bigotimes}_x \mathcal D_\gamma^{\otimes 2}\; \r
  \; \l \, \mathop{\textstyle\prod}_{x\ \mathrm{even}} \mathbb T_x\;
  \mathop{\textstyle\bigotimes}_x \mathcal D_\gamma^{\otimes 2} \r. 
  \label{eq:T_brickwork_depol_explicit}
\end{equation}

A key simplification is that the local doubled operator space closes on a small invariant subspace. To make the local dephasing mechanism transparent, it is convenient to work directly in an occupation basis. On each replica we introduce the projectors
\[
P^\uparrow \equiv \tfrac12(\mathbb{1}+\sigma^z),
\qquad
P^\downarrow \equiv \tfrac12(\mathbb{1}-\sigma^z),
\]
and use these to define four diagonal two-replica states
\[
\begin{aligned}[t]
\mathbf{uu} &\equiv P^\uparrow\boxtimes P^\uparrow
\qquad &
\mathbf{ud} &\equiv P^\uparrow\boxtimes P^\downarrow \\
\mathbf{du} &\equiv P^\downarrow\boxtimes P^\uparrow
\qquad &
\mathbf{dd} &\equiv P^\downarrow\boxtimes P^\downarrow
\end{aligned}
\]
together with the charged coherence states
\[
\mathbf p \equiv 2\,\sigma^+\boxtimes \sigma^-,
\qquad
\mathbf m \equiv 2\,\sigma^-\boxtimes \sigma^+.
\]
The local two-replica dynamics therefore closes on a six-state space $\{ \mathbf{dd},\mathbf{du},\mathbf{ud},\mathbf{uu},\mathbf p,\mathbf m\}$, as noted previously for $U(1)$-conserving random circuits~\cite{McCulloch2023,StretchedExp,PhysRevLett.122.250602,Khemani_2018,Rakovszky2018}.

The diagonal sector has a simple interpretation: on each replica, $\mathbf u$ and $\mathbf d$ behave as occupied and empty sites, respectively. Apart from the coherences $\mathbf p,\mathbf m$, local charge conservation means that the bond dynamics is SSEP-like on each replica, together with short-ranged inter-replica couplings discussed below. What is special about the charged sector is the local action of the transfer matrix on coherences $\mathbf p$ and $\mathbf m$. In particular, when a coherence sits next to a replica disagreement, the circuit average kills the configuration:
\begin{equation}
(\mathbf p,\mathbf{ud}),\ (\mathbf p,\mathbf{du}),\ (\mathbf m,\mathbf{ud}),\ (\mathbf m,\mathbf{du})
\;\to\; 0,
\label{eq:coh_killed_by_disagreement}
\end{equation}
and similarly for site-swapped inputs. The filtering out of these replica-disagreement configurations is a direct dephasing effect. For example, in the local configuration $(\mathbf p,\mathbf{ud})$, the two forward contours of the doubled gate lie in different total-charge sectors, with $N\equiv n_x+n_{x+1}=2$ in replica 1 and $N=0$ in replica 2, while the two backward contours both lie in the $N=1$ sector. Since a $U(1)$-conserving Haar gate acts independently in each charge block, these sectors acquire independent random phases. Averaging over the gate therefore gives zero, and the same reasoning applies to all configurations in Eq.~\eqref{eq:coh_killed_by_disagreement}. This is the microscopic origin of the branch conditioning introduced in the effective theory: the charge operator can propagate with phase coherence between the two replicas only when its local neighborhood is replica aligned.

The remaining bond updates take the form~\cite{von_Keyserlingk_2022}
% \textcolor{red}{[Make nice table.]}
% \begin{align}
% (\mathbf{uu},\mathbf{dd}) &\to \tfrac16\Big[
% 2(\mathbf{uu},\mathbf{dd})
% +(\mathbf{ud},\mathbf{du})
% +(\mathbf p,\mathbf m)
% +\text{(swap)}
% \Big], \nonumber\\
% (\mathbf{ud},\mathbf{du}) &\to \tfrac16\Big[
% 2(\mathbf{ud},\mathbf{du})
% +(\mathbf{uu},\mathbf{dd})
% -(\mathbf p,\mathbf m)
% +\text{(swap)}
% \Big], \nonumber\\
% (\mathbf p,\mathbf m) &\to \tfrac16\Big[
% 2(\mathbf p,\mathbf m)
% +(\mathbf{uu},\mathbf{dd})
% -(\mathbf{ud},\mathbf{du})
% +\text{(swap)}
% \Big], \nonumber\\
% (\mathbf{ud},\mathbf{dd}) &\to \tfrac12(\mathbf{ud},\mathbf{dd}) + \tfrac12(\mathbf{du},\mathbf{dd}), \nonumber\\
% (\mathbf{dd},\mathbf{dd}) &\to (\mathbf{dd},\mathbf{dd}),
% \label{eq:non_ssep_rules_ud_basis}
% \end{align}
%JAJ: How is this
\begin{mdframed}[roundcorner=10pt, backgroundcolor=gray!15]
\footnotesize
\centering
\renewcommand{\arraystretch}{1.6}
\begin{tabular}{@{} l c l @{}}
\hline
\textbf{State} & & \textbf{Transition rule} \\
\hline
$(\mathbf{uu},\mathbf{dd})$ & $\to$ &
  $\tfrac{1}{6}\bigl[2(\mathbf{uu},\mathbf{dd})+(\mathbf{ud},\mathbf{du})+(\mathbf{p},\mathbf{m})+(\text{swap})\bigr]$ \\
$(\mathbf{ud},\mathbf{du})$ & $\to$ &
  $\tfrac{1}{6}\bigl[2(\mathbf{ud},\mathbf{du})+(\mathbf{uu},\mathbf{dd})-(\mathbf{p},\mathbf{m})+(\text{swap})\bigr]$ \\
$(\mathbf{p},\mathbf{m})$ & $\to$ &
  $\tfrac{1}{6}\bigl[2(\mathbf{p},\mathbf{m})+(\mathbf{uu},\mathbf{dd})-(\mathbf{ud},\mathbf{du})+(\text{swap})\bigr]$ \\
$(\mathbf{ud},\mathbf{dd})$ & $\to$ &
  $\tfrac{1}{2}(\mathbf{ud},\mathbf{dd})+\tfrac{1}{2}(\mathbf{du},\mathbf{dd})$ \\
$(\mathbf{dd},\mathbf{dd})$ & $\to$ &
  $(\mathbf{dd},\mathbf{dd})$ \\
\hline
\end{tabular}
\label{tab:non_ssep_rules_ud_basis}
\end{mdframed}
together with the corresponding site-swapped rules, and those obtained by swapping $\mathbf u \leftrightarrow \mathbf d$. The first three rules are the only local processes that mix the diagonal sector with the charged coherence sector.

The on-site depolarizing channel is also simple in this basis. On diagonal states it acts via stochastic resets,
\begin{equation}
\mathcal D_\gamma(\mathbf u)
=
(1-\gamma)\mathbf u
+\gamma(\mathbf u + \mathbf d)/2,
\label{eq:depol_u_d_reset}
\end{equation}
and similarly for $\mathcal D_\gamma(\mathbf d)$. Therefore, each replica undergoes independent $\mathbf u \leftrightarrow \mathbf d$ randomization at rate $\gamma$. On the charged coherence symbols, depolarization acts as pure attenuation,
\begin{equation}
\mathbf p \to (1-\gamma)^2 \mathbf p,
\qquad
\mathbf m \to (1-\gamma)^2 \mathbf m.
\label{eq:depol_pm_reset}
\end{equation}

The remaining question is how strongly these locally generated coherence pairs feed back on the slow diagonal environment.

\subsection{Dilute coherence approximation}
\label{subsec:dilute-coherence}
The transition rules above also make the microscopic origin of the gradient bottleneck transparent. Local coherence pairs are not produced from generic diagonal configurations, but only from the antisymmetric (singlet) bond combination on each replica. In this sense, the same constraint that made smooth diagonal charge profiles slow in the earlier hydrodynamic discussion now appears directly in the microscopic transfer matrix: the production of additional off-diagonal operator weight is itself gradient-suppressed. This observation motivates a controlled simplification of the dynamics, which we describe below.

Writing $n_x^{(a)}\in\{0,1\}$ for the occupation variable associated with $\mathbf u$ on replica $a$, one has
\[
\mathbf u^{(a)}_x \mathbf d^{(a)}_{x+1}-\mathbf d^{(a)}_x \mathbf u^{(a)}_{x+1}
=
n_x^{(a)}-n_{x+1}^{(a)},
\]
so the statistical weight for the creation of $(\mathbf p,\mathbf m)$ is proportional to the product of the signed bond gradients on the two replicas,
\[
(n_x^{(1)}-n_{x+1}^{(1)})(n_x^{(2)}-n_{x+1}^{(2)}).
\]
%some rewriting by JAJ with original above
Locally flat or symmetric charge configurations therefore do not generate coherences. In this work, we adopt a \emph{dilute-coherence} or no-backflow approximation~\cite{von_Keyserlingk_2022}: we neglect the generation of short-lived coherence pairs and their subsequent recombination back into the slow diagonal sector. This is natural on hydrodynamic scales, where both the production of such pairs and their reinjection into the charge-neutral operator sector are gradient-suppressed. At leading order, integrating out these fast coherence-pair modes simply renormalizes the inter-replica coupling, which is itself irrelevant in the diffusive scaling regime as we discuss next (the coherence pair propagator is derived in Appendix~\ref{app:pm_propagator} and integrated out in Appendix~\ref{app:NLSM}).

%This is the microscopic version of the hydrodynamic statement made earlier: charged off-diagonal weight is sourced by charge gradients. In this work, we adopt a \emph{dilute-coherence} or no-backflow approximation~\cite{von_Keyserlingk_2022}: the generation of short-lived coherence pairs followed by their recombination and contribution to the hydrodynamic (charge-neutral operator) profile, is considered negligible. Both coherence generation from and reinjection into the charge neutral operator sector is gradient-suppressed and therefore remains weak over smooth (\textit{i.e.}, hydrodynamic) density profiles. The leading order physics of this backflow is captured by, essentially, a Schrieffer-Wolff transformation: integrating out these fast coherence-pair modes contributes to the effective Hamiltonian a renormalization to the interreplica term, which is irrelevant in the diffusive scaling regime (discussed further in \autoref{sec:bulk-decoupling}, \autoref{app:NLSM}, and \autoref{app:pm_propagator}).

To implement this simplified model, we modify the transition rules described above by deleting the channels that create local coherence pairs $(\mathbf p,\mathbf m)$ or $(\mathbf m,\mathbf p)$ from diagonal configurations. In the charged sector of principal interest, $\mathcal Q=1$, the resulting approximate microscopic model is therefore a substochastic dynamics for a single tagged coherence propagating in a diagonal environment: away from the coherence, the two replicas evolve as coupled stochastic chains with excluded hopping of the symbols $\mathbf u/\mathbf d$ and weak depolarizing resets, while at the coherence itself trajectories are killed whenever the tagged coherence encounters a replica disagreement.

As an a posteriori check, the void-profile numerics shown in Sec.~\ref{sec:optimal_void_profiles} already show that neglecting the short-lived coherence-pair backflow channels does not affect the macroscopic optimal void profile. In Figs.~\ref{fig:screened-void-profile} and \ref{fig:aging_void_profile}, the curves labeled ``dilute'' were obtained using the approximate sub-stochastic dynamics defined here. The excellent agreement with the full transfer matrix evolution indicates that the dilute-coherence approximation captures the same coarse-grained hydrodynamic large-deviation physics in both the weak-noise and noiseless regimes.

\subsection{Bulk replica decoupling}
\label{sec:bulk-decoupling}
We now determine the bulk dynamics in the dilute-coherence approximation, which closes on diagonal symbols $\{\mathbf{uu},\mathbf{ud},\mathbf{du},\mathbf{dd}\}$. It is convenient to pass to a continuous-time formulation, replacing the brickwork update by independent Poisson clocks on each bond. This does not affect the long-wavelength physics, but yields a time-translation invariant Markov generator
\begin{equation}
  \mathcal L \;=\; \sum_x \mathcal L_{x,x+1} ,
\end{equation}
where $\mathcal L_{x,x+1}$ is the local bond generator obtained by Poissonizing the discrete bond update.

Using the occupation variables $n_x^{(a)}\in\{0,1\}$ for the $\mathbf u$ symbol for replicas $a=1,2$, the diagonal sector is a two-replica interacting symmetric exclusion process. The key point is that the local bond dynamics is strongly constrained by charge conservation. In a single replica, the flat charge configurations are precisely the bond-triplet states, while the nontrivial hopping dynamics is carried by the singlet sector. Since the triplet states are eigenoperators of the local bond update, whenever one replica is locally in a flat charge configuration, the other replica undergoes the usual SSEP hopping with the same rate as in the single-replica problem. The hopping rate on one replica is modified only when the other replica is itself in the singlet sector.

A useful representation of this structure is obtained by introducing spin-$1/2$ operators $\vec S=\tfrac12\vec\sigma$ acting on the classical bits $n_x^{(a)}$. On a bond $(x,x+1)$, the single-replica SSEP exchange generator is proportional to the singlet projector
\begin{equation}
  P^{(a)}_{x,x+1}
  \;\equiv\;
  \tfrac14-\vec S^{(a)}_x\cdot \vec S^{(a)}_{x+1},
\end{equation}
so that, up to an overall choice of time units, the single replica SSEP generator is given by $\mathcal L_x^{(a)} \propto -P^{(a)}_{x,x+1}$. The only possible local inter-replica correction is one that is activated precisely when \emph{both} replicas are in the singlet sector. The unique local operator with this property is the product of singlet projectors,
\begin{equation}
  \mathcal L_{x,x+1}
  = \sum_a \mathcal L_{x,x+1}^{(a)}
  + \mathcal V_{x,x+1},
  \ \
  \mathcal V_{x,x+1} \propto P^{(1)}_{x,x+1}P^{(2)}_{x,x+1}.\nonumber
\end{equation}
This is indeed the inter-replica interaction generated by the explicit diagonal-sector rules in Sec.~\ref{subsec:stationary_ensemble}, once the coherence-creation channels have been removed.

The hydrodynamic implication is immediate. In a coarse-grained description, the only slow bulk fields are the conserved densities $\rho_a(x,t)\equiv \langle P^{\up (a)}_x\rangle$. The single-replica terms $\mathcal L_{x,x+1}^{(a)}$ generate the usual diffusive propagator, so each singlet projector $P_{x,x+1}$ carries the same derivative counting as the leading diffusive gradient term. The inter-replica product $P^{(1)}P^{(2)}$ therefore enters only at higher order in gradients. Schematically, it contributes terms of the form
\begin{equation}
  \mathcal V(x)\sim
  \big(\partial_x\rho_1\big)^2\big(\partial_x\rho_2\big)^2+\cdots,
\end{equation}
which under diffusive scaling $x\to bx$, $t\to b^2 t$ are RG-irrelevant compared to the leading single-replica diffusive terms~\cite{McCulloch2023}. We therefore neglect this bulk replica coupling in what follows, so that the two replicas flow to independent hydrodynamic fluids in the infrared. A detailed argument starting from the microscopic generator and passing to the spin coherent-state path integral is given in Appendix~\ref{app:NLSM}. Here, the action on each replica is a $\mathbb{C}{\rm P}^{1}$ nonlinear sigma model, with interreplica coupling terms which are transparently irrelevant. This formalism also allows us to account for loop corrections from $\mbf{pm}$ pairs, which enter the diagonal sector at large scales by dressing the coefficient of the inter-replica term (the Green's function for these pairs is evaluated in Appendix~\ref{app:pm_propagator}). The $\mathbb{C}{\rm P}^{1}$ model then recovers the MSR action under a simple field redefinition.

Our aim now is to recover, from this microscopic sub-stochastic rule, the same continuum structure proposed heuristically in Sec.~\ref{sec:slow-manifold}: a Dirichlet condition at the coherence together with a bias on the outward density gradient.

%In the next section we show that the continuum description of this conditioned stochastic process is precisely the fluctuating hydrodynamic theory with a Dirichlet boundary condition and boundary-gradient Doob tilt proposed on general grounds in Sec.~\ref{sec:slow-manifold}.

%\section{From microscopic filtering to conditioned hydrodynamics}

%We now coarse-grain the approximate tagged-coherence process derived in Sec.~\ref{RUCsection}. Within the dilute-coherence approximation, the only nontrivial two-replica structure is localized at the coherence itself: away from it, the two replicas reduce to diffusive diagonal fluids with weak depolarization, while at the coherence histories are pruned whenever the neighboring site differs between replicas. It is convenient to first hold the coherence fixed at the origin and analyze the diagonal environment on either side. This isolates the local conditioning in its simplest form. We will then restore the full problem with a mobile coherence. Our aim is to recover, from this microscopic sub-stochastic rule, the same continuum structure proposed heuristically in Sec.~\ref{sec:slow-manifold}: a Dirichlet condition at the coherence together with a Doob tilt on the outward density gradient.

\subsection{Branch projection and boundary depletion}

In this section, we see the microscopic origin of the Dirichlet condition in the putative branch-restricted quasi-stationary state. It is convenient to first hold the coherence fixed at the origin and analyze the diagonal operators on either side. This isolates the local conditioning in its simplest form. We will then restore the full problem with a mobile coherence. %Our aim is to recover, from this microscopic sub-stochastic rule, the same continuum structure proposed heuristically in Sec.~\ref{sec:slow-manifold}: a Dirichlet condition at the coherence together with a Doob tilt on the outward density gradient.

We begin with the half-line $x>0$, since once the coherence is held fixed the two sides decouple. Away from the coherence, each replica undergoes ordinary SSEP dynamics together with weak local depolarization at rate $\gamma$. The distinctive feature of the $\mathcal Q=1$ sector is entirely local: a history is discarded whenever the site adjacent to the coherence carries different charge in the two replicas. In a generic finite-entropy background such boundary mismatches would be generated at $O(1)$ rate, since diffusion continually refills the boundary from the bulk. Long-lived charged histories must therefore strongly suppress replica disagreement at the coherence neighbor, which is precisely what is achieved by forming a locally polarized void.

To discuss this local conditioning it is convenient to keep depolarized sites explicit. Depolarization maps both $P^\uparrow$ and $P^\downarrow$ to the same equal-weight diagonal mixture, which we denote by
\[
\hat{\mathbb 1}\equiv \frac{\mathbb 1}{2}=\frac12\left(P^\uparrow+P^\downarrow\right).
\]

In the bulk a depolarized spin simply participates in the same symmetric exchange dynamics as the polarized charge states. The only nontrivial step is the local filtering at the coherence. If only one replica carries $\hat{\mathbb 1}$, the filter projects that local mixture onto the component that agrees with the other replica, with the corresponding factor $1/2$. If both replicas carry $\hat{\mathbb 1}$, the filter retains the two agreeing channels and discards the disagreeing ones, so that
\[
\hat{\mathbb 1}\boxtimes \hat{\mathbb 1}
\longrightarrow
\frac14 P^\downarrow\boxtimes P^\downarrow
+
\frac14 P^\uparrow\boxtimes P^\uparrow,
\]
while the mixed $P^\uparrow \boxtimes P^\downarrow$ and $P^\downarrow \boxtimes P^\uparrow$ components are removed. This local rule is exact and branch-neutral: microscopically, the filter only enforces replica agreement.

Assume now that the system is already in the down-polarized quasi-stationary void branch. The two surviving channels of the local filter then have very different fates. The $P^\downarrow \boxtimes P^\downarrow$ channel simply restores the established void polarization; this is the branch-compatible channel. The $P^\uparrow \boxtimes P^\uparrow$ channel instead places a \emph{hazard}---a wrongly polarized $P^\uparrow$---next to the coherence in each replica; this is the hazard channel. The distinction is emergent: the local filter itself does not prefer one over the other, but after the $P^\uparrow \boxtimes P^\uparrow$ outcome is picked, the two $P^\uparrow$ diffuse independently. If either one returns to the coherence while the other is elsewhere, the two replicas differ (with high probability) and the history is pruned. For the hazard channel to persist, both wrongly polarized spins must therefore avoid returning until they are erased by depolarization, i.e., before a time $\tau\sim \gamma^{-1}$. In one dimension the probability of no return up to time $\tau$ is $O(\tau^{-1/2})$. This gives a factor $O(\gamma^{1/2})$ for each $P^\up$ hazard, and hence an overall suppression $O(\gamma)$ for the hazard channel. To leading order, one therefore retains only the branch-compatible channel.

Equivalently, the local conditioning at the coherence-neighbor bond factorizes across replicas and acts as a pure projection onto $P^\down$ on both replicas,
\begin{equation}
    A\boxtimes B \to P^\down A P^\down \boxtimes P^\down B P^\down,
\label{eq:hard-repair-projector}
\end{equation}
up to $O(\gamma)$ corrections from the suppressed hazard channel. This is precisely the projective conditioning anticipated in Eq.~\eqref{eq:branch-projection}. After each update of the coherence-neighbor bond, the neighboring site is therefore restored to the polarization of the pre-existing void. In the down-polarized branch this means
\begin{equation}
n^{(1)}_1=n^{(2)}_1=0,
\label{eq:hard-repair}
\end{equation}
which we refer to as the branch-projection limit. In particular, $P^\uparrow$ occupation at the boundary site is $O(\gamma)$-suppressed.

\subsection{Void metastability and branch restriction}
\label{sec:metastability}

The conditioned dynamics admits two symmetry-related void polarizations, corresponding to down- and up-polarized voids. At weak depolarization these define metastable branches. A void can only change polarization through a boundary-mediated conversion process: 
converting a down-void into an up-void requires repeatedly selecting the branch-incompatible outcome at the coherence, rather than the branch-compatible one that preserves the established void polarization, while still avoiding replica disagreements at the coherence neighbor.

Since depolarization events occur at rate $\gamma$, successfully converting the void polarization requires maintaining a boundary layer of mixed local charge for a time of order
\[
\tau \sim \gamma^{-1}.
\]

Such a switching attempt is strongly suppressed. Once the boundary region is no longer well polarized, replica mismatches are generated at $O(1)$ rate.  An attempted conversion must survive a pruning rate $r_{\rm kill}(t)$ that remains $O(1)$ throughout a time window of order $\tau\sim\gamma^{-1}$. Its success probability therefore scales as
\[
P_{\rm switch}\sim \exp\!\left[-\int_0^\tau dt\, r_{\rm kill}(t)\right]
\sim e^{-c/\gamma},
\]
with $c>0$ nonuniversal.

The corresponding switching time is exponentially large,
\[
\tau_{\rm switch}\sim \gamma^{-1} e^{c/\gamma},
\]
whereas all branch-internal timescales remain power-law in $\gamma$, including the screening time and the quasi-stationary decay time. It is therefore self-consistent, throughout the weak-noise regime of interest, to work within a single symmetry-broken void branch. Unless stated otherwise, we choose the down-polarized branch.

At asymptotically long times the exact quasi-stationary state restores particle-hole symmetry and is given by the equal-weight mixture of the two branches, up to corrections that are nonperturbative in $\gamma$ and controlled by the exponentially slow switching dynamics.

\subsection{Conditioning as a boundary gradient tilt}
\label{subsec:selection-reset-current}

We now connect the local filtering rule to the hydrodynamic description. In the weak-noise regime the void is parametrically large, $\xi\sim \gamma^{-1/2}\gg 1$, so the distinction between the coherence position and the nearest-neighbor site is an unimportant microscopic offset. We therefore identify the coherence neighbor at lattice site $1$ with the continuum boundary at $x=0$. In the branch-projection limit, Eq.~\eqref{eq:hard-repair}, this gives the Dirichlet condition $\rho_a(0,t)=0$ to leading order in $\gamma$. The remaining conditioning may then be imposed indirectly, on the precursor configurations that attempt to refill that site.

In the branch-projection limit the replicas decouple at the boundary, so it is enough to first discuss a single replica. We therefore suppress replica indices for the moment. Let $n_2$ be the occupation of site $2$. The configuration $n_2=1$ is precisely the local precursor that can refill the coherence neighbor at  site $1$. Under the conditioning in Eq.~\eqref{eq:hard-repair-projector}, the outcome in which the charge hops onto the coherence neighbor is discarded. For the present microscopic rule---bulk SSEP evolution with bond exchange probability $1/2$---this leaves an effective weight $1/2$ for each precursor configuration with $n_2=1$.

Let $b(t)\equiv \langle n_2(t)\rangle$ denote the expected density at site $2$ for a coarse-grained profile $\rho(x,t)$. Reinstating the replica labels, the effective rate of loss of probability weight is therefore
\begin{equation}
  r_{\rm loss}[{\rho}]
  =
  s\,(b_1+b_2) + \cdots,
  \label{eq:pmismatch-b}
\end{equation}
with $s=1/2$ for the present brickwork circuit rule. With $\rho_a(0,t)=0$, a derivative expansion gives
\[
b_a(t)=\rho_a(1,t)-\rho_a(0,t)\approx \partial_x\rho_a(0^+,t),
\]
so Eq.~\eqref{eq:pmismatch-b} becomes
\begin{equation}
r_{\rm loss}[\rho]
=
s\sum_{a=1}^2 \partial_x\rho_a(0^+,t).
\label{eq:effective-gradient-tilt}
\end{equation}
This is the origin of the boundary-gradient bias in the tilted measure in Eq.~\eqref{eq:gradient_tilt}.

It remains to check that the neglected occupation of site $1$ is indeed subleading. Relaxing the branch-projection approximation, a nonzero occupation at the coherence neighbor requires both replicas to place charge there in the same update, and therefore has expectation
\[
\langle n^{(a)}_1\rangle=\mathcal O\bigl(b_1 b_2\bigr).
\]
Across a quasi-stationary void, the density changes by $O(1)$ over a distance $\xi\sim \gamma^{-1/2}$, so $\partial_x\rho_a(0^+)\sim \sqrt{\gamma}$, and hence
\begin{equation}
\big\langle n^{(a)}_1\big\rangle=\mathcal O\bigl((\partial_x\rho)^2\bigr)=\mathcal O(\gamma).
\label{eq:branch-proj-subleading-correction}
\end{equation}
The Dirichlet condition is therefore self-consistent to leading order.

The coarse-grained effect of the local filtering in Eq.~\eqref{eq:effective-gradient-tilt} over an evolution time $T$ is thus a multiplicative weight
\[
\exp\!\bigg(-\int_0^T dt\,r_{\rm loss}[{\rho}]\bigg)
\]
attached to the density history $\rho(t)$. Equivalently, the unnormalized two-replica partition sum for $x>0$ (and fixed coherence) may be written as
\begin{equation}
  \mathcal{Z}^{\rm 2-rep}_T
  =
  \Big\langle
  \exp\!\Big[
  -s\int_0^T dt \sum_{a=1}^2 \partial_x\rho_a(0^+,t)
  \Big]
  \Big\rangle,
  \label{eq:ZT-def-gradient}
\end{equation}
where $\langle\cdot\rangle$ denotes an expectation with respect to the untilted fluid measure, together with the Dirichlet condition $\rho_a(0,t)=0$. This is precisely the boundary-gradient tilt anticipated in Sec.~\ref{sec:slow-manifold}. The complete continuum description then follows by applying standard SSEP macroscopic fluctuation theory~\cite{Bertini_2015,PhysRevLett.129.040601} to the bulk dynamics.

%From here on we will use two equivalent languages interchangeably. Microscopically, Eq.~\eqref{eq:hard-repair-projector} is a local conditioning of the microscopic fluid configurations filter acting on depolarized spins that reaches the coherence: only the branch-compatible component is retained, while the incompatible component is pruned. Coarse-grained, the same process appears as the boundary-gradient Doob tilt in Eq.~\eqref{eq:ZT-def-gradient}.

\subsection{Recovering the one-replica action}
\label{subsec:recovering_one_replica_action}

Up to this point we fixed the coherence at the origin in order to isolate the local conditioning of the diagonal occupation operators. Restoring its motion is simple. In the dilute-coherence approximation, coherences are neither created nor destroyed in the bulk. Thus a single coherence simply propagates by nearest-neighbor hopping under the same bond updates that move a single particle. Between conditioning events (i.e., in a perfect void), the coherence coordinate $X(t)$ therefore undergoes an unbiased random walk with bare diffusion constant $D_X=D$.

For a prescribed trajectory $X(t)$, the two replicas evolve as independent conditioned fluids on the full line punctured at the coherence position, i.e., with the moving Dirichlet condition $\rho_a(X(t),t)=0$ and boundary Doob tilt $\sum_{a=1,2} S_{\rm bdy}[\rho_a,X]$, with $S_{\rm bdy}[\rho,X]$ given in Eq.~\eqref{eq:gradient_tilt}. The resulting unnormalized two-replica partition sum is therefore
\begin{equation}
\mathcal G^{(2)}(x_T,x_0;T)
=
\int^{X(T)=x_T}_{X(0)=x_0} \mathcal D X\, e^{-S_X[X]} \mathcal Z_T^{(2)}[X],
\label{eq:two_rep_moving_action}
\end{equation}
with
\begin{equation}
\mathcal Z_T^{(2)}[X]
=
\prod_{a=1}^2
\int_{\rho_a(X)=0}
\mathcal D\rho_a\,\mathcal D\pi_a\,
e^{-S_{\rm bulk}[\rho_a,\pi_a]-S_{\rm bdy}[\rho_a;X]},
\label{eq:2-rep-fluid-action}
\end{equation}
where the bulk MFT action $S_{\rm bulk}[\rho,\pi]$ is as given in Eq.~\eqref{eq:Sbulk_1rep_general}, where $\rho_a(X)=0$ is shorthand for $\rho_a(X(t),t)=0$, and where $S_X[X]$ is the two-replica coherence action
\begin{equation}
S_X[X]
=
\frac{1}{4D}\int_0^T dt\,\dot X(t)^2.
\label{eq:SX_weak_noise_mobile}
\end{equation}

Equation~\eqref{eq:two_rep_moving_action} is the phase-averaged two-replica statistical mechanics partition function obtained after circuit averaging. To recover the physical one-replica action, we now undo this phase averaging. For a single circuit realization, a coherence trajectory carries a quantum mechanical amplitude for its walk in, e.g., the down branch,
\begin{equation}
A[X]=e^{-S_X[X]/2}\,e^{i\Phi[X]},
\end{equation}
where $\Phi[X]$ is the realization-dependent phase accumulated along the coherence worldline. The key point is that circuit averaging suppresses interference between distinct coherence trajectories,
\begin{align}
\mathbb E_U\!\left[A[X]A[X']^*\right]
&=
e^{-\frac12 S_X[X]-\frac12 S_X[X']}
\mathbb E_U e^{i\Phi[X]-i\Phi[X']}\nonumber\\
&\propto
e^{-S_X[X]}\,\delta_{X,X'}.
\label{eq:path_diagonal_phase_average}
\end{align}
The phase-averaged two-replica theory is the trajectory-diagonal part of the one-replica quantum path sum: the amplitude and its conjugate are forced onto the same worldline, producing the classical random-walk weight $e^{-S_X[X]}$ in Eq.~\eqref{eq:two_rep_moving_action}.

Thus the one-replica hybrid coherence-fluid path sum for the raising-operator Green's function $\langle \sigma^+_{x_t}(t)\sigma^-_{x_0}(0)\rangle$ is
\begin{equation}
\mathcal G(x_t,x_0;t)
=
\sum_{X:x_0\to x_t}
e^{-S_X[X]/2}\,e^{i\Phi[X]}\,\mathcal Z_t[X],
\label{eq:recovered_hybrid_path_sum}
\end{equation}
as previously given in Eq.~\eqref{eq:G1rep_full_general} on general grounds.

\section{Dynamics of the coherence-void polaron: weak noise case $\gamma > 0$}
\label{sec:polaron-dynamics}

We now restore the motion of the coherence and study its dynamics. The relevant observable is again the coherence Green's function $\mathcal G(x,t)\equiv \langle \sigma_x^+(t)\sigma_0^-(0)\rangle$, whose effective coherence--fluid path sum (Eq.~\eqref{eq:G1rep_full_general}) was derived microscopically in Sec.~\ref{RUCsection}. In Sec.~\ref{sec:optimal_void_profiles}, we held the coherence fixed in order to isolate the conditioned fluid profile that supports its survival: in the weak-noise case this profile is a quasi-stationary void of finite size, while in the noiseless case it is an aging void that broadens diffusively. Once the coherence motion is restored, the natural object is therefore a dressed one, consisting of the coherence together with the self-generated void that supports its survival. We will refer to this composite object as a coherence--void polaron. Unlike conventional polarons, however, the dressing here is not a small equilibrium polarization cloud, but a far-from-equilibrium hydrodynamic large deviation corresponding to a near-complete local depletion of charge.

In a perfectly polarized background, a single raising operator simply undergoes one-body propagation; in spatio-temporally inhomogeneous settings such as random circuits and Brownian Hamiltonian models, this propagation has diffusive scaling~\cite{Ahlbrecht2012,Joye2011}. More precisely, the wavefunction carries realization-dependent phases and fluctuations, while its norm develops a deterministic self-averaging envelope that controls the exploration scale in the absence of fluid conditioning.

The essential new ingredient is that the background is not a perfect charge vacuum. As the coherence moves, it conditions the surrounding fluid, and the conditioned fluid in turn biases the coherence trajectory. Trajectories that remain near previously depleted regions benefit from an environment that has already been made dilute, whereas trajectories that wander into fresh regions must pay the additional cost of creating and sustaining a new void there. The resulting trajectory \(X(t)\) therefore experiences a retarded self-attraction.

In the weak-noise regime, this self-attraction produces a two-scale dynamics. On short scales, the coherence moves rapidly inside the void, exploring an internal region of width \(\gamma^{-1/3}\). On longer scales, this internal motion crosses over to the slow diffusion of the entire coherence--void polaron, whose translation mode has an effective diffusion constant \(D_{\rm eff}\sim \sqrt{\gamma}\). Equivalently, at fixed \(\gamma>0\) and asymptotically long times, the weak-noise coherence propagator acquires the small-momentum form
\refstepcounter{equation}\label{eq:weaknoise_scaling_form_opening}
\begin{wideeqbox}
\vspace{4pt}
\noindent
\makebox[\linewidth][l]{%
  \makebox[3em][l]{}%
  \makebox[\dimexpr\linewidth-8em\relax][c]{%
    $\displaystyle
    |\mathcal G(k,t)|^2
\sim
\exp\!\left[
-\sqrt{\gamma}
\bigl(\Lambda+\tilde D_{\rm eff} k^2 + \cdots \bigr)\, t
\right],
    $%
  }%
  \makebox[5em][r]{(\theequation)}%
}
\vspace{0pt}
\end{wideeqbox}
\noindent with \(\Lambda=O(1)\), and $\tilde D_{\rm eff}\equiv {D_{\rm eff}}/{\sqrt{\gamma}}=O(1)$. In the rest of this section we derive this retarded polaron description by integrating out the fluid, then characterize both the translation mode and the internal polaron motion, as well as localization of the coherence-void polaron to a bond impurity.

\subsection{Non-local action for the coherence}

We now seek an effective action for the coherence trajectory \(X(t)\) by integrating out the fluid degrees of freedom. The relevant object is the phase-averaged norm-squared propagator \(\overline{|\mathcal G(x,t)|^2}\), whose coarse-grained description is given by the two-replica path sum Eq.~\eqref{eq:two_rep_moving_action}. In the weak-noise regime, the screened void relaxes on the timescale \(\gamma^{-1}\) and has spatial extent \(\xi\sim \gamma^{-1/2}\). The effective diffusion constant for the coherence--void polaron is therefore parametrically small in $\gamma$ since translating the dressed object requires dragging a void of size \(\xi\). As a result, the displacement of the polaron over one void-relaxation time is parametrically smaller than the void size. This justifies an adiabatic approximation in which the conditioned fluid profile remains close, locally in time, to the optimal void profile identified by the pinned-coherence saddle of Sec.~\ref{sec:optimal_void_profiles}. Since that saddle is replica symmetric, we likewise restrict to replica-symmetric moving histories,
\begin{equation}
\rho_1=\rho_2\equiv \rho,\qquad \pi_1=\pi_2\equiv \pi.
\end{equation}

For fixed $X(t)$, the fluid is determined by the conditioned MFT saddle equations
\begin{align}
\partial_t \rho
&=
D\,\partial_x^2 \rho
-
\partial_x\!\left[\sigma(\rho)\,\partial_x \pi\right]
-
\frac{\gamma}{2}
\Big[(1-\rho)e^{\pi}-\rho e^{-\pi}\Big],
\nonumber
\\
\partial_t \pi
&=
- D\,\partial_x^2 \pi
-
\frac{\sigma'(\rho)}{2}\,(\partial_x \pi)^2
+
\frac{\gamma}{2}\Big(e^{\pi}-e^{-\pi}\Big),
\label{eq:mobile_saddle_maintext}
\end{align}
with moving Dirichlet data
\begin{equation}
\rho(X(t),t)=0, \qquad \pi(X(t),t)=-s,
\label{eq:mobile_dirichlet_bcs_maintext}
\end{equation}
together with
\begin{equation}
\rho(|x|\to\infty,t)=\frac12, \qquad \pi(|x|\to\infty,t)=0.
\end{equation}

In the dilute-fluid regime appropriate to the void core, exclusion effects may be neglected and the conditioned fluid problem becomes linearizable under a Cole--Hopf transformation. The fluid degrees of freedom can then be integrated out exactly for a fixed coherence trajectory $X(t)$. As shown in Appendix~\ref{app:retarded_volterra_from_one_replica}, this yields a retarded non-local action for the coherence alone,
\refstepcounter{equation}\label{eq:Seff_maintext_corrected}
\begin{wideeqbox}
\noindent
\makebox[\linewidth][l]{%
  \makebox[3em][l]{}%
  \makebox[\dimexpr\linewidth-8em\relax][c]{%
    $\displaystyle
    S_{\rm eff}[X]=S_X[X]+2\big(e^s-1\big)\int_0^T dt\, r_X(t).
    $%
  }%
  \makebox[5em][r]{(\theequation)}%
}
\end{wideeqbox}
\noindent where the free kinetic term $S_X[X]$ is given in Eq.~\eqref{eq:SX_weak_noise_mobile}. Here $r_X(t)$ is the instantaneous fluid-filtering rate associated with the gradient tilt (Eq.~\eqref{eq:gradient_tilt}) along the prescribed coherence trajectory $X(t)$. It is given by the outward gradient jump
\begin{equation}
r_X(t)
\equiv
-\,D\Big[\partial_x \rho_X(x,t)\Big]_{x=X(t)^-}^{x=X(t)^+},
\label{eq:rX_maintext}
\end{equation}
evaluated on the optimal density profile $\rho_X(x,t)$ for the given trajectory $X(t)$. Legal histories of the fluid-filtering rate $r_X(t)$ are those consistent with the implicit Volterra equation
\refstepcounter{equation}\label{eq:volterra_maintext}
\begin{wideeqbox}
\noindent
\makebox[\linewidth][l]{%
  \makebox[3em][l]{}%
  \makebox[\dimexpr\linewidth-8em\relax][c]{%
    $\displaystyle
    \int_0^t d\tau\,
    G_\gamma\!\left(X(t)-X(\tau),\,t-\tau\right)\,r_X(\tau)
    =
    \frac{e^{-s}}{2},
    $%
  }%
  \makebox[5em][r]{(\theequation)}%
}
\end{wideeqbox}
with screened kernel
\begin{equation}
G_\gamma(\Delta x,\tau)
=
\frac{e^{-\gamma \tau}}{\sqrt{4\pi D \tau}}
\exp\!\left[-\frac{(\Delta x)^2}{4D\tau}\right].
\label{eq:screened_kernel_maintext}
\end{equation}

The dependence of the filtering rate $r_X(t)$ on the past history of the coherence position may be understood as follows. A filtering event at time \(\tau\) induces a small additional depletion in the conditioned density field near the coherence position \(X(\tau)\). This incremental depletion then spreads diffusively until it is cut off by depolarization on the timescale \(\gamma^{-1}\), so its contribution to the depletion at the coherence position at time \(t\) is weighted by the screened propagator \(G_\gamma(X(t)-X(\tau),t-\tau)\). The depletion at the coherence is therefore given by a causal weighted sum of past filtering events. Since the coherence sits at a point of maximal depletion (\(\rho(X(t),t)=0\)), this weighted sum is fixed, thereby closing a first-kind Volterra equation for the filtering-rate history \(r_X(t)\). (The full derivation, which properly accounts for the conjugate field, is given in Appendix~\ref{app:retarded_volterra_from_one_replica} and yields Eq.~\eqref{eq:volterra_maintext}.)

\subsection{Feynman polaron ansatz}
\label{subsec:feynman-polaron}

We now adopt a Feynman-style effective description of the weak-noise coherence--void polaron. As in the conventional Fr\"ohlich polaron problem~\cite{Frohlich1954,FeynmanPolarCrystal}, we model the retarded self-attraction by introducing an explicit slow coordinate for the translation of the dressed object together with a faster internal degree of freedom. We therefore decompose the coherence coordinate as
\begin{equation}
X(t)=Y(t)+\zeta(t),
\label{eq:polaron-coordinate}
\end{equation}
where \(Y(t)\) describes the motion of the coherence together with its self-generated void, while \(\zeta(t)\) is an internal displacement, heuristically the separation of the coherence from the void center.

The corresponding Gaussian Feynman ansatz is~\cite{FeynmanPolarCrystal}
\begin{equation}
S_{\rm F}[Y,\zeta]
=
\int_0^T dt\,
\left[
\frac{\dot Y(t)^2}{4D_{\rm eff}}
+
\frac{\dot\zeta(t)^2}{4D}
+
\frac{D}{4\ell_\zeta^4}\,\zeta(t)^2
\right].
\label{eq:feynman-ansatz-maintext}
\end{equation}
Here \(D_{\rm eff}\) is the long-time diffusion constant of the translational mode, while \(\ell_\zeta\) sets the width of the internal bound state. We first determine \(D_{\rm eff}\) from the large-deviation cost of translating the dressed object, and then determine \(\ell_\zeta\) variationally below.

\textbf{Translation mode.}
Assuming that the internal mode remains bound, with finite width \(\ell_\zeta\), the asymptotic low-frequency, small-momentum behaviour of the coherence Green's function is governed by the translation mode \(Y\). The coefficient \(D_{\rm eff}\) in Eq.~\eqref{eq:feynman-ansatz-maintext} may therefore be extracted from the large-deviation cost of a low-velocity boost of the dressed object. In particular, a ballistic trajectory $Y(T)=vT+o(T)$ implies $X(T)=vT+o(T)$, so the leading extensive-in-time cost is the same as that of a rigidly translating coherence trajectory. For \(X(t)=vt\), the Volterra equation Eq.~\eqref{eq:volterra_maintext} becomes time-translation invariant, so that $r_X(t)\to r_v$ with $r_v$ defined by
\begin{equation}
r_v
\int_0^\infty du\,
\frac{e^{-\gamma u}}{\sqrt{4\pi D u}}
\exp\!\left[-\frac{v^2u}{4D}\right]
=
\frac{e^{-s}}{2},
\end{equation}
with \(u=t-\tau\). The integral is elementary, yielding
\begin{equation}
r_v=e^{-s}\sqrt{D\gamma+\frac{v^2}{4}},
\qquad
r_0=e^{-s}\sqrt{D\gamma},
\label{eq:rv_maintext}
\end{equation}
where \(r_0\) is the fluid filtering rate for a pinned coherence.

The large-deviation cost per unit time then follows from Eq.~\eqref{eq:Seff_maintext_corrected}:
\begin{equation}
\frac{I(v)-I(0)}{T}
=
\frac{v^2}{4D}
+
2(e^s-1)(r_v-r_0).
\label{eq:Iv_maintext}
\end{equation}
Expanding Eq.~\eqref{eq:rv_maintext} at small \(v\) gives
\begin{align}
\frac{I(v)-I(0)}{T}
&=
\left[
\frac{1}{4D}
+
\frac{1-e^{-s}}{4\sqrt{D\gamma}}
\right]v^2
+\mathcal O(v^4)\nonumber \\
&\equiv
\frac{v^2}{4D_{\rm eff}}+\mathcal O(v^4).
\end{align}
Thus
\begin{equation}
D_{\rm eff}(\gamma)^{-1}
=
D^{-1}
+
\frac{1-e^{-s}}{\sqrt{D\gamma}},
\qquad
D_{\rm eff}(\gamma)\sim \sqrt{\gamma}.
\label{eq:Deff_maintext}
\end{equation}
The same weak-noise scaling is derived in two complementary ways in Appendix~\ref{app:weak_noise_polaron_diffusivity}: from the low-frequency limit of the retarded \(X\)-only action, and independently from the large-deviation rate function of the full nonlinear MFT fluid action without invoking the dilute-fluid approximation.

\textbf{Internal mode.}
We now use the same Gaussian ansatz Eq.~\eqref{eq:feynman-ansatz-maintext} as a trial action to estimate the width \(\ell_\zeta\) of the internal bound state. At this stage \(D_{\rm eff}\) is already fixed by the translation-mode analysis above, so the only remaining variational parameter is \(\ell_\zeta\). Under the corresponding trial measure the confinement width and internal exploration time are given by
\begin{equation}
\langle \zeta^2\rangle_{\rm tr}=\ell_\zeta^2,
\qquad
\tau_\zeta=\frac{\ell_\zeta^2}{D},
\label{eq:lzeta-def}
\end{equation}
where $\langle\cdots\rangle_{\rm tr}$ denotes expectation with respect to the trial measure. 

The optimal \(\ell_\zeta\) is determined by extremizing the Feynman--Jensen variational functional~\cite{FeynmanPolarCrystal}
\begin{equation}
\mathcal F_{\rm var}(\ell_\zeta)
=
\mathcal F_{\rm tr}(\ell_\zeta)
-
\big\langle S_{\rm F}[Y,\zeta]\big\rangle_{\rm tr}
+
\big\langle S_{\rm eff}[X]\big\rangle_{\rm tr},
\label{eq:FJ-functional}
\end{equation}
where
\begin{equation}
\mathcal F_{\rm tr}(\ell_\zeta)\equiv -\log Z_{\rm tr}(\ell_\zeta),
\ \
Z_{\rm tr}(\ell_\zeta)=
\int \mathcal D Y\,\mathcal D\zeta\,e^{-S_{\rm F}[Y,\zeta]}.
\end{equation}
In the present problem, the true action Eq.~\eqref{eq:Seff_maintext_corrected} is non-Gaussian. As such, the non-local action $S_{\rm eff}[X]$ cannot be evaluated under the trial measure exactly. We therefore proceed approximately by annealing the Volterra kernel under the Gaussian trial process. This gives
\begin{align}
\mathcal F_{\rm var}(\ell_\zeta)
\approx\
&\mathcal F_{\rm tr}(\ell_\zeta)
-
\big\langle S_{\rm F}[Y,\zeta]\big\rangle_{\rm tr}
+
\big\langle S_X[X]\big\rangle_{\rm tr}
\nonumber\\
&+
2(e^s-1)\,T\,\overline r_{\ell_\zeta},
\label{eq:FJ-functional-approx}
\end{align}
where \(\overline r_{\ell_\zeta}\) is the annealed filtering rate.

The Gaussian contribution is straightforward. Its \(Y\)-dependent part is independent of \(\ell_\zeta\), while the remaining \(\ell_\zeta\)-dependence is set by the confinement cost of the internal mode,
\begin{equation}
\frac{1}{T}
\left(
\mathcal F_{\rm tr}
-
\langle S_{\rm F}\rangle_{\rm tr}
+
\langle S_X\rangle_{\rm tr}
\right)
\sim
\omega_\zeta
\sim
\frac{D}{\ell_\zeta^2}.
\label{eq:trial-part-scaling}
\end{equation}

It remains to estimate the \(\ell_\zeta\)-dependence of the fluid-conditioning term. The annealed evaluation of the Volterra kernel, given in Appendix~\ref{app:annealed_trial_expectation}, yields
\begin{equation}
\overline r_{\ell_\zeta}
=
e^{-s}\sqrt{D\gamma}
+
\mathcal O(\gamma \ell_\zeta),
\label{eq:rbar_scaling_statement}
\end{equation}
where the first term is just the pinned-coherence filtering rate. The same scaling can be understood heuristically as follows.

Under the trial measure, the internal coordinate \(\zeta\) repeatedly explores a core of width \(\ell_\zeta\) on the timescale $\tau_\zeta$. If \(\ell_\zeta\ll \xi\sim \sqrt{D/\gamma}\), then \(\tau_\zeta\ll \gamma^{-1}\), so during one sweep depolarization creates only a parametrically small density inside the explored core,
\begin{equation}
\bar\rho_{\rm in}\sim \gamma\tau_\zeta
\sim
\frac{\gamma \ell_\zeta^2}{D}
\sim
\left(\frac{\ell_\zeta}{\xi}\right)^2
\ll 1.
\label{eq:rhoin_scaling_main}
\end{equation}
Thus the repeatedly explored inner region remains asymptotically dilute, and the outer fluid profile is, to leading order, just the pinned quasi-stationary void shifted to the edges of this explored core. The leading additional conditioning cost therefore comes from filtering the depolarized spins created inside the explored core itself. Since these are generated at rate \(\gamma\) per unit length, a core of width \(\ell_\zeta\) contributes an excess filtering rate
\begin{equation}
\delta r(\ell_\zeta)\sim \gamma \ell_\zeta,
\label{eq:delta_r_heuristic_main}
\end{equation}
in agreement with Eq.~\eqref{eq:rbar_scaling_statement}.

Combining Eq.~\eqref{eq:trial-part-scaling} with Eq.~\eqref{eq:delta_r_heuristic_main}, the \(\ell_\zeta\)-dependent part of the approximate variational functional has the parametric form
\begin{equation}
\frac{\mathcal F_{\rm var}}{T}
\sim
\frac{D}{\ell_\zeta^2}
+
\gamma \ell_\zeta.
\label{eq:Fvar-balance}
\end{equation}
Extremizing with respect to \(\ell_\zeta\) gives
\begin{equation}
\ell_\zeta \sim \left(\frac{D}{\gamma}\right)^{1/3},
\qquad
\omega_\zeta \sim D^{1/3}\gamma^{2/3}.
\label{eq:lzeta_scaling_screened}
\end{equation}
Thus the coherence is bound to the slow polaron coordinate on a scale \(\ell_\zeta\ll \xi\), so the internal wandering remains parametrically narrower than the full screened void. This in turn justifies the rigid-translation estimate used above to extract \(D_{\rm eff}\).

\subsection{Localization to a slow bond}
\label{sec:slow-bond-localization}

Despite its extended, slowly varying density profile, the coherence--void polaron is distinctly non-hydrodynamic. This is evident not only in its weak-noise spectral structure, where its decay rate has a different \(\gamma\)-dependence from that of ordinary hydrodynamic modes, but also in its response to impurities. We now consider the coherence--void polaron in the presence of a single slow bond of hopping rate \(q<1\), with all other bonds having rate \(1\). Whereas an ordinary hydrodynamic mode remains delocalized in the presence of such a local defect, the coherence--void polaron is trapped by it, with a localization length \(\ell_{\rm loc}\sim \gamma^{-1/3}\), as we show below.

Impurities are well known to localize polarons~\cite{PhysRevB.5.3029,PhysRevB.63.184304,PhysRevB.79.180301,PhysRevB.35.7533}. For the coherence--void polaron, the key point is that the quasi-stationary decay rate is set by the large-deviation cost of maintaining the surrounding void. This cost is given by the fluid action Eq.~\eqref{eq:Sbulk_1rep_general} evaluated on the optimal density and response fields, and is local: for a spatially inhomogeneous profile, the local cost density is $\mathcal{O}(D(\partial_x\rho)^2)$. A hopping impurity, i.e. a local reduction of the diffusivity $D\to D(x)$, therefore lowers the large-deviation cost whenever it lies inside the polaron core. The resulting trapping potential is therefore extended across the full polaron core.

To make this precise, fix the coherence at position $y$ relative to the slow bond and consider the corresponding pinned coherence quasi-stationary problem as in Sec.~\ref{subsec:weak_noise_stationary_saddle}. The slow bond perturbs the pinned-coherence QSS decay rate, which we determine to leading order in perturbation theory around the homogeneous (pinned) saddle.

For a local hopping defect, the local diffusivity and mobility are perturbed together,
\begin{equation}
D\to D(x)=D+\delta D(x),
\ \
\sigma(\rho)\to \sigma(x,\rho)=D(x)\chi(\rho),
\nonumber
\end{equation}
with $\delta D(x)$ localized on an $O(1)$ region around the defect and of total weight proportional to $1-q$. Since the clean pinned profile is already a saddle solution, the first-order shift of the QSS eigenvalue is obtained by differentiating the stationary action with respect to this local perturbation and evaluating the result on the unperturbed pinned saddle solution $(\rho_y,\pi_y)$. This gives
\begin{align}
\Delta \lambda_{\rm QSS}(y)
=
&\int dx\,\delta D_q(x)
\Big[
\partial_x\pi_y(x)\,\partial_x\rho_y(x)\nonumber\\
&\qquad \qquad -\frac{\chi(\rho_y(x))}{2}\big(\partial_x\pi_y(x)\big)^2
\Big]+\cdots,
\label{eq:delta-lambda-pert}
\end{align}
where the $\cdots$ indicates terms of order $\mathcal{O}\big((1-q)^2\big)$. Since the defect is local, Eq.~\eqref{eq:delta-lambda-pert} samples the unperturbed profile only in an $\mathcal{O}(1)$ region around the defect.

In the weak-noise stationary saddle, both $\partial_x\rho_y$ and $\partial_x\pi_y$ are of order $\xi^{-1}e^{-|y|/\xi}$, so the term in the square brackets in Eq.~\eqref{eq:delta-lambda-pert} is of order $\xi^{-2}e^{-2|y|/\xi}$. We therefore obtain
\begin{equation}
\Delta \lambda_{\rm QSS}(y)
\sim
-\frac{\varrho(q)}{\xi^2}\,e^{-2|y|/\xi}
\sim
-\varrho(q)\gamma\,e^{-2|y|/\xi},
\label{eq:slowbond-total}
\end{equation}
to leading order in perturbation theory (and where $\varrho(q)\sim 1-q$ as $q\to 1$). Thus the defect acts as an attractive potential well for the coherence--void polaron, with a range $\mathcal{O}(\xi)$ inherited from the size of the polaron core.

Having now identified the defect-induced shift of the pinned-coherence quasi-stationary decay rate, it remains to specify the kinetic term for the dressed object. In the following, we ignore the internal motion of the coherence inside the polaron core and retain only rigid translations of the coherence--void polaron, i.e., the motion of the slow polaron coordinate introduced in Sec.~\ref{subsec:feynman-polaron}. This mode is simply diffusive, with parametrically small effective diffusion constant $D_{\rm eff}(\gamma)\sim \sqrt{\gamma}$ (Eq.~\eqref{eq:Deff_maintext}).

Its effective action is therefore
\begin{equation}
S_{\rm tr}[y]=\int_0^T dt\,\frac{\dot y(t)^2}{4D_{\rm eff}(\gamma)},
\end{equation}
while the pinned-QSS shift provides the effective potential
\begin{equation}
V(y)\equiv -\Delta\lambda_{\rm QSS}(y).
\end{equation}
The localization problem is therefore an imaginary-time Schr\"odinger problem,
\begin{equation}
-D_{\rm eff}(\gamma)\,\psi''(y)+V(y)\psi(y)=E\,\psi(y),
\label{eq:airy_schrod}
\end{equation}
where $\psi(y)$ is an eigenfunction of the Schr\"odinger operator. The long-time quasi-stationary distribution of polaron positions is governed by its ground state.

For $|y|\ll \xi$, the effective potential expands as
\begin{equation}
V(y)=V(0)+\varrho(q)\,\frac{\gamma}{\xi}\,|y|
+O\!\left(\frac{\gamma y^2}{\xi^2}\right).
\label{eq:wedge_slope}
\end{equation}
Dropping the irrelevant additive constant $V(0)$, this is a linear wedge potential at leading order. Rescaling $y=\ell_{\rm loc}z$ with
\begin{equation}
\ell_{\rm loc}
\equiv
\left(
\frac{D_{\rm eff}(\gamma)}{\varrho(q)\gamma^{3/2}}
\right)^{1/3}
\sim
\gamma^{-1/3}\,\varrho(q)^{-1/3},
\label{eq:ell_loc_main}
\end{equation}
reduces Eq.~\eqref{eq:airy_schrod} to the universal Airy form
\begin{equation}
(-\partial_z^2+|z|)\psi=\varepsilon_0\psi,
\label{eq:Airy}
\end{equation}
(where $\varepsilon_0$ is the ground state energy) so the localization length is set by the Airy scale Eq.~\eqref{eq:ell_loc_main}. Thus a single slow bond localizes the coherence--void polaron on the scale $\ell_{\rm loc}\sim \gamma^{-1/3}$. This is the same scaling as the internal binding length of the coherence inside the coherence--void polaron. The detailed localization profile is therefore modified by the internal motion, but the localization-length scaling itself is robust.

\begin{figure*}[t]
    \centering
    \includegraphics[width=0.99\textwidth]{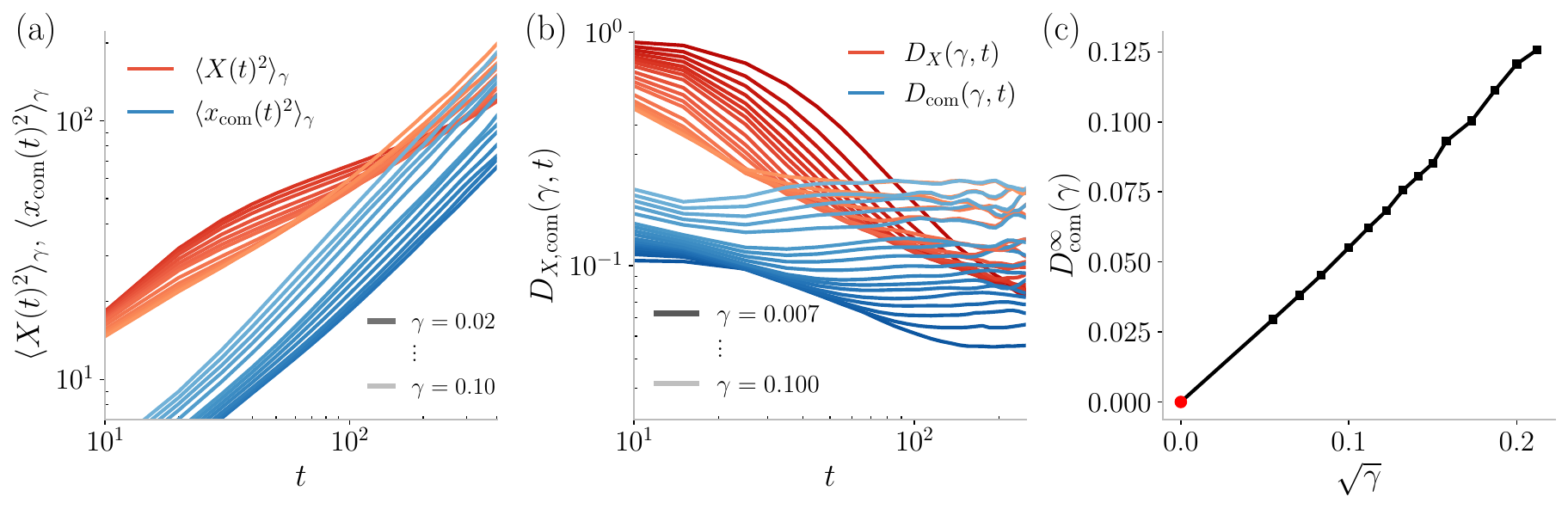}
\caption{
\textbf{Weak noise coherence--void polaron translation mode.}
(a) Mean-squared displacements of the coherence, \(\langle X(t)^2\rangle\), and of the void center of mass, \(\langle x_{\rm com}(t)^2\rangle\), for several values of \(\gamma\). At early times the coherence moves rapidly inside an already well-polarized quasi-stationary void, while the void center of mass diffuses much more slowly. At later times the coherence motion crosses over to match the void center of mass, indicating binding of the coherence to the slow polaron coordinate.
(b) Corresponding time-dependent diffusion constants \(D_X(\gamma,t)=\tfrac12\partial_t\langle X(t)^2\rangle\) and \(D_{\rm com}(\gamma,t)=\tfrac12\partial_t\langle x_{\rm com}(t)^2\rangle\). For those values of \(\gamma\) where a clear plateau is reached, the late-time agreement of \(D_X\) and \(D_{\rm com}\) shows directly that the coherence and void center move together as a single dressed object.
(c) Long-time diffusivity extracted from the void center-of-mass motion, \(D_{\rm com}^\infty(\gamma)\), plotted against \(\sqrt{\gamma}\). The approximately linear dependence is consistent with the weak-noise prediction \(D_{\rm eff}(\gamma)\sim \sqrt{\gamma}\).
Numerical data are obtained from large-scale population-dynamics sampling~\cite{Giardina2011} of the dilute-coherence approximate sub-stochastic evolution (Sec.~\ref{subsec:dilute-coherence}) as described in the main text. Simulations were performed on a chain of length \(L=600\) for \(\gamma\in\{0.007,\,0.01,\cdots,\,0.09,\,0.1\}\), using \(200\) independent runs with \(2\times10^5\) clones for each \(\gamma\), after a burn-in time \(t_{\rm burn}=5/\gamma\). The values \(D_{\rm com}^\infty(\gamma)\) in (c) are obtained by averaging \(D_{\rm com}(\gamma,t)\) over the time window \(t\in[150,400]\).}
\label{fig:no7_screened_X_COM_MSD}
\end{figure*}

\subsection{Numerical evidence}
\label{subsec:screened_Deff_numerics}

\textbf{Translation mode.}
We now test the predicted slow diffusion of the coherence--void polaron in Eq.~\eqref{eq:Deff_maintext}, with diffusion constant \(D_{\rm eff}(\gamma)\sim \sqrt{\gamma}\). For this numerical test we work within the dilute-coherence approximation (Sec.~\ref{subsec:dilute-coherence}). The reason is practical: the void center of mass is straightforward to extract from trajectory-based sampling of the approximate sub-stochastic dynamics, but is much harder to access directly in the full two-replica transfer-matrix evolution. As a check on this approximation, Appendix~\ref{app:weak_noise_polaron_diffusivity} compares direct TEBD evolution of the full two-replica statistical mechanics model with that of the dilute-coherence approximation. Although direct TEBD simulations cannot reach the late-time plateau  in diffusivity accessible to population dynamics, they show that the time-dependent coherence diffusion constant agrees between the full and dilute models at small \(\gamma\) over the accessible late-time window.

To define the void center of mass it is convenient to use an overcomplete local basis in which the identity operator \(\hat{\mathbb{1}}\) is kept as an explicit symbol: depolarizing events then replace polarized spins \(\mathbf{u/d}\) in either replica by \(\hat{\mathbb{1}}\), with the \(\hat{\mathbb{1}}\)'s subsequently evolved under the same SSEP excluded hopping. Along each sampled trajectory the system spontaneously selects one of the two void polarizations. We therefore identify, at each time, the majority polarized symbol (\(\mathbf u\) or \(\mathbf d\)) and define the void center of mass \(x_{\rm com}(t)\) as the center of mass of that majority species. Since these majority symbols are sourced by the coherence and removed only by depolarization, they provide a direct microscopic proxy for the void.

Figure~\ref{fig:no7_screened_X_COM_MSD}(a) shows the coherence mean-squared displacement \(\langle X(t)^2\rangle\) together with the void center-of-mass MSD \(\langle x_{\rm com}(t)^2\rangle\), while Fig.~\ref{fig:no7_screened_X_COM_MSD}(b) shows the corresponding time-dependent diffusion constants
\begin{equation}
D_X(\gamma,t)=\tfrac12\,\partial_t\langle X(t)^2\rangle,
\qquad
D_{\rm com}(\gamma,t)=\tfrac12\,\partial_t\langle x_{\rm com}(t)^2\rangle.
\nonumber
\end{equation}
The data show a clear two-stage dynamics. At early times the coherence moves rapidly, with an \(O(1)\), approximately \(\gamma\)-independent diffusion constant, since it initially explores the interior of an already well-polarized void. By contrast, the void center of mass diffuses much more slowly from the outset. At later times, once the coherence has wandered far enough to probe the denser edges of the screened void, its motion incurs the conditioning cost associated with refilling the void, and its effective diffusion constant crosses over toward that of the void center of mass. For those values of \(\gamma\) where a clear plateau is reached, the agreement of \(D_X(\gamma,t)\) and \(D_{\rm com}(\gamma,t)\) at late times shows directly that the coherence is bound to the void core and that the two move together as a single dressed object.

As \(\gamma\) is reduced, the crossover in \(D_X(\gamma,t)\) moves to progressively later times, preventing a direct extraction of the asymptotic dressed diffusion constant from the coherence MSD alone. By contrast, \(D_{\rm com}(\gamma,t)\) saturates much earlier and therefore provides a much cleaner estimate of the long-time diffusivity. The resulting values of \(D_{\rm com}^\infty(\gamma)\), shown in Fig.~\ref{fig:no7_screened_X_COM_MSD}(c), display a clear linear dependence on \(\sqrt{\gamma}\), consistent with Eq.~\eqref{eq:Deff_maintext}.

\textbf{Internal motion.}
We test this prediction using the same Monte Carlo trajectory sampling introduced in the previous section, now measuring the relative-coordinate MSD, $\big\langle \big(X(t)-x_{\rm com}(t)\big)^2 \big\rangle$.

\begin{figure}[t]
    \centering
    \includegraphics[width=1.\linewidth]{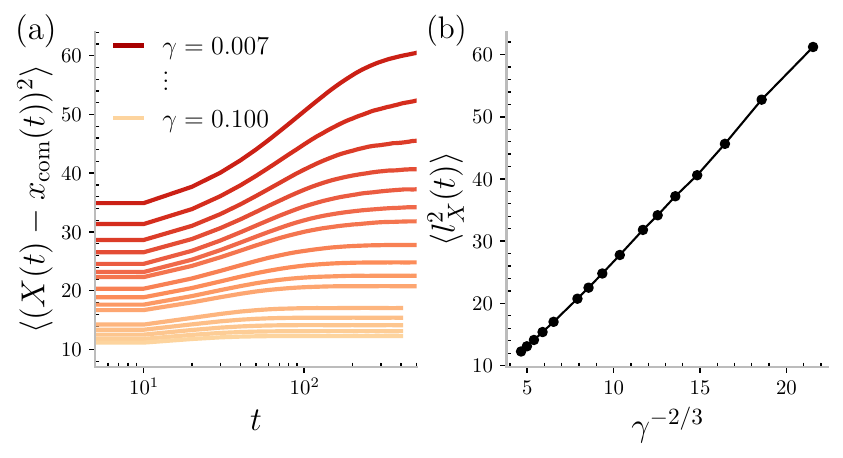}
    \caption{
\textbf{Internal motion of the weak noise coherence--void polaron.}
(a) Mean-squared displacement of the relative-coordinate of the coherence and the void center-of-mass, \(\langle (X(t)-x_{\rm com}(t))^2\rangle\), for \(\gamma=0.007,\,0.01,\ldots,\,0.1\). In each case the relative-coordinate MSD approaches a late-time plateau, showing that the coherence remains confined inside the void, with the confinement length increasing as \(\gamma\) is reduced. (b) Plateau value \(\ell_X^2(\gamma)\), extracted from the late-time relative-coordinate MSD, plotted against \(\gamma^{-2/3}\). The near-linear dependence is consistent with the Feynman-polaron prediction \(\ell_\zeta\sim \gamma^{-1/3}\). These data are obtained from the same dilute-coherence population-dynamics sampling as in Fig.~\ref{fig:no7_screened_X_COM_MSD}.
}
    \label{fig:screened_relative_MSD}
\end{figure}

As shown in Fig.~\ref{fig:screened_relative_MSD}(a), the relative-coordinate MSD approaches a plateau for every $\gamma$, demonstrating that the coherence remains confined inside the screened void. The plateau height increases as $\gamma$ is reduced, consistent with the broadening of the internal exploration region. Defining $\ell_X^2(\gamma)$ from the late-time plateau value, Fig.~\ref{fig:screened_relative_MSD}(b) shows a near-linear dependence on $\gamma^{-2/3}$, in good agreement with Eq.~\eqref{eq:lzeta_scaling_screened}.

These simulations are obtained in the dilute approximation (see Sec.~\ref{subsec:dilute-coherence}). The full transfer-matrix numerics in the next subsection provide an additional indirect check: the slow-bond localization length scales in the same way, $\ell_{\rm loc}\sim \gamma^{-1/3}$, consistent with the conclusion that the internal wandering scale is not parametrically larger than this scale.

\begin{figure*}[t]
    \centering
    \includegraphics[width=1.\linewidth]{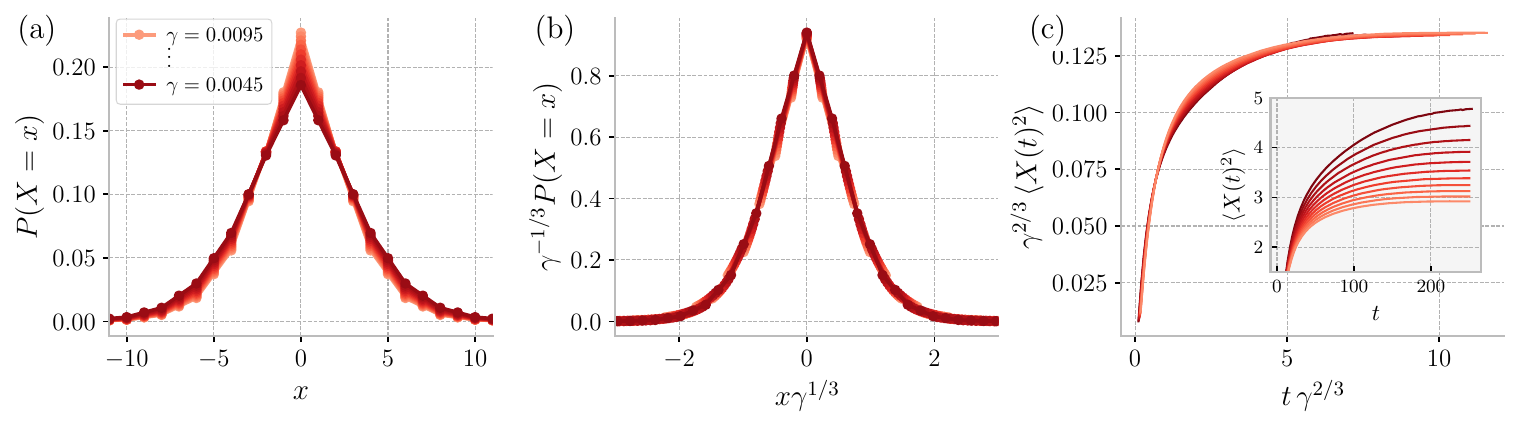}
    \caption{
\textbf{Localization of the coherence--void polaron to a slow bond.}
(a) Late-time stationary coherence-position distribution \(P(X=x)\) in the presence of a single slow bond at \(x=0\). 
(b) Rescaled late-time position distributions plotted as \(\gamma^{-1/3}P(X=x)\) against \(\gamma^{1/3}x\). The collapse is consistent with the scaling form \(P(X=x)\sim \gamma^{1/3} f(\gamma^{1/3}x)\), corresponding to a localization length \(\ell_{\rm loc}\sim \gamma^{-1/3}\). 
(c) Rescaled mean-squared displacement \(\gamma^{2/3}\langle X(t)^2\rangle\) plotted against \(\gamma^{2/3}t\). The late-time saturation is consistent with the Airy scaling predicted by Eq.~\eqref{eq:airy_schrod}: each curve saturates at \(\langle X(t)^2\rangle\sim \gamma^{-2/3}\) on a timescale \(t\sim \gamma^{-2/3}\), corresponding to \(\ell_{\rm loc}\sim \gamma^{-1/3}\) and \(\tau_{\rm loc}\sim \gamma^{-2/3}\).
Numerical data are obtained from TEBD evolution~\cite{tenpy2024} of the full two-replica transfer matrix with a single slow bond of hopping rate \(q=0.2\) (all other bonds having rate \(1\)), on an open chain of length \(L=40\) with bond dimension \(\chi=500\), for \(\gamma=0.0035,\,0.004,\ldots,0.0085\). The initial state is an equal-weight superposition of coherence states on the two sites adjacent to the defect, and the dynamics is evolved to time \(t=250\).
}
    \label{fig:screened_slowbond_localization}
\end{figure*}

\textbf{Localization to a slow bond.}
Equation~\eqref{eq:airy_schrod} predicts a localization length $\ell_{\rm loc}(\gamma)\sim \gamma^{-1/3}$ and time $\tau_{\rm loc}\sim \gamma^{-2/3}$, and equivalently the small $\gamma$ scaling forms
\begin{equation}
P(X=x)\sim \gamma^{1/3}f(\gamma^{1/3}x),
\quad
\gamma^{2/3}\,\langle X(t)^2\rangle = g(\gamma^{2/3}t),\nonumber
\end{equation}
for the normalized coherence-position distribution relative to the slow bond at \(x=0\).

To test this prediction, we simulate the full two-replica transfer-matrix dynamics in the presence of a single slow bond and extract the normalized coherence-position distribution \(P(x,t)\). The late-time distribution is shown in Fig.~\ref{fig:screened_slowbond_localization}(a), with the corresponding rescaled profiles in Fig.~\ref{fig:screened_slowbond_localization}(b). The data show that the coherence remains localized near the defect, with a width consistent with \(\ell_{\rm loc}\sim \gamma^{-1/3}\). Figure~\ref{fig:screened_slowbond_localization}(c) shows the rescaled MSD, \(\gamma^{2/3}\langle X(t)^2\rangle\), plotted against \(\gamma^{2/3}t\).  For each \(\gamma\), the MSD saturates at \(\langle X(t)^2\rangle\sim \gamma^{-2/3}\) on a timescale \(t\sim \gamma^{-2/3}\), in agreement with the Airy scaling form above.

Since the measured coordinate is the physical coherence position relative to the slow bond, this observable includes both motion of the slow polaron coordinate and internal wandering of the coherence inside the void. The observed \(\gamma^{-1/3}\) localization scale therefore also supports the conclusion that the internal wandering is not parametrically larger than \(O(\gamma^{-1/3})\).

\section{Dynamics of the coherence-void polaron: zero-noise case $\gamma = 0$}
\label{sec:aging-polaron}

We now turn to the noiseless limit of the coherence propagator \(\mathcal G(x,t)\equiv \langle \sigma_x^+(t)\sigma_0^-(0)\rangle\). %As in the weak-noise case, we quantify its typical exploration scale through the phase-averaged norm-squared propagator \(\overline{|\mathcal G(x,t)|^2}\), whose coarse-grained description is obtained by integrating out the fluid degrees of freedom.
As in the weak-noise case, the coherence trajectory \(X(t)\) is governed by a self-attractive walk, since motion through regions already depleted by its past history incurs a lower fluid-conditioning cost than motion into fresh, unconditioned fluid. At \(\gamma=0\), however, this same self-attractive mechanism leads to qualitatively different long-time dynamics. As shown already in Sec.~\ref{subsec:noiseless_aging_saddle}, the optimal void now grows diffusively, \(\xi_{\gamma=0}(t)\sim \sqrt{Dt}\), rather than saturating at a finite size. The resulting dressed object is thus an \emph{aging} coherence--void polaron. Because the self-attraction is unscreened, the future trajectory remains coupled to its full history back to the insertion point of the Green's function, \(X(0)=0\). It is therefore not organized around a freely translating center, and there is no analogue of the weak-noise soft translation mode. We accordingly estimate the exploration scale of this aging polaron directly from a Feynman-style trial process for \(X(t)\), with a time-dependent width \(\ell_X(t)\) as the single variational parameter.

Within the Feynman-style variational closure developed below, one finds subdiffusive wandering of the coherence, \(X(t)\sim t^{1/3}\), corresponding to a dynamical exponent \(z=3\). Together with the leading hydrodynamic large-deviation cost of sustaining the aging void, the leading spatio-temporal structure of the coherence Green's function is captured by the scaling form
\refstepcounter{equation}\label{eq:noiseless_scaling_form_opening}
\begin{wideeqbox}
\vspace{3pt}
\noindent
\makebox[\linewidth][l]{%
  \makebox[3em][l]{}%
  \makebox[\dimexpr\linewidth-8em\relax][c]{%
    $\displaystyle
    |\mathcal G(x,t)|^2
    \sim
    \exp\!\left[-c\sqrt{t}\right]\,
    F\bigg(\frac{x}{t^{1/3}}\bigg).
    $%
  }%
  \makebox[5em][r]{(\theequation)}%
\vspace{6pt}
}
\end{wideeqbox}
\noindent with \(c>0\) nonuniversal. We now derive this aging-polaron description from the retarded coherence action and then analyze the motion of the void center and the relative coordinate.

The weak-noise result in Eq.~\eqref{eq:weaknoise_scaling_form_opening} and the zero-noise result above correspond to different, non-commuting orders of limits. Equation~\eqref{eq:weaknoise_scaling_form_opening} describes the long-time limit at fixed \(\gamma>0\), for which \(t\gg\gamma^{-1}\) and the void has reached its screened quasi-stationary form, followed by the weak-noise limit \(\gamma\to0\). By contrast, Eq.~\eqref{eq:noiseless_scaling_form_opening} is obtained by setting \(\gamma=0\) first and then taking \(t\to\infty\). At small but finite \(\gamma\), the noiseless aging regime therefore persists for \(1\ll t\ll\gamma^{-1}\), before crossing over at \(t\sim\gamma^{-1}\) to the stationary weak-noise regime.
(Indeed, at the crossover \(t\sim\gamma^{-1}\), the noiseless void lengthscale \(\xi(t)\sim t^{1/2}\) and coherence wandering lengthscale \(X(t)\sim t^{1/3}\) become \(\gamma^{-1/2}\) and \(\gamma^{-1/3}\), respectively, matching the weak-noise void size and coherence confinement length.)

\subsection{Feynman-style ansatz for the aging polaron}
The effective weight for a trajectory $X(t)$ is given by the same non-local non-Gaussian action given by Eq.~\eqref{eq:Seff_maintext_corrected}, but with the filtering rate \(r_X(t)\) now determined by the noiseless Volterra equation
\begin{equation}
\int_0^t d\tau\,
G\!\left(X(t)-X(\tau),\,t-\tau\right)\,r_X(\tau)
=
\frac{e^{-s}}{2},
\label{eq:volterra_maintext_noiseless}
\end{equation}
with conservative diffusion kernel
\begin{equation}
G(\Delta x,\tau)
=
\frac{1}{\sqrt{4\pi D \tau}}
\exp\!\left[-\frac{(\Delta x)^2}{4D\tau}\right].
\label{eq:kernel_maintext_noiseless}
\end{equation}

For the purpose of estimating the exploration scale, we approximate this history-dependent attraction by an effective attraction toward the center of the growing void $x=0$ using a Feynman-style variational ansatz of a time-dependent harmonic well~\cite{FeynmanPolarCrystal}. Concretely, we take
\begin{equation}
S_{\rm tr}[X]
=
\int_0^T dt\,
\left[
\frac{\dot X(t)^2}{4D}
+
\frac{D}{4\ell_X(t)^4}\,X(t)^2
\right],
\label{eq:trial_action_X_main}
\end{equation}
which treats the coherence as moving in a time-dependent harmonic potential whose width is set by \(\ell_X(t)\). At time \(t\) the trial process has instantaneous confinement width
\begin{equation}
\langle X(t)^2\rangle_{\rm tr}\sim \ell_X(t)^2,
\end{equation}
and the associated exploration timescale is
\begin{equation}
\tau_X(t)\sim \frac{\ell_X(t)^2}{D}.
\label{eq:tauX_main}
\end{equation}
As in the weak-noise polaron problem, the Feynman--Jensen functional involves the expectation of the true non-local \(X\)-action Eq.~\eqref{eq:Seff_maintext_corrected} under the trial measure, which here is not available in closed form. We therefore use the same annealed approximation as in the weak-noise case: we average the kernel in Eq.~\eqref{eq:volterra_maintext_noiseless} under the trial process, and use the resulting approximate retarded functional to determine \(\ell_X(t)\) variationally. The corresponding annealed estimate, derived in
Appendix~\ref{app:annealed_trial_expectation}, gives
\begin{equation}
\delta r_X(t)\equiv r_X(t)-r_0(t)\sim \frac{\ell_X(t)}{t},
\label{eq:deltar_scaling_main}
\end{equation}
where \(r_0(t)\) is the pinned \(X(t)=0\) filtering rate. We now argue for this scaling on physical grounds.

Under the trial measure, the coherence explores an inner core region of width \(\ell_X(t)\) on the timescale \(\tau_X(t)\sim \ell_X(t)^2/D\). Assuming \(\ell_X(t)\sim t^{\alpha}\) with $\alpha < 1/2$, this explored region remains parametrically narrower than the diffusively growing void and is rapidly swept over by the coherence on the timescale over which the outer profile evolves. In the pinned-coherence zero-noise saddle, the baseline fluid current at the coherence is \(J_0(t)\sim t^{-1/2}\), so the total number of depolarized spins entering the explored core during one sweep is \(N_{\rm in}(t)\sim J_0(t)\tau_X(t)\sim t^{-1/2}\ell_X(t)^2/D\). The corresponding mean density is therefore \(\bar\rho_{\rm in}(t)\sim N_{\rm in}(t)/\ell_X(t)\sim \ell_X(t)/\sqrt{Dt}\), which vanishes as $t\to \infty$. Thus, as in the weak-noise problem, the outer void sees, to leading order, an effectively empty inner core, so that the outer profile is governed by the same pinned coherence saddle equations except with the Dirichlet condition shifted to the edges of the expanding explored region.

In the weak-noise problem, the explored region has a fixed size, and the leading excess fluid-conditioning cost is simply the cost of filtering the additional depolarization events generated inside the explored core. In the noiseless problem there are no such bulk depolarizing events. The excess cost above the pinned \(X(t)=0\) reference must therefore come from the additional filtering required to expand the explored region in time, namely from the slow outward motion of the effective Dirichlet boundary.

We now estimate this growth-induced cost directly. We assume an algebraic exploration law
\begin{equation}
\ell_X(t)\sim t^\alpha,\quad \tau_X(t)\sim \frac{\ell_X(t)^2}{D}\sim t^{2\alpha},
\quad
\alpha<\frac12.
\label{eq:alpha_assumption_main}
\end{equation}
To extract the excess filtering rate, we work on the right half-line and move to the coordinate $y=x-\ell_X(t)$, so that the slowly moving Dirichlet boundary is fixed at \(y=0\). Writing the density as
\begin{equation}
\rho(y,t)=\rho_0(y,t)+\rho_1(y,t)+\cdots,
\end{equation}
with \(\rho_0\) the pinned-coherence saddle solution and \(\rho_1\) the first correction induced by the slow boundary motion, the linearized diffusion equation becomes
\begin{equation}
(\partial_t-D\partial_y^2)\rho_1(y,t)
=
-\dot\ell_X(t)\,\partial_y\rho_0(y,t),
\label{eq:rho1_linear_main}
\end{equation}
with $\rho_1(0,t)=0$, %Thus the slow outward motion of the explored core appears as a source proportional to the boundary gradient of the void profile for a pinned coherence. 
which is solved using the half-line Dirichlet heat kernel
\begin{equation}
G_D(y,z,t)
=
\frac{1}{\sqrt{4\pi Dt}}
\left[
e^{-(y-z)^2/(4Dt)}
-
e^{-(y+z)^2/(4Dt)}
\right].
\label{eq:halfline_dirichlet_kernel_main}
\end{equation}
The excess boundary gradient on the right half-line is therefore
\begin{align}
\partial_y\rho_1(0,t)
=
-\int_0^t d\tau\,\dot\ell_X(\tau)
\int_0^\infty dz &\,
\partial_y G_D(0,z,t-\tau)\nonumber\\
&\times \partial_z\rho_0(z,\tau).
\label{eq:boundary_gradient_response_main}
\end{align}

The unperturbed optimal density profile $\rho_0$ supplies an \(\mathcal{O}(\tau^{-1/2})\) fluid gradient over a diffusive range \(z\sim \sqrt{\tau}\). Meanwhile, \(\partial_y G_D(0,z,t-\tau)\) contributes a factor \((t-\tau)^{-3/2}\) and cuts off the \(z\)-integral at \(z\sim \sqrt{t-\tau}\). Thus the \(z\)-integral in Eq.~\eqref{eq:boundary_gradient_response_main} scales as $\left[\tau(t-\tau)\right]^{-1/2}$. Hence
\begin{equation}
\partial_y\rho_1(0,t)
\sim
\frac{1}{D}\int_0^t d\tau\,
\frac{\dot\ell_X(\tau)}{\sqrt{\tau}\sqrt{t-\tau}}\sim \frac{\ell_X(t)}{Dt},
\label{eq:boundary_gradient_scaling_main}
\end{equation}
where the second equality follows for \(\ell_X(t)\sim t^\alpha\). Since the filtering rate is set by the outward gradient jump, the full excess rate has the same scaling as Eq.~\eqref{eq:deltar_scaling_main}, $\delta r_X(t)\sim {\ell_X(t)}/{t}$.

As in the weak-noise polaron calculation, the remaining pieces of the Feynman--Jensen functional are those that can be evaluated exactly under the Gaussian trial measure. Their \(\ell_X\)-dependence is simply the confinement cost of the trial harmonic well, so the variational functional (up to \(\ell_X\)-independent contributions) takes the form
\begin{equation}
\mathcal F_{\rm var}[\ell_X(t),T]
\sim
\int_0^T \! dt\left(c_g\frac{D}{\ell_X(t)^2}
+
c_r\frac{\ell_X(t)}{t}\right),
\label{eq:Fvar_reduced_noiseless_main}
\end{equation}
with \(c_r,c_g=\mathcal O(1)\).

Extremizing pointwise then gives
\begin{equation}
\ell_X(t)\sim (Dt)^{1/3},
\label{eq:ellX_scaling_main}
\end{equation}
which self-consistently remains parametrically smaller than the diffusive void scale \(\xi \sim \sqrt{Dt}\). The coherence therefore explores a region whose characteristic width grows as
\begin{equation}
X(t)\sim t^{1/3}.
\label{eq:X_t13_main}
\end{equation}
Combining this subdiffusive exploration scale with the leading-order stretched exponential decay of Eq.~\eqref{eq:noiseless_correlator_scaling} (for the stationary coherence $X(t)=0$), we are led to the leading order scaling form in Eq.~\eqref{eq:noiseless_scaling_form_opening} for the phase-averaged norm-squared raising-operator Green's function.

We also note that the void center of mass and the relative coordinate inherit the same wandering exponent as the coherence itself. The aging void is built up by the history of filtering events along the coherence worldline, so its center is determined by a weighted time-average of the past coherence trajectory, $x_{\rm com}(t)\approx N(t)^{-1}\int_0^t d\tau\, r(\tau) X(\tau)$, where \(r(t)\sim t^{-1/2}\) is the filtering rate and $N(t)=\int_0^t d\tau\, r(\tau)\sim \sqrt{t}$ is the total number of filtering events. For \(X(t)\sim t^{1/3}\), this weighted average has the same scaling, \(x_{\rm com}(t)\sim t^{1/3}\). The same exponent also follows for the relative coordinate \(X(t)-x_{\rm com}(t)\) (see Appendix~\ref{appsec:aging_com}).

\begin{figure*}
    \centering
    \includegraphics[width=1.\linewidth]{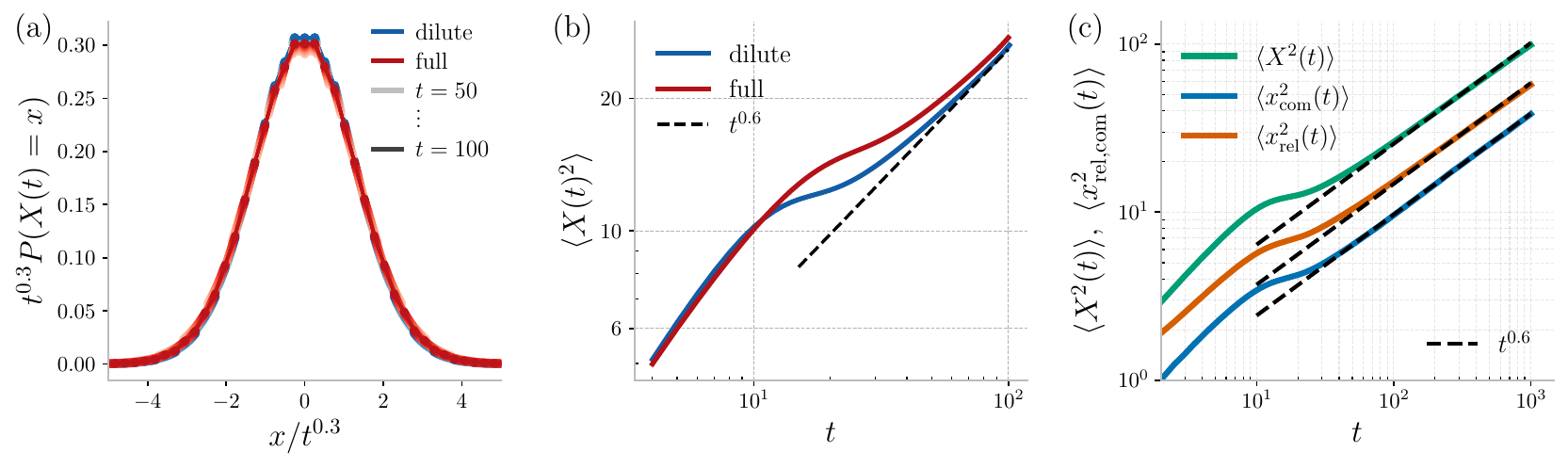}
    \caption{
\textbf{Subdiffusion of the zero--noise aging coherence--void polaron.}
(a) Phase-averaged coherence envelope \(\overline{|\mathcal G(x,t)|^2}\), normalized at each time and plotted against \(x/t^\alpha\). A good collapse is obtained for \(\alpha\simeq 0.3\), with close agreement between the full two-replica statistical mechanics model (``full'') and the dilute-coherence approximation (``dilute'').
(b) Second moment of the normalized coherence envelope, \(\langle X(t)^2\rangle\), showing subdiffusive growth \(\langle X(t)^2\rangle\sim t^{0.60}\) over the accessible late-time window, again with close agreement between the full model and the dilute-coherence approximation.
(c) Mean-squared displacements of the coherence, the void center of mass \(x_{\rm com}(t)\), and the relative coordinate \(x_{\rm rel}(t)\equiv X(t)-x_{\rm com}(t)\). All three observables display essentially the same subdiffusive growth, showing that the subdiffusive polaron scaling is robust.
Panels (a,b) show TEBD simulations~\cite{tenpy2024} on an open chain of length \(L=60\) with bond dimension \(\chi=500\) and evolved up to time \(t=100\). The data in panel (c) is obtained using population dynamics sampling in the dilute-coherence approximation.
}
    \label{fig:aging_3panel_wide}
\end{figure*}

\subsection{Numerical evidence}

We test the predicted subdiffusive wandering using TEBD simulations of the two-replica statistical-mechanics transfer matrix, both for the full microscopic model and for the dilute-coherence approximation introduced in Sec.~\ref{subsec:dilute-coherence}. The quantity directly accessed in the two-replica evolution is the phase-averaged Green's function \(\overline{|\mathcal G(x,t)|^2}\), obtained by overlapping the evolved statistical state with the local coherence basis states \(|(\sigma^+\otimes \sigma^-)_{x}\rangle\). We normalize \(\overline{|\mathcal G(x,t)|^2}\) to a probability distribution over positions to isolate its spatial broadening.

Figure~\ref{fig:aging_3panel_wide}(a) shows the resulting rescaled profiles plotted against \(x/t^\alpha\). We find a good collapse for \(\alpha\simeq 0.3\), with close agreement between the full statistical mechanics model and the dilute-coherence approximation, and in reasonably good agreement with the predicted scaling form \(\overline{|\mathcal G(x,t)|^2}\sim e^{-c\sqrt{t}}F(x/t^{1/3})\).

In Fig.~\ref{fig:aging_3panel_wide}(b) we show the second moment
\begin{equation}
\langle X(t)^2\rangle \equiv \sum_x x^2\,\overline{|\mathcal G(x,t)|^2}/\mathcal N(t),
\ \
\mathcal N(t)\equiv \sum_x \overline{|\mathcal G(x,t)|^2}.
\nonumber
\end{equation}
Over the accessible late-time window the MSD is well described by a power law \(\langle X(t)^2\rangle\sim t^{0.6}\) (again with close agreement between the full model and the dilute-coherence approximation). The observed wandering is slightly more subdiffusive than the simple \(t^{2/3}\) estimate from the approximate Markovian variational closure, but nevertheless shows a clear subdiffusive wandering scale.

%Since the Feynman-style aging-polaron ansatz is only an approximate closure of the full non-Markovian self-attraction, we do not expect it to determine the exponent quantitatively exactly.

We also numerically calculate the motion of the void center of mass \(x_{\rm com}(t)\) and the relative coordinate \(x_{\rm rel}(t)\equiv X(t)-x_{\rm com}(t)\) of the void center and the coherence by following the strategy introduced in Sec.~\ref{subsec:screened_Deff_numerics}, where depolarized spins \(\mathbb{1}\) are retained as an explicit local symbol and the instantaneous polarized population provides a direct microscopic proxy for the fluid depletion in the void. 
%Along each sampled trajectory we define \(x_{\rm com}(t)\) as the center of mass of the majority species and also measure \(x_{\rm rel}(t)\equiv X(t)-x_{\rm com}(t)\). 
The resulting mean-squared displacements are shown in Fig.~\ref{fig:aging_3panel_wide}(c). All three observables display essentially the same subdiffusive growth over the accessible time window, with \(\langle x_{\rm com}(t)^2\rangle\) and \(\langle x_{\rm rel}(t)^2\rangle\) tracking the coherence MSD up to multiplicative offsets. Although the measured growth is slightly more subdiffusive, at \(\mathcal O(t^{0.6})\), than the simple \(t^{2/3}\) estimate of the Feynman-style closure, the common subdiffusive scale across the coherence wandering, the void center-of-mass motion, and the relative coordinate supports a robust subdiffusive aging polaron.

%\section{Hydrodynamic transport modes}
% \jaj{[this is pre-v1 so it will require heavy editing-- should be considered a placholder]}

% \jaj{working on this section in another document.}
% \textcolor{red}{[Story about bimodal operators (hydro source plus non-hydro blob), the diffuson cascade, Alex's numerics]}

\section{Spectral structure in the weak-noise limit}
\label{sec:spectral-structure}

\subsection{Random circuits with depolarization}

So far, we have focused on the mean-square correlation function $\mathbb{E}_\gamma(|\langle \sigma^+_i(t) \sigma^-_0(0)\rangle|^2)$, where $\mathbb{E}_\gamma$ denotes an ensemble average over random charge-conserving unitary circuits interrupted by depolarizing noise of strength $\gamma$. This mean-square correlator decays at a rate $\sim \sqrt{\gamma}$; its evolution is governed by a transfer matrix that is invariant under spacetime translations, and conserves the ``operator charge'' $\mathcal{Q}$. We now discuss the spectral properties of this transfer matrix, which is block-diagonal in momentum and $\mathcal{Q}$. The leading eigenvalues of this transfer matrix in each sector determine the mean-square decay rates of operators in that sector. Although the channel corresponding to a \emph{typical} circuit is not translation invariant, we anticipate that the squared magnitude of the correlation function in such a circuit will decay at the mean-square rate: i.e., that the leading eigenvalues of the mean-square transfer matrix will be the leading \emph{Lyapunov exponents} of the corresponding operators in a typical instance. Moreover, we anticipate that the operators associated with the leading Lyapunov exponent in a typical instance will share the operator weight distribution of the leading transfer-matrix eigenvector. (This conjecture can be verified by exploring higher moments of the correlation function, but we do not attempt this in the present work.) 

In the $\mathcal{Q} = 1$ sector, we have already estimated the leading eigenvalues as a function of $k$: specifically, they scale as $\lambda(k) = \sqrt{\gamma}(\Lambda + \tilde{D}_{\rm eff} k^2)$, where $D_{\rm eff } = \sqrt{\gamma} \tilde{D}_{\rm eff}$ is the diffusion constant of the polaron. This leading dependence also holds for sectors with larger (but $O(1)$) values of $\mathcal{Q}$: the leading eigenmode involves multiple coherences bound to the same void. The corresponding eigenvalue is set by the probability of forming the void, and by the collective motion of the void and coherence(s), which are both $\mathcal{Q}$-independent, so to order $\sqrt{\gamma}$ the gap is the same in all of these sectors. (However, there will be corrections that are subleading in $\gamma$ or $k$.) In addition to the ``source'' part of the eigenvector, which consists of single-coherence configurations, it will typically also contain a ``tail'' of generic operators of characteristic operator weight $\sim 1/\sqrt{\gamma}$. This dependence follows directly from the arguments of Ref.~\cite{SpectralGaps}: the source operator decays by turning into shorter-lived operators with a higher coherence density, and these operators grow to a size $w^*$ where the typical decay rate $\gamma w^*$ matches the rate at which they are sourced by the one-coherence sector. 

\begin{figure}[tb]
    \centering
    \includegraphics[width=0.45\textwidth]{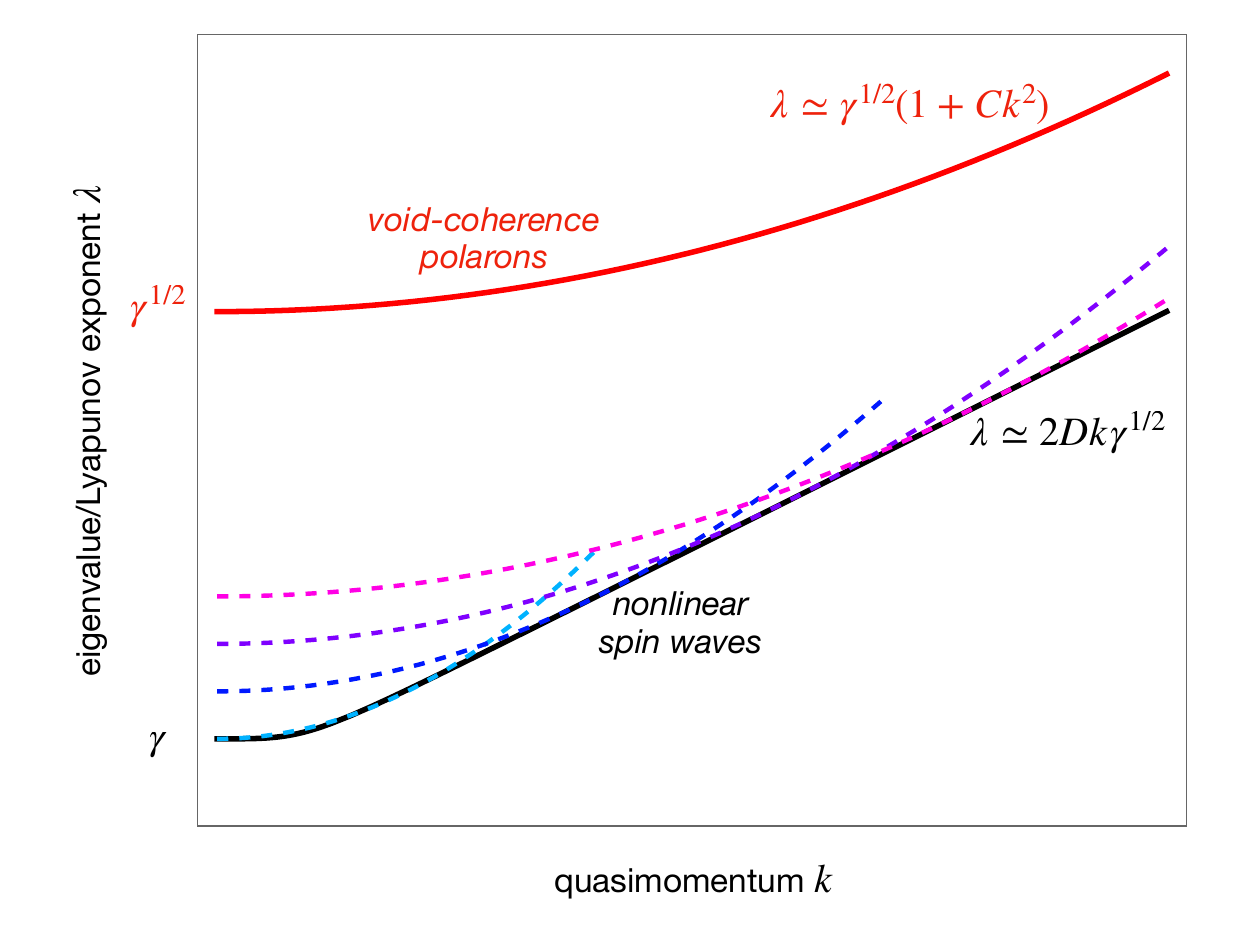}
    \caption{\textbf{Structure of leading eigenvalues.} Eigenvalues of the transfer matrix for $\mathcal{Q} \neq 0$ and $\mathcal{Q} = 0$, as a function of quasimomentum $k$. The gaps in the two types of sectors scale parametrically differently at $k = 0$; however, for any nonzero $k$ and weak enough noise, both gaps scale as $\sqrt{\gamma}$.}
    \label{vcp}
\end{figure}

The physics of the neutral sector $\mathcal{Q} = 0$ is different, because it contains hydrodynamic modes. We now estimate the leading $k$-dependent eigenvalues in this sector. At $k = 0$, one can see by inspection that the leading eigenvector is the charge $Q$ itself, and its eigenvalue is $\gamma$. At $k \neq 0$, a natural ansatz for the ``source manifold'' is the set of operators with no coherences, i.e., products of $\mathbb{1}$ and $\sigma^z$. Define $Z_k \equiv \sum\nolimits_k e^{i k x} \sigma^z_x$, i.e., the operator that measures a density wave at wavevector $k$. The operator $Z_k$ decays into non-hydrodynamic operators at a rate $D k^2$, in addition to decaying directly into the bath at rate $\gamma$. Therefore the total decay rate is $\gamma + D k^2$. In general, however, this linear density wave will not be an eigenvector: it will decay via a ``diffuson cascade'' into multiple density waves at smaller wavevectors~\cite{delacretaz2020heavy}. The corresponding operators can be written as $\prod_{j = 1}^n Z_{q_j}$, where $\sum_j q_j = k$. An operator of this form decays out of the source manifold at a rate $\sum_{j = 1}^n D q_j^2 + n \gamma$. Optimizing over $n$, one finds that for any finite $k$, as $\gamma \to 0$, the optimal decay rate is achieved at $n^{*} \sim k\l D /\gamma\r^{1/2}$, with all the momenta chosen as $q = k/n$. The associated decay rate is $2\sqrt{\gamma D} k$. 
%jaj (april 29) -- there were some errors/typos in this derivation that I think I have fully corrected now. We should be cognizant of whether these propagated.
Thus, remarkably, at fixed $k$ in the $\gamma \to 0$ limit, the leading eigenvalues in the $\mathcal{Q} = 0$ and $\mathcal{Q} \neq 0$ both scale the same way with $\gamma$. In the $\mathcal{Q} = 0$ sector, a further crossover happens for $k \sim \sqrt{\gamma}$: at yet smaller $k$, the slowest source operator is $Z_k$, and the decay rate is $\gamma + D k^2$. These scaling forms are plotted in Fig.~\ref{vcp}.

We have confirmed this scaling of the $\mathcal{Q} = 0$ gap. %, as well as the prediction that the leading eigenvector has size $\sim k/\sqrt{\gamma}$. jaj - I actually have this fig, but there are already a lot of figs going on in the relevant appendix, I don't think it's necessarily saying all that much.
Since the dynamics in the source manifold is purely classical, these $\mathcal{Q} = 0$ predictions can be checked using a classical stochastic model. This model, and the simulation details, are presented in App. \ref{app:charge_neutral_modes}. In addition, we have checked that (in the quantum model) the distribution of operator sizes has a ``tail'' at large sizes.% $\sim 1/\sqrt{\gamma}$, as claimed above. 

% \jaj{just to be clear the figure just shows emission into the tail right now. It's (surprisingly) hard to get the tail fully converged because the emission rate is very low, so there isn't a scaling analysis of the tail size. I have just fixed some convergence issues tonight (april 19th) so I can add a scaling analysis of the tail size from numx. Can only probe the regime where the tail size scales as $1/\gamma$ though because you can't fix momentum in TEBD.} \sg{I am fine with just softening the claim of what we show. I don't think getting the tail right should be a focus of this paper. We can just say we show the emergence of a tail.}
% \jaj{yeah agreed but already have this running and it probably won't take longer than the other cleanup to dump it in. It'd be nice to show some converged tails, but I agree the scaling analysis is whatever. We also just can't get the $\gamma^{-1/2}$ in TEBD (that I think is a lost cause, not worth trying).} 

\subsection{Translation-invariant models}

So far, we have taken the unitary part of the evolution to be random in spacetime. We now comment on the spectrum of the channel (or Lindbladian) associated with a \emph{fixed} evolution operator. In this case we can directly consider eigenvalue spectra rather than Lyapunov spectra. 

In such deterministic systems, we do not have the same microscopic control over many-body dephasing as in the random-circuit setting, where collisions involving the coherence acquire independent random phases at distinct spacetime points. A natural expectation is that dephasing instead accumulates gradually over many collisions. By the same considerations as in Sec.~\ref{sec:slow-manifold}, contributions to the coherence Green's function are still biased against histories with many such collisions. This motivates an effective coarse-grained fluid conditioning in which the projective conditioning of Eq.~\eqref{eq:branch-projection} is replaced by a soft local occupation tilt around the coherence,
\begin{equation}
    \mathcal{Z}_T^{\rm det}[X]
    =
    \left\langle
    \exp\left[
        -\lambda
        \int_0^T dt
        \int_{X(t)-a/2}^{X(t)+a/2} dx\,\rho(x,t)
    \right]
    \right\rangle.
    \label{eq:soft_ti_tilt}
\end{equation}
Here \(a\) is a microscopic collision range and \(\lambda\) is a phenomenological dephasing strength set by the deterministic collision dynamics. Equation~\eqref{eq:soft_ti_tilt} should be viewed as a coarse-grained ansatz for how deterministic many-body dephasing reweights fluid histories for a prescribed coherence trajectory \(X(t)\). It implements one natural local model of the effective fluid conditioning, in which histories with a larger particle density near the coherence are suppressed more strongly. Another natural coarse-grained conditioning would be to penalize collision events directly; this differs from Eq.~\eqref{eq:soft_ti_tilt}, which weights the total residence time of particles near the coherence. In what follows we restrict to the simple occupation-tilt conditioning. A more microscopic accounting of many-body dephasing in deterministic systems is left for future work.

Importantly, dynamics under a fixed evolution operator differs from a random one in the low-density limit: single-particle motion is either localized (if there is any disorder), or ballistic (if the dynamics is translation invariant). In both cases, the dynamics inside the void differs from that in random circuits, where the diffusion constant has a finite low-density limit: instead, for a fixed evolution operator, the diffusion constant scales as \(1/\rho\) as \(\rho \to 0\) in the ballistic case, or vanishes in the localized case. The dynamics of the void in the localized case is similar to the case we have considered: the spreading of a typical region into a void is still diffusive, with a different tail to low density~\cite{PhysRevX.7.041047}. Thus the leading behavior of the correlation function and the RP resonance will be unchanged in this case. However, the optimal density profile will change; also, as \(\gamma \to 0\), the coherence becomes localized inside the void, so the diffusion constant of the void--coherence polaron will be parametrically suppressed. Understanding the \(k\)-dependence in this case, e.g., using ensembles of charge-conserving random Floquet circuits, is an interesting task for future work.

For translation-invariant systems, our conclusions are more strongly modified. Since the diffusion constant diverges inside the void, particles that enter the void can travel to the coherence (and dephase it) parametrically faster. Previously we argued~\cite{StretchedExp} that autocorrelation functions decay as $\exp(-t^{2/3})$ in translation-invariant systems in the absence of noise. Based on the phenomenological tilted fluid measure in Eq.~\eqref{eq:soft_ti_tilt}, we now argue that the spectral gap in non-hydrodynamic sectors vanishes as $\gamma \to 0$ even in this case, and in particular that it vanishes at least as fast as $\gamma^{1/4}$. 

In what follows, we assume that the dominant contribution to the (non-hydrodynamic) spectral gap is the quasi-stationary large-deviation cost of the conditioned fluid, as in the spacetime-random case. We also assume that coherence motion gives only subleading corrections to this leading cost, so that the gap can be estimated from the fixed-coherence problem, \(X(t)=0\). To argue that the gap is at most $\mathcal{O}(\gamma^{1/4})$, we optimize a survival strategy based on repeatedly reconstituting the void around the coherence. We follow histories in which a void of size $\ell$ is centered at the coherence and is reconstituted by post-selection once every $\tau$ time steps. Thus, every $\tau$ time steps we pay an additional probability cost of order $\exp(-c\ell)$~\footnote{This assumes that the probability cost of a void is not super-exponential in $\ell$---an assumption that is satisfied in any generalized Gibbs state at finite temperature and density, as well as, trivially, in the void itself.}, while between renewals the coherence is dephased by particles that reach the center of the void.

We fix $\ell$ and $\tau$ by minimizing the cost per unit time of this resampled void protocol. Consider the dynamics inside a large void at early times. Particles that enter the void move, at worst, ballistically toward the coherence. Thus the density at the coherence generated by the noise is of order $\gamma t$: a particle reaching the coherence at time $t$ could have been created a distance $t'$ away at a time of order $t'$ in the past, for any $t'\leq t$. There is also an influx from the edges of the void. The hydrodynamic solution in Ref.~\cite{StretchedExp} gives a central density bounded by $t/\ell^2$ for a void of size $\ell$, before the void has filled in.

The accumulated dephasing cost over one renewal period is therefore of order $(\gamma+\ell^{-2})\tau^2$, while reconstituting the void costs $c\ell$. Thus the variational decay rate is
\[
    \Gamma_{\rm var}(\ell,\tau)
    \sim
    {c\ell\over \tau}
    +
    b(\gamma+\ell^{-2})\tau ,
\]
where $b$ and $c$ are nonuniversal positive constants. Minimizing over $\tau$ at fixed $\ell$ gives $\tau^2\sim \ell/(\gamma+\ell^{-2})$, and hence $\Gamma_{\rm var}(\ell)\sim [\ell(\gamma+\ell^{-2})]^{1/2}$. The remaining minimization over $\ell$ balances the two terms $\gamma\ell$ and $\ell^{-1}$, so that $\ell\sim \gamma^{-1/2}$. At this optimum, $\tau\sim\gamma^{-3/4}$ and $\Gamma_{\rm var}\sim \gamma^{1/4}$. Equivalently, at the optimum each renewal costs a factor $\exp[-O(\gamma^{-1/2})]$, while the accumulated dephasing over one renewal period is also $\exp[-O(\gamma^{-1/2})]=\exp[-O(\gamma^{1/4}\tau)]$. Thus, even under pessimistic assumptions about the motion of noise-generated particles inside the void and the cost of reconstituting the void, the local coherence survives for a timescale that diverges as $\gamma\to0$. This renewed-void construction is only a variational lower bound on the survival probability; the true conditioned dynamics can optimize over smoother density histories and may decay more slowly.

\begin{figure}
    \centering
    \includegraphics[width=1.0\linewidth]{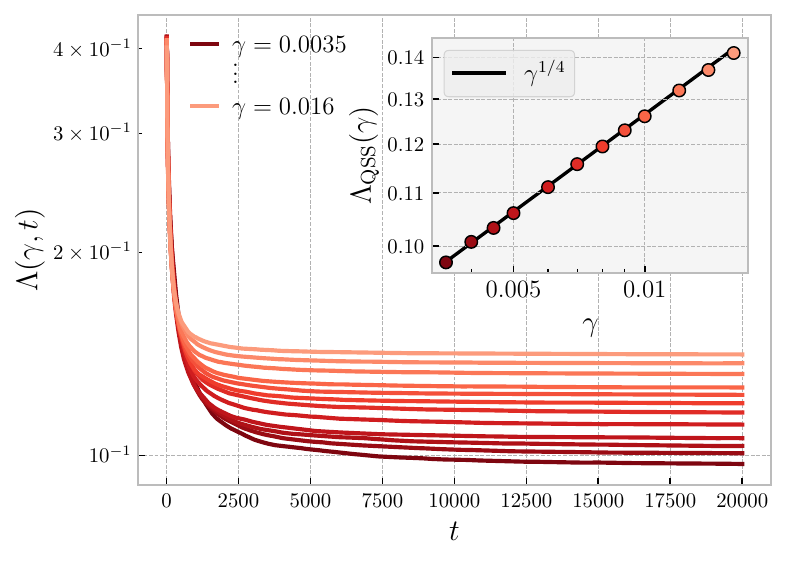}
    \caption{\textbf{Quasi-stationary decay rate estimate in translation-invariant systems}. We simulate a one-dimensional gas of ballistic point particles with velocity randomizing collisions and a weak particle reservoir with strength $\gamma$ (described in the main text) on a system of size \(L=360\). The tilted measure in Eq.~\eqref{eq:soft_ti_tilt} is sampled with \(\lambda=1/2\) and \(a=1\). Main panel: instantaneous decay rate \(\Lambda(\gamma,t)=-\partial_t \log{\mathcal{Z}^{\rm det}(\gamma,t)}\), obtained from population-dynamics simulations with \(N=10^4\) clones up to time \(t=2\times 10^4\), for \(\gamma=0.0035,\ldots,0.016\). Inset: quasi-stationary plateau value \(\Lambda_{\rm QSS}(\gamma)\), obtained by averaging the late-time instantaneous decay rate, plotted against \(\gamma\). The dashed guide shows \(\gamma^{1/4}\).}
    \label{fig:ballistic-gas-tilted-measure}
\end{figure}

We now numerically compute the quasi-stationary decay rate, i.e., the leading eigenvalue of the tilted fluid generator associated with Eq.~\eqref{eq:soft_ti_tilt}, specialized to a fixed coherence, \(X(t)=0\). The underlying fluid is a nonlinear stochastic model designed to capture the relevant low-density hydrodynamics of translation-invariant systems~\cite{StretchedExp,2026arXiv260102475M}. Specifically, we use a one-dimensional gas of ballistic point particles with velocities \(|v|\leq 1\). Particles move ballistically between collisions, and their velocities are randomized upon collision. Since momentum is relaxed only by two-body collisions, the mean free time at density \(\rho\) scales as \(\rho^{-1}\), and hence the effective diffusion constant diverges as \(D(\rho)\sim 1/\rho\) in the low-density limit. We model depolarizing noise, or more generally weak particle-number-nonconserving noise, by coupling this gas to a bath at density \(1/2\): particles are injected uniformly in space at rate \(\gamma/2\) per unit length and are removed independently at rate \(\gamma\). We denote the corresponding tilted partition sum for evolution time \(t\) and noise strength \(\gamma\) by \(\mathcal{Z}^{\rm det}(\gamma,t)\), and evaluate it using population-dynamics sampling~\cite{Giardina2011}.

The instantaneous decay rate of this tilted fluid problem is \(\Lambda(\gamma,t)\equiv-\partial_t \log \mathcal{Z}^{\rm det}(\gamma,t)\), and its long-time plateau defines the quasi-stationary rate \(\Lambda_{\rm QSS}(\gamma)=\lim_{t\to\infty}\Lambda(\gamma,t)\). 
Figure~\ref{fig:ballistic-gas-tilted-measure} shows \(\Lambda(\gamma,t)\) for several values of \(\gamma\), together with the plateau values \(\Lambda_{\rm QSS}(\gamma)\) in the inset.  The data show a clear approach to a quasi-stationary regime, and the extracted plateau rates are consistent with \(\Lambda_{\rm QSS}(\gamma)\sim \gamma^{1/4}\).

\subsection{Charge-conserving noise}

We now briefly comment on the case where the noise is charge-conserving, e.g., pure dephasing at rate $\gamma'$ in the $\sigma^z$ basis. First, we discuss the mean-square correlation function. In contrast to depolarizing noise, dephasing noise cannot ``fill in'' the void by adding particles. Thus, conditional on the coherence surviving to time $t$, it is likeliest to have been surrounded by a void of size $\sim \sqrt{D t}$, as in the ``aging saddle'' discussed in Sec. \ref{sec:aging-polaron}. The main difference from the noiseless case is that the stretched exponential decay of the correlator is multiplied by an overall factor $e^{-\gamma' t}$, since this is the rate at which the site containing the coherence is dephased. 

The structure of eigenoperators is more complex in this case, since the transfer matrix is block-diagonal not only in $\mathcal{Q}$, but also separately in the charge on the bra and the ket. Based on the decay of the correlation function, however, we conjecture that the gap of the transfer matrix in a typical sector with operator charge $\mathcal{Q}$ should scale as $\mathcal{Q} \gamma'$. For $\mathcal{Q} = 0$, there is a subspace of exact zero modes, which can be represented as moments of $Q$ or as projectors onto the maximally mixed state with any fixed value of $Q$. The leading eigenmodes above this subspace are again sourced by operators of the form $Z_k$ discussed above. However, in this case the characteristic lifetime of each $Z_k$ scales purely as $\sum_{j = 1}^n D q_j^2$. Thus the longest-lived excitation is achieved by splitting the momentum $k$ arbitrarily finely, i.e., into $\sim k L$ excitations of wavevector $\sim 1/L$. The gap for a typical $\mathcal{Q} = 0$ sector with wavevector $k$ is then $\sim k/L$, and is independent of $\gamma'$.

\section{Discussion}
\label{sec:discussion}

This work has laid out a systematic framework for describing how quantum coherences between charge sectors can survive for anomalously long times, even under chaotic high-temperature evolution, by binding to hydrodynamic large deviations. This binding is a nonperturbative effect that is neglected in standard field-theory treatments that linearize about some vacuum, e.g., the Luttinger liquid at finite temperature. These treatments therefore come to the conclusion that coherences (and therefore, e.g., the single-particle Green's function) decay exponentially. As we have seen, however, this standard conclusion is qualitatively wrong in one dimension; instead, the decay of the Green's function is governed by large deviations that are neglected by any linearized theory, but are captured within our framework. Moreover, in the presence of noise, this bound state of the coherence and its accompanying void is the leading eigen-operator of the open-system dynamics outside the hydrodynamic sectors. Thus the well-defined gapped Ruelle-Pollicott resonances that occur in the absence of conservation laws are entirely absent in the presence of charge conservation---not, as conjectured in Ref.~\cite{Mori_24}, present once hydrodynamic operators are projected out. (These conclusions assume the existence of zero-entropy charge sectors, such as the fully charge polarized states considered here~\footnote{By contrast, if the system contains neutral ancillary degrees of freedom that remain locally active even in an extremal charge sector, a charge void no longer protects a simple local coherence such as $\sigma^+$. In that case we expect the non-hydrodynamic sector to behave more like a system without conservation laws, with well-defined gapped RP resonances and exponentially decaying local correlations.}, and in the Hamiltonian setting, where the third law guarantees zero-entropy ground-state sectors.)

Many directions remain to be explored, both theoretically and experimentally. Our main predictions regarding the Green's function and its sensitivity to noise can be directly explored in present-day experiments with spinful ultracold atoms, superconducting qubit arrays, and other forms of synthetic matter. Our work also yields sharp predictions for condensed matter experiments involving photoemission spectroscopy or scanning-tunneling microscopy, both techniques that probe the single-particle Green's function. Applied to this context, our results for the Green's function $\langle c^\dagger(x,t) c(0,0) \rangle$ lead to sharp, non-Lorentzian predictions for the lineshape at finite temperature. An important task for future work is to identify parameters for which this non-Lorentzian behavior is most visible. In conventional metals, the probability of spontaneously forming a void is very low; however, doped semiconductors, Moir\'e materials, and other low-density electronic systems are promising candidates for observing these effects. 

More broadly, the framework developed here provides a concrete route toward a general description of quantum-coherent finite-temperature dynamics in systems with conservation laws. We focused on the simplest charged object, a single local coherence, and on the two-point Green's function that probes its survival. However, the construction is not tied to this particular observable. One can instead consider several coherences inserted at different spacetime points, higher-point functions involving products of charged operators, post-quench quantum correlations, or quantum-coherent correlations in nonequilibrium steady states. In each case, the same basic structure should apply: local coherences propagate within and condition a fluctuating hydrodynamic fluid, while the relevant correlation functions are controlled by atypical fluid histories. Extending the present formalism to such settings would provide a systematic way to describe genuinely quantum-coherent phenomena at finite temperature and filling.%, beyond the single-particle Green's function studied here.

Finally, it would be interesting to identify signatures of these nonperturbative effects that persist beyond one dimension. %In $d$ dimensions, the probability cost of maintaining a void that lives for a time $t$ scales as $\exp(t^{(d+1)/2})$. Therefore, the void contribution does not parametrically dominate the typical exponential decay. However, o
Our analysis implies that multiple coherences inserted into the same void are (up to power-law prefactors) just as long-lived as single coherences. Meanwhile, standard field-theory approaches would conclude that an $n$th order coherence decays $n$ times as fast as an elementary one. These scenarios can potentially be distinguished by computing ratios of correlation functions, e.g.,
\begin{equation}
R_2(x,x',y;t) \equiv \frac{\langle c^\dagger(x,t) c^\dagger(x',t) c(y,0)c(0,0)\rangle}{\langle c^\dagger(x,t) c(y,0)\rangle \langle c^\dagger(x',t) c(y,0)\rangle} \nonumber.
\end{equation}
It would be interesting to generalize our results to develop sharp predictions for such ratios of correlators, and to explore these experimentally, e.g., in NMR experiments probing multiple quantum coherences~\cite{PhysRevA.80.052323}. 

Since the first version of this manuscript appeared, two developments have emerged. One has found closely related void physics, including in Rényi entanglement growth, by mapping the replicated dynamics to effective ferromagnetic Hamiltonians~\cite{2026arXiv260811297L}. The other has developed, in a distinct mesoscopic scaling limit in which the interaction strength is taken to zero in the thermodynamic limit, a similar hydrodynamic description of quantum coherence in diffusive quantum systems~\cite{2026arXiv260728255B}.

\begin{acknowledgments} 
    The authors thank Curt von Keyserlingk, Luca Delacr\'etaz, David Huse, Adam Nahum, Friedrich H\"ubner, Shinsei Ryu, Romain Vasseur, and Shuolong Yang for helpful discussions.
    E.M. and S.G. acknowledge support from the Co-design Center for Quantum Advantage (C2QA) under contract number DE-SC0012704. J.A.J. was supported by the National Science Foundation Graduate Research Fellowship Program under Grant No. DGE-2039656. E.M. also acknowledges support from the European Union (ERC, STAQQ, 101171399). Views and opinions expressed are however those of the authors only and do not necessarily reflect those of the European Union or the European Research Council Executive Agency. Neither the European Union nor the granting authority can be held responsible for them.

    The simulations presented in this article were performed on computational resources managed and supported by Princeton Research Computing, a consortium of groups including the Princeton Institute for Computational Science and Engineering (PICSciE) and the Office of Information Technology's High Performance Computing Center and Visualization Laboratory at Princeton University. Any opinions, findings, and conclusions or recommendations expressed in this material are those of the authors and do not necessarily reflect the views of the U.S. National Science Foundation.
\end{acknowledgments}

\bibliography{Refs.bib}

\appendix

\section{Coherent state path integral and replica decoupling}\label{app:NLSM}
In this section, we employ a coherent state path-integral formalism to argue for the replica decoupling hypothesis at the core of the dilute coherence approximation. Additionally, we rederive the MSR action used in the main text as the field theory limit of the two-copy microscopic RUC model.

\subsection{Spin coherent state formalism}
First, we provide a brief and self-contained summary of the coherent state path integral, as it exhibits minor variations upon more standard calculations. We begin by adopting the operator-to-state mapping $P^{\up/\down} \to \ket{\up/\down}$ and work on a replicated spin chain, which can be thought of as a ladder geometry since all couplings will remain inter-replica local. The state of each spin is parameterized by angles $\theta\in\lb  0, \pi \rb$ and $\varphi \in \left[ - \pi , \pi \right)$, which also defines a vector on $S^{2}$, $\hat{n}$. To this basic parameterization we add a replica index, $\alpha$, so that we have $\theta^{\l\alpha \r}\l x_{i}, \tau\r$ and $\varphi^{\l \alpha\r}$ with associated unit vector $\hat{n}^{\l \alpha\r}\l x_{i},\tau\r$. This parameterization introduces a gauge redundancy resulting from overall rotations of an initial reference vector. Here, we will ignore any subtleties resulting from this redundancy as they are not directly relevant to the problem at hand; detailed expositions of a very similar quantization procedure can be found in \cite{Auerbach_94,Fradkin}. In this choice of parameterization, the state is 
\begin{equation}
    \ket{\hat{n}} = 
    \begin{pmatrix}
        \cos\l\theta/2\r \\
        \sin\l \theta/2\r e^{i \varphi}
    \end{pmatrix}
    \quad {\rm and } \quad
    \rho\l \hat{n}\r = \frac{\mbf{1} + \hat{n}\cdot \vec{\sigma}}{2}, 
\end{equation}
from which we may read off $\expval{\vec{\sigma}}{\hat{n}} = \hat{n}$.

We begin with the partition function,
\begin{equation}\label{eq:coherentz}
    \mc{Z}= \tr{e^{-\mc{H}t}} = \lim_{\delta \tau\to 0 }\tr{ \prod_{k = 1 }^{t/\delta \tau} e^{-\delta \tau\mc{H}}}.
\end{equation}
We can then employ the resolution of the identity on a single spin
\begin{equation}\label{eq:identity_resolution}
    \int_{S^{2}} d\mu\l \hat{n}\r \, \rho\l\hat{n}\r= \int_{S^{2}}\frac{d\varphi}{2\pi}d\cos\l\theta \r \, \frac{\mbf{1}+\hat{n}\cdot \vec{\sigma}}{2} = \mbf{1}.
\end{equation}
This is lifted to the path integration measure 
\begin{eqnarray}
    && \mathcal{D}\, \hat{n}\l x, \tau\r = \prod_{\alpha = 1,2}\prod_{i,k}d\mu\l \hat{n}^{\l \alpha \r} \l x_{i}, \tau_{k}\r\r \nonumber \\
    & & = \prod_{\alpha = 1,2}\prod_{i,k}\frac{d\varphi^{\alpha}\l x_{i},\tau_{k}\r}{2\pi}d\cos\l\theta \l x_{i} , \tau_{k}\r\r
\end{eqnarray}
where we will introduce discretized time $\tau_{k} = k\delta \tau$.

One can therefore evaluate the Boltzmann weight to first order in $\delta \tau$ as 
\begin{eqnarray}
    &&\bra{ \hat{n}^{(\vec{\alpha})}\l  \tau_{k+1}\r }\l 1 - \delta \tau\mc{H}\r\ket{ \hat{n}^{(\vec{\alpha})}\l \tau_{k}\r }  \\
    &=& \l 1 - \frac{\delta \tau \bra{ \hat{n}^{(\vec{\alpha})}\l \tau_{k+1}\r } \mc{H}\ket{ \hat{n}^{(\vec{\alpha})}\l \tau_{k}\r }}{\braket{ \hat{n}^{(\vec{\alpha})}\l \tau_{k+1}\r }{\hat{n}^{(\vec{\alpha})}\l \tau_{k}\r}}\r\nonumber \\
    & & \qquad \times \braket{ \hat{n}^{(\vec{\alpha})}\l \tau_{k+1}\r }{\hat{n}^{(\vec{\alpha})}\l \tau_{k}\r} \nonumber \\
    &=& \exp\!\l \rm log\braket{ \hat{n}^{(\vec{\alpha})}\l \tau_{k+1}\r }{\hat{n}^{(\vec{\alpha})}\l \tau_{k}\r}  -\delta \tau \mc{H}\l  \tau_{k}\r \r\nonumber
\end{eqnarray}
where we have suppressed the spatial argument (in the sense of an implicit product). In the next subsection, we will evaluate the latter of these two terms, but the former we simply quote and explain:
\begin{eqnarray}
    &&\braket{ \hat{n}^{(\vec{\alpha})}\l x, \tau_{k+1}\r }{\hat{n}^{(\vec{\alpha})}\l x, \tau_{k}\r}  \\
    &=& \exp\! \l -\frac{i \delta \tau}{2} \,  \sum_{\alpha} \l 1+ \cos\l \theta^{( \alpha )}\l x ,\tau\r \r\partial_{\tau} \varphi^{(\alpha)}\l x, \tau \r\r \r \nonumber .
\end{eqnarray}
This is the Wess-Zumino (sometimes called Berry phase) term, up to an additive total derivative which is not important for the bulk action. It appears due to a phase representing the area swept out on $S^{2}$ between the three vectors $\hat{n}\l x, \tau_{k+1}\r$, $\hat{n}\l x, \tau_{k}\r$, and a reference which fixes the gauge. The dynamics appear in the action somewhat unconventionally (only via the Wess-Zumino term). A more traditional kinetic term is also generated, but the mass scales with $\delta \tau $ and it thus does not survive the continuous time limit.

\subsection{Evaluation of Hamiltonian in the dilute coherence approximation and continuum limit}
Now, we recall the two-replica Hamiltonian density for continuous time $U(1)$ random circuits (note that $\mc{H} = -\mc{L}$) in the dilute coherence approximation is written as 
\begin{equation} \label{eq:moment_hamiltonian}
    \mc{H}_{i} =    P^{(1)}_{i,i+1} + P^{(2)}_{i,i+1} + \frac{4}{3} P^{(1)}_{i,i+1}P^{(2)}_{i,i+1} 
\end{equation}
where $P^{\alpha}_{i,j } = \frac{1}{4}\l 1 - \vec{\sigma}^{(\alpha)}_{i}\cdot\vec{\sigma}^{(\alpha)}_{j}\r$ is the standard singlet projector.

\begin{widetext}
First, we define the replicated, two-site coherent state at time $\tau_{k}$ as 
\begin{equation}
    \ket{i,j; k} = \ket{\hat{n}^{(1)}\l x_{i}, \tau_{k}\r \hat{n}^{(1)}\l x_{j}, \tau_{k}\r\hat{n}^{(2)}\l x_{i}, \tau_{k}\r \hat{n}^{(2)}\l x_{j}, \tau_{k}\r}
\end{equation}
The Hamiltonian density can then be written (with some benevolent abuse of notation) as
\begin{eqnarray}
    \delta \tau  \mc{H}\l x_{i}, \tau_{k}\r &=& \delta \tau \frac{\mel{i,i+1,k+1}{ P^{(1)}_{i,i+1} + P^{(2)}_{i,i+1} + \frac{4}{3} P^{(1)}_{i,i+1}P^{(2)}_{i,i+1} }{i,i+1; k} }{\braket{i,i+1; k +1}{i,i+1; k}}\nonumber \\
    & & = \frac{\delta \tau}{4} \l \l \sum_{\alpha} + \frac{1}{3}\prod_{\alpha}\r\l 1 - \hat{n}^{(\alpha)}\l x_{i},\tau_{k}\r \cdot\hat{n}^{(\alpha)}\l x_{i+1},\tau_{k}\r \r\r  + O\l \l \delta\tau\r^{2}\r
\end{eqnarray}
Within the spatial continuum limit, this expression can be evaluated in the expansion
\begin{equation}
    1- \hat{n}^{(\alpha)}\l x_{i}, \tau_{k}\r\cdot \hat{n}^{(\alpha)}\l x_{i+1}, \tau_{k}\r =  \frac{1}{2}\l\hat{n}^{(\alpha)}\l x_{i+1}, \tau_{k}\r -\hat{n}^{(\alpha)}\l x_{i}, \tau_{k}\r  \r^{2} \approx \left. \frac{a^{2}}{2}\l\partial_{x} \hat{n}^{(\alpha)}\l x, \tau_{k}\r\r^{2}\right|_{x = x_{i}} 
\end{equation}
at leading order where $a$ is the lattice spacing. A short calculation yields $\l \partial_{x} \hat{n}^{(\alpha)}\l x, \tau\r\r^{2} = \l\partial_{x}\theta^{(\alpha)}\l x , \tau \r\r^{2} + \sin^{2}\l \theta^{(\alpha)}\l x , \tau \r \r \l\partial_{x} \varphi^{(\alpha)}\l x , \tau \r \r^{2}$
and we can then write the Hamiltonian term 
\begin{equation}
    \mc{H}\l x_{i}, \tau_{k}\r = \left. \frac{1}{8}\l a^{2}\sum_{\alpha} + \frac{a^{4}}{6} \prod_{\alpha}\r \l \l\partial_{x}\theta^{(\alpha)}\l x , \tau \r\r^{2} + \sin^{2}\l \theta^{(\alpha)}\l x , \tau \r \r \l\partial_{x} \varphi^{(\alpha)}\l x , \tau \r \r^{2}  \right|_{x = x_{i}}\r
\end{equation}

At this point, we may carry out the continuum approximations in $x$ and $\tau$ with $\sum_{\tau_{k}}\to \int \frac{d\tau}{\delta \tau} $ and $\sum_{x_{i}} \to \int \frac{dx}{a}$ to yield the following action
\begin{equation}
    \mc{S}\lb \theta, \varphi\rb =\int dx d\tau \lb \frac{i}{2a}\sum_{\alpha} \l 1 + \cos\l\theta^{(\alpha)}\r\r\partial_{\tau} \varphi^{(\alpha)}  + \frac{1}{8}\l a\sum_{\alpha} + \frac{a^{3}}{6} \prod_{\alpha}\r \l\l \partial_{x} \theta^{(\alpha)} \r^{2} + \sin^{2}\l \theta^{(\alpha)}\r \l \partial_{x} \varphi^{(\alpha)}\r^{2} \r\rb
\end{equation}
where we drop the explicit spatial and temporal dependencies, so that $\theta^{(\alpha)} \equiv \theta^{(\alpha)}\l x, \tau \r$ and $\varphi^{(\alpha)}= \varphi^{(\alpha)}\l x , \tau\r$. We henceforth set $a = 1$.

Now, one can make the field redefinition $\frac{1}{2}\l 1 + \cos \l \theta \l x, \tau\r\r\r \equiv \rho\l x, \tau\r$ to introduce the density, under which the Wess-Zumino term trivially becomes $i \rho \partial_{\tau} \varphi$ revealing the canonical conjugate structure between $\rho$ and $\varphi$. The other terms (again, after a brief calculation) become
\begin{subequations}
    \begin{eqnarray}
        \sin^{2} \l\theta^{(\alpha)}\r \l \partial_{x}\varphi^{(\alpha)}\r^{2} &\to & 4 \rho^{(\alpha)} \l 1 - \rho^{(\alpha)} \r \l \partial_{x}\varphi^{(\alpha)}\r^{2} = 2 \sigma\l \rho^{\alpha}\r \l \partial_{x} \varphi^{(\alpha)}\r^{2} \\ 
        \l \partial_{x} \theta^{(\alpha)} \r^{2}&\to &\frac{\l\partial_{x} \rho^{(\alpha)}\r^{2}}{\rho^{(\alpha)}\l 1 - \rho^{(\alpha)}\r} = 2 \frac{\l\partial_{x} \rho^{(\alpha)}\r^{2}}{\sigma \l\rho^{(\alpha)}\r}
    \end{eqnarray}
\end{subequations}
Now, the action in the dilute coherence approximation (neglecting all coherence loop corrections) can be written as 
\begin{equation}
    S\lb \rho , \varphi \rb = \int dxd\tau \, \lb i \rho\partial_{\tau} \varphi + \frac{1}{4}\l \sum_{\alpha} + \frac{1}{6}\prod_{\alpha}\r\l \frac{\l\partial_{x} \rho^{(\alpha)}\r^{2}}{\sigma\l \rho^{(\alpha)}\r} + \sigma\l \rho^{(\alpha)}\r \l \partial_{x}\varphi^{(\alpha)}\r^{2}\r \rb.
\end{equation}
A comment on the measure is in order: when we take the continuum limit and perform our field redefinition to variables $\rho$ and $\varphi$ we get
\begin{equation}
    \prod_{\alpha, i,k}d\mu\l \hat{n}^{\alpha}\l x_{i}, \tau_{k}\r\r \propto d\varphi^{(\alpha)}\l x_{i}, \tau_{k}\r d\cos\l \theta^{(\alpha)}\l x_{i}, \tau_{k}\r\r \propto \mc{D}\varphi \, \mc{D} \rho
\end{equation}
where we neglect an unimportant constant factor at each spacetime point. In this choice of field variables, the metric \textit{determinant} is trivialized and the target manifold ``appears'' flat, despite the intrinsic curvature of $S^{2}$.
\end{widetext}

\subsection{Scaling argument for replica decoupling}
We begin by adopting a diffusive scaling, $x\to bx$ and $\tau \to b^{2}\tau$. We will take $b\to \infty$ and therefore work in units of distance much larger than $a$ with units of $\tau$ scaled commensurately for diffusive transport. From the Wess-Zumino and single-replica terms, we can establish the dimensions of the fields $\Delta_{\rho}$ and $\Delta_{\varphi}$, defined by $\rho \to b^{\Delta_{\rho}}\rho$ and $\varphi \to b^{\Delta_{\varphi}}\varphi$. A slight subtlety arises here as only the derivatives of $\rho$ scale in this problem -- we must expand $\rho\l x ,\tau\r = \rho_{0} + \delta \rho\l x, \tau \r$ around a nontrivial background. Essentially, this sets $\sigma \l \rho \r$ constant while the hydrodynamics occur over top of $\rho_{0}$. One term appears potentially problematic: $\rho_{0}\partial_{\tau} \varphi$, but this becomes a total derivative and vanishes from the bulk action with only $\delta \rho \partial_{\tau} \varphi $ contributing. The integration measure yields three units of $b$ and therefore the equations become 
\begin{subequations}
    \begin{eqnarray}
         0 &=& 3 + \Delta_{\rho} +  \Delta_{\varphi} - 2  \\
         0 &=& 3 + 2\Delta_{\rho}-2  \\
         0 &=& 3 + 2\Delta_{\varphi} - 2.
    \end{eqnarray}
\end{subequations}
These equations are overdetermined but consistent, and solved by $\Delta_{\rho} = \Delta_{\varphi} = -1/2$. Then the scaling dimension of the inter-replica coupling becomes $3 + 4\Delta_{\rho/\varphi} - 4 = -3$, and therefore the term is irrelevant.

\subsection{Bilocal corrections at first coherence-pair loop-order}
We recall the transition matrix elements to create $\mbf{pm}$ pairs, $\ket{\partial_{x}\rho^{(1)} \partial_{x} \rho^{(2)}}\bra{\mbf{pm}^{+}}$ where $\mbf{pm}^{+} = \frac{1}{8}\l \ket{\mbf{pm}} + \ket{\mbf{mp}}\r$ and its hermitian conjugate. These matrix elements are explicitly evaluated in Appendix \ref{app:pm_propagator}. We evaluate the first loop-order corrections by evaluating the (density dependent) propagator of the $\mbf{pm}^{+}$ pair and inserting it into a bilocal term in the effective action, or essentially a Schrieffer-Wolff transformation on the Hamiltonian given in \autoref{eq:moment_hamiltonian}. The evaluation of the propagator is conducted in \autoref{app:pm_propagator}. We may evaluate the exit and re-entry matrix elements in terms of the fields $\rho$ and $\varphi$ by noticing that the two-site state $\ket{\partial_{x}\rho^{(1)} \partial_{x} \rho^{(2)}}$ can be rewritten as $\ket{0^{(1)} 0^{(2)}}$ where $\ket{0_{nm}} =\ket{\up_{n}\down_{m}} -\ket{\down_{n}\up_{m}}$ is the singlet state on sites $n$ and $m$. Then we can directly evaluate the overlap 
\begin{equation}\label{eq:singlet_cspi}
    \braket{0}{\hat{n}\l x_{i}\r\hat{n}\l x_{i+1}\r}= \frac{a}{2}\l i \sin\l \theta \r \partial_{x}\varphi  + \partial_{x}\theta \r \equiv V(x),
\end{equation}
at leading order in the derivative expansion and neglecting the time argument. Here, we have employed a symmetric wavefunction gauge,
\begin{equation}
    \ket{\hat{n}\l \theta, \varphi\r} =
    \begin{pmatrix}
        \cos\l \theta/2\r e^{- i \varphi/2} \\
        \sin\l \theta /2\r e^{i\varphi/2}
    \end{pmatrix}
\end{equation}
which can only result in temporal boundary terms (and which we therefore neglect in the context of the bulk action). This result is notably consistent with the evaluation of $P_{i,i+1}$ in that its modulus squared matches.

\begin{widetext}
Then (setting $a=1$) the contribution of $\mbf{pm}^{+}$ pairs in the sense of the effective action is proportional to
\begin{eqnarray}
    S' &\propto&  \int_{\tau_{2}>\tau_{1}}  d\tau_{1} d\tau_{2} dx_{1} dx_{2}   \    V^{(1)}(x_{1},\tau_{1}) V^{(2)}\l x_{1},\tau_{1}\r G\l \tau_{2} - \tau_{1}, x_{2} - x_{1}\r \tilde{V}^{(1)}\l x_{2}, \tau_{2}\r \tilde{V}^{(2)}\l x_{2},\tau_{2}\r \nonumber \\
    &\ \ &= \int_{\tau_{-}>0} d\tau dx d\tau_{-} dx_{-} \ V^{(1)}( x-x_{-}, \tau  -\tau_{-})V^{(2)}( x-x_{-}, \tau  -\tau_{-}) G \l x_{-}, \tau_{-}\r  \nonumber  \\
    && \hspace{3.5cm}  \times  \tilde{V}^{(1)}\l x + x_{-}, \tau +  \tau_{-} \r \tilde{V}^{(2)}\l x + x_{-}, \tau + \tau_{-} \r \nonumber \\
    & \ \ & \approx \l \int_{\tau_{-}>0} dx_{-}d\tau_{-} G\l x_{-}, \tau_{- }\r\r \int dxd\tau \ V^{(1)}\l x, \tau \r V^{(2)}\l x , \tau\r \tilde{V}^{(1)}\l x,\tau \r \tilde{V}^{(2)}\l x, \tau\r \nonumber \\
    &\ \  & \propto  \int dx d\tau \, \,  \prod_{\alpha} \l  \frac{\l\partial_{x} \rho^{(\alpha)}\r^{2}}{\sigma\l \rho^{(\alpha)}\r} +  \sigma\l \rho^{(\alpha)}\r \l \partial_{x}\varphi^{(\alpha)}\r^{2}\r
\end{eqnarray}
where $\tilde{V} = V^{*}$ is the complex conjugate. In the second line we change variables to $\tau = \l \tau_{1} + \tau_{2}\r /2 $, $\tau_{-}= \tau_{2} - \tau_{1}$, $x = \l x_{1} + x_{2}\r / 2 $, and $x_{-} = \l x_{2} - x_{1}\r/2$ and work strictly in the spatiotemporal bulk (we will ignore the domain of $\tau_{-}$). Then, in the last line, we expand $V^{(\alpha)}\l x \pm x_{-}, \tau \pm \tau_{-}\r \approx V^{(\alpha)}\l x ,\tau\r + O( \tau_{-}\partial_{t}V^{(\alpha)}\l x , \tau\r) + O( x_{-}\partial_{x}V^{(\alpha)}\l x , \tau\r)$. These derivative terms can be discarded because the Green's function $G$ sets an $O(1)$ scale on $\tau_{-}\sim m^{-1}_{\rm eff} \approx \l 2\sigma\l \rho \r\r^{-1}$ and $x_{-}\sim m^{-1/2}_{\rm eff} \approx \l 2\sigma\l\rho \r\r^{-1/2}$; $G$ cuts off on parametrically shorter length and time scales than those over which the fields vary. The product term itself, mirroring the structure of the inter-replica coupling, renders this correction irrelevant. Some further details about the structure of the $\mbf{pm}^{+}$ Green's function are given in Appendix~\ref{app:pm_propagator}.

\end{widetext}

\subsection{Recovery of MSR action}

Now that we have established that the action in the diffusive scaling regime becomes replica decoupled (including coherence-loop corrections), we can recover the standard action applied in MFT by another field redefinition. Since the replicas are decoupled and identical, we drop replica indices and treat this as a single-replica problem. It is also helpful (though in no way necessary) to rescale the gate application rate/coupling constant so that potential terms in the action become
\begin{equation}
    V\l x, \tau\r = \frac{\l \partial_{x}\rho\r^{2}}{2\sigma\l \rho\r} + \frac{\sigma\l\rho\r}{2} \l \partial_{x} \varphi \r^{2}.
\end{equation}

Starting from the MSR definition of the current, $j = -\partial_{x} \rho + \sigma \l \rho\r \partial_{x }\pi $, we can substitute into the (one-dimensional) continuity equation, $\dot{\rho} = - \partial_{x}j = \partial_{x}^{2} \rho -  \partial_{x}\l \sigma\l \rho \r \partial_{x }\pi \r$. Combined with the variation in $\varphi$, which yields  $ \dot{\rho} =  i\partial_{x} \l \sigma\l\rho \r \partial_{x} \varphi\r $, we get 
\begin{equation}
    0 =  \partial_{x}\l \partial_{x}\rho-  \sigma \l \rho\r\lb \partial_{x} \pi +i\partial_{x}\varphi \rb\r.
\end{equation}
This equation is a requirement to recover the MSR action, but the solution is non-unique. We can look to the path integral measure to guide our intuition further. An obvious method in which to keep the measure ``flat'' is a transformation of the form $\pi = - i \varphi + f\l \rho, \varphi\r $ (we will compute the Jacobian after solving for $f$). A non-unique equation for $f$ is 
\begin{equation}
    \partial_{x} f = \frac{\partial_{x}  \rho}{\sigma} =\partial_{x} \int d\rho \, \frac{1}{\sigma\l \rho \r} .
\end{equation}
The above integral can be solved with a partial fraction decomposition to yield the (again non-unique) solution $f= \frac{1}{2}\log\!\big[\rho/\l 1 - \rho\r\big]$, which defines the response field
\begin{equation}
    \pi = -i\varphi + \frac{1}{2}\log{\frac{\rho}{1-\rho}}.
\end{equation}

Under this change of field variables, we must evaluate the transformation of the path-integral measure at each spacetime point:
\begin{equation}
    {\rm det }\lb \mc{J}\rb=
    \begin{vmatrix}
        1 & 0 \\
        \frac{1}{2\sigma\l\rho\r} & -i
    \end{vmatrix} = -i,
\end{equation}
which is trivial as expected. We can then rewrite
\begin{equation}
    \l \partial_{x}\varphi \r^{2} = - \l \partial_{x} \pi \r^{2} + 2\frac{\l\partial_{x}\pi\r\l\partial_{x}\rho\r}{\sigma\l \rho\r} - \frac{\l \partial_{x} \rho \r^{2}}{\sigma^{2}\l \rho\r}
\end{equation}

Therefore, the action becomes
\begin{equation}
    S = \int dx d\tau\, \l \pi \dot{\rho} -  \frac{\sigma\l \rho \r}{2}\l\partial_{x}\pi \r^{2} + \l \partial_{x} \pi \r\l\partial_{x}\rho\r \r.
\end{equation}
This is exactly the SSEP MSR action postulated in the main text. We note that an additional term appears from substituting the conjugate field into the Wess-Zumino term, but it is a total derivative. Namely,  $ i \rho \partial_{\tau} \varphi = - \rho\partial_{\tau}\l \pi - \frac{1}{2} \log{\frac{\rho}{1-\rho}} \r $ and we easily see $\frac{\dot{\rho}}{2\l 1- \rho\r} = \frac{-1}{2}\partial_{\tau} \log \!\big[1- \rho\big]$, so $i \rho \partial_{\tau} \varphi = \dot{\rho} \pi $ up to total derivatives.

\section{Coherence pair production matrix elements and evaluation of coherence pair propagator}\label{app:pm_propagator}

In this section, we evaluate the propagator for $\mbf{pm}$ coherence pairs.
\subsection{Transition matrix elements}
We begin with the transition rules into and out of the $\mbf{pm}^{+}$ sector (recalling that they are symmetric). The off-diagonal elements of the two-site transition matrix have a single non-zero singular value (and similarly one singular vector on the coherent and hydrodynamic sectors each). The relevant singular vector on the hydrodynamic sector can be expressed as
\begin{equation}
        2\ket{ \l \partial_{x}\rho\r^{2}} = \l\mbf{uu}, \mbf{dd}\r + \l\mbf{dd}, \mbf{uu}\r - \l\mbf{ud}, \mbf{du}\r - \l\mbf{du}, \mbf{ud}\r,
\end{equation}
where $\l \cdot, \cdot \r$ indicate the states on site $i$ and $i+1$. The consecutive letters correspond to different replicas, and we use $\ket{\l\partial_{x} \rho\r^{2}}$ as shorthand for $\ket{\partial_{x}\rho^{(1)} \partial_{x} \rho^{(2)}}$. Using the microscopic occupation number, we can re-express 
\begin{eqnarray}\label{eq:derivative_stochastic rule}
    \l \mbf{uu}, \mbf{dd}\r  &\to& n^{(1)}_{i}n_{i}^{(2)} \l 1 - n_{i+1}^{\l 1 \r}\r\l 1 - n_{i+1}^{\l 2 \r}\r \nonumber \\
    \l \mbf{dd}, \mbf{uu}\r  &\to& \l 1 - n_{i}^{\l 1 \r}\r\l 1 - n_{i}^{\l 2 \r}\r n^{(1)}_{i+1}n_{i+1}^{(2)}  \nonumber \\
    \l \mbf{ud}, \mbf{du}\r & \to & n_{i}^{\l 1 \r} \l 1 - n_{i}^{\l 2 \r}\r\l  1- n^{(1)}_{i+1} \r n_{i+1}^{(2)} \nonumber \\
    \l \mbf{du}, \mbf{ud}\r & \to & \l 1 - n_{i}^{\l 1 \r}\r  n_{i}^{\l 2 \r} n^{(1)}_{i+1}\l 1- n_{i+1}^{(2)}\r.
\end{eqnarray}
Summed together, we can organize the terms as 
\begin{equation}
    \ket{\l \partial_{x} \rho\r^{2}} = \frac{1}{2}\l n_{i+1}^{(1)} - n_{i}^{(1)}\r \l  n_{i+1}^{(2)} - n_{i}^{(2)} \r.
\end{equation}

Now, we notice that $\mbf{pm}$ pairs can only be created in a spatially symmetric fashion and define the two-site states
\begin{equation}
    \ket{\mbf{pm}^{\pm}} = \frac{1}{8}\l\ket{\mbf{pm}} \pm  \ket{\mbf{mp}}\r.
\end{equation}
Then the state $\ket{\mbf{pm}^{-}}$ is identically dynamically decoupled from the hydrodynamics. We can then express the mixing matrix elements for the $\mbf{pm}$ and hydrodynamic sectors symmetrically (since we have not introduced dissipation all processes must be reversible) as
\begin{equation}
    \mc{H}_{\rm od}= \frac{1}{3}\lb \ket{\l \partial_{x}\rho\r^{2}}\bra{\mbf{pm}^{+} \vphantom{\l \partial_{x}\rho\r^{2}}}+ {\rm h.c.}\rb.
\end{equation}

\subsection{Equation of motion}
After the $\mbf{pm}^{+}$ pair is created (let us momentarily assume over a background with no replica disagreements), the $\mbf{p}$ and $\mbf{m}$ diffuse freely and independently up to a contact interaction. We can then work in the center of mass and relative coordinates for the $\mbf{pm}^{+}$ pair, defining $x\equiv \l x_{1} + x_{2} \r /2$ and $y \equiv \l x_{1} - x_{2} \r/2$ (here $x_{1}$ and $x_{2}$ are the locations of the symmetrized $\mbf{p}/\mbf{m}$ and $\mbf{m}/\mbf{p}$). Then the dynamics of the $\mbf{pm}^{+}$ pair absent replica disagreements can be written as the two-particle sector of a Wick-rotated Lieb-Liniger model:
\begin{equation}
    \dot{\Phi}\l x, y , t \r = \l \partial_{x}^{2} +\partial_{y}^{2} - c \delta\l y \r  \r \Phi\l x,y, t \r.
\end{equation}
The delta function interaction results from the transition rule
\begin{equation}
    \ket{\mbf{pm}^{+}} \to \frac{2}{3} \ket{\mbf{pm}^{+}} + \frac{1}{3}\ket{\l \partial_{x}\rho\r^{2}}
\end{equation}
where we add and subtract $\frac{1}{3}\ket{\mbf{pm}^{+}}\bra{\mbf{pm}^{+}}$ to get the delta function $\mbf{pm}$ interaction over an otherwise well-behaved random walk (the $\ket{\mbf{pm}^{+}}$ state must be stationary under the two-site random walk transition). We have absorbed the factor of $1/3$ into the definition of $c$. We will continue onward largely disregarding specific values of constants as the result is insensitive to them.

Now, we must also treat the attenuation of $\mbf{pm}^{+}$ due to the background. We accomplish this in the continuous time setting by recognizing that $\mbf{ud}$ and $\mbf{du}$ introduce an $O(1)$ decay rate when adjacent to $\mbf{p}$ or $\mbf{m}$ (in the discrete time circuit, they merely annihilate $\mbf{p}$ and $\mbf{m}$). We can identically rewrite the density of such operators as
\begin{equation}
    \rho_{ \rm RD} = \rho_{\mbf{ud}} + \rho_{\mbf{du}} = 2\lb \sigma\l \rho \r - R^{12}_{c}\rb \equiv m_{\rm eff}
\end{equation}
where $\sigma \l \rho\r = \rho\l 1- \rho \r$ is the standard mobility and $R^{12}_{c}\l x_{i} \r = \left< n_{i}^{(1)}n_{i}^{(2)}\right> - \left< n_{i}^{(1)}\right>\left< n_{i}^{(2)}\right> = \left< n_{i}^{(1)}n_{i}^{(2)}\right> - \rho^{2}\l x_{i}\r$ is the connected inter-replica correlator. We expect the connected inter-replica correlator to be negligible in the bulk since the bulk actions for either replica decouple at long wavelengths.

Now, we shall make our most substantial approximation, treating $m_{\rm eff}$ as essentially spatially constant. This approximation is obviously not strictly true but, since $m_{\rm eff}$ directly depends only on $\rho$ after removing the inter-replica correlator, the chain rule implies the gradients of $m_{\rm eff}$ can only enter our equations at higher density-gradient order (and are therefore subleading). Then, the homogenized equations of motion (\textit{i.e.}, neglecting the source term from the hydrodynamic sector) become
\begin{equation}
    \dot{\Phi} = \l \partial_{x}^{2} - m_{\rm eff} + \partial_{y}^{2} - c\delta\l y \r \r \Phi.
\end{equation}
The Green's function of this equation can be evaluated analytically within the single-loop approximation (\textit{i.e.}, neglecting $\mbf{pm}^{+}$ interactions) as we do below. However, a qualitative analysis offers almost as much insight, so we summarize the results of the calculation here.

The main result: the dynamics of the $\mbf{pm}^{+}$ pair is essentially diffusive, with an $O(1)$ timescale cutoff introduced by $m_{\rm eff} \approx  2 \sigma \l \rho \r$ and a commensurate spatial scale $m^{-1/2}_{\rm eff}$. Therefore these dynamics act as an $O(1)$ factor and low-pass filter on the exit and re-entry factors, which become very small in the hydrodynamic limit. The injection and re-entry terms carry a schematic factor of $ \l \partial_{x}\rho\r^{2}$ (the precise factors are calculated in Appendix~\ref{app:NLSM}). Together, this forces the correction to ``look like'' the standard inter-replica coupling term up to a factor and the aforementioned low-pass filter (at $k\sim \sqrt{1/m_{\rm eff}}$ which is very large compared to the hydrodynamic scale).

\subsection{\titlemath{\mbf{pm}^{+}} propagator}
Remarkably, the $\mbf{pm}^{+}$ propagator is possible to derive in closed integral form under our previous assumption that the effective mass varies on large spatial scales. We begin with the homogeneous equation of motion for $\Phi$:
\begin{equation}
    0 = \l \partial_{y}^{2} + \partial_{x}^{2}- c\delta\l y \r  - m_{\rm eff}  - \partial_{t}  \r \Phi .
    \label{eq:phi_homogeneous_eqm}
\end{equation}
We shift to Fourier space in $x$ and $t$ and note that the  propagator for this problem holding $k$ constant and setting $c=0$ is
\begin{equation}
    G_{0}\l y_{i}, y_{f};k, \omega \r = \int \dbar q  \   \frac{e^{iq\l y_{f} - y_{i}\r}}{i\omega - q^{2} -k^{2} - m_{\rm eff}}.
\end{equation}
It is helpful to define the position space representation of this object as 
\begin{equation}
    \mbf{G}_{0} \l k,\omega \r =  \int dy_{i} dy_{f}\, \dbar q \ \frac{\ket{y_{i}} e^{iq\l y_{f} - y_{i}\r}\bra{y_{f}}}{i\omega - q^{2} -k^{2} - m_{\rm eff}} .
\end{equation}

Now, we recall that $\Phi$ remains completely uncoupled from the hydrodynamic sector away from $y=0$. Dynamical generation of $\mbf{pm}^{+}$ only occurs at $y=0$ and flows back at $y=0$ so we only care about evaluating the propagator starting and ending at position $y=0$. Therefore, we can apply the Sherman-Morrison formula \cite{ShermanMorrison} to evaluate the propagator at $y_{i}=0$ and $y_{f}=0$
\begin{eqnarray}
    G\l k , \omega \r &=& \bra{0} \mbf{G}\l k, \omega \r \ket{0} = f\l k , \omega\r\l  1 + \frac{ c f\l k , \omega \r}{1 -c f\l k , \omega \r}\r \nonumber \\
    & = & \frac{f\l k, \omega\r}{1 - cf\l k , \omega \r} = \frac{1}{f^{-1}\l k , \omega\r - c}
\end{eqnarray}
where we define the helpful function $f\l k, \omega \r = G_{0}\l 0,0; k, \omega \r$. Now, we must evaluate $f$,
\begin{eqnarray}
    f &=& \lim_{\delta y \rightarrow 0 ^{+} }\int \dbar q \ \frac{e^{i q \delta y }}{i\omega - k^{2} - m_{\rm eff} - q^{2}} \nonumber \\
    &=& \frac{i}{2\sqrt{ i\omega -k^{2} - m_{\rm eff}}}.
\end{eqnarray}
Now, (absorbing the factor of $i$ into the square root), we can write
\begin{equation}
    G\l k , \omega \r = \frac{1}{2\sqrt{k^{2} + m_{\rm eff} - i \omega } - c}.
\end{equation}
Then, we can write propagator in position space in de-dimensionalized form as
\begin{equation}
    G\l x,  t\r = m_{\rm eff}\int \dbar\Gamma\dbar \eta \frac{-e^{-\Gamma \l t / m^{-1}_{\rm eff}\r  + i\eta \l x / m_{\rm eff}^{-1/2}\r}}{2\sqrt{\Gamma - 1 - \eta^{2}} + i c/\sqrt{m_{\rm eff}}}
\end{equation}
Unfortunately, this propagator cannot be evaluated directly by contour integration due to a non-trivial analytic structure (though it can, of course, be expressed in integral form). However, setting $c=0$ we clearly recover the $2d$ heat kernel with a gap, which enforces the length and time scales described above. This limit overestimates the response at $y=0$ and should not affect any qualitative differences in the dynamics with respect to $x$ (since we argue for the RG irrelevance of this term, an overestimate is inconsequential). Thus, under the scaling $x\to bx$ and $t\to b^{2} t $ we see that the memory time is small and the memory kernel becomes, essentially, a delta function, so we simply recover the standard interreplica interaction, which we previously argued to be RG irrelevant in this scaling.

\section{Retarded non-local coherence action}
\label{app:retarded_volterra_from_one_replica}

In this appendix we derive the effective $X$-only description of a moving coherence by integrating out the conditioned fluid degrees of freedom. Our starting point is the phase-averaged two-replica representation of the norm-squared coherence Green's function,
\begin{equation}
\overline{|\mathcal G(x_T,x_0;T)|^2}
=
\int_{X(0)=x_0}^{X(T)=x_T}\!\!\mathcal D X\;
e^{-S_X[X]}\,
\mathcal Z_T[X]^2,
\label{eq:green_function_path_sum_appendix}
\end{equation}
where $S_X[X]=\frac{1}{4D}\int_0^T dt\,\dot X^2$ is the bare coherence action and $\mathcal Z_T[X]$ is the one-fluid conditioned partition sum for a prescribed coherence trajectory $X(t)$. Throughout this appendix we restrict to the replica-symmetric saddle
\(
\rho_1=\rho_2\equiv \rho
\),
\(
\pi_1=\pi_2\equiv \pi
\),
and work in the branch-projection limit established in the main text, Eq.~\eqref{eq:hard-repair}, for which the coherence sits, to leading order, in a perfect down-polarized void (the size of the subleading correction is given in Eq.~\eqref{eq:branch-proj-subleading-correction}).

For fixed $X(t)$, the one-fluid weight entering \eqref{eq:green_function_path_sum_appendix} is a finite-time MFT path integral,
\begin{equation}
\mathcal Z_T[X]
=
\int_{\rho(X(t),t)=0}^{\pi(x,T)=0}
\mathcal D\rho\,\mathcal D\pi\;
e^{-S_{\rm fluid}[\rho,\pi;X]-\Phi_0[\rho(\cdot,0)]},
\label{eq:Z_X_fixed_path_appendix}
\end{equation}
where the moving Dirichlet condition $\rho(X(t),t)=0$ enforces the empty coherence core, while the terminal condition $\pi(x,T)=0$ reflects the fact that in the charged Green's function the final time involves a contraction with the identity operator at each site $x\neq X(T)$ at the final time. The term $\Phi_0[\rho(\cdot,0)]$ is an explicit initial-time functional, determined by the initial fluid state in the coherence Green's function (e.g., the half-filling fluid state).

For an infinite temperature, half-filled SSEP fluid, the initial ensemble is the Bernoulli product measure at density $1/2$. Its macroscopic fluctuation functional is
\begin{align}
\Phi_0^{\rm SSEP}[\rho_0]
=
&\int\! dx
\Big[
\rho_0(x)\log\! \big[2\rho_0(x)\big]\nonumber\\
&\qquad +
\bigl(1\! -\! \rho_0(x)\bigr)\log\! \big[2(1\! -\!\rho_0(x))\big]
\Big].
\label{eq:Phi0_SSEP_appendix}
\end{align}

Our aim is to integrate out the fluid to obtain the non-local path measure for $X$ alone used in the main text. We will first rewrite the moving Dirichlet problem as an equivalent fast-sink problem, then pass to a bilinear Cole--Hopf representation of the dilute theory for which we can exactly integrate out the fluid degrees of freedom.

\subsection{Equivalent fast-sink formulation and dilute fluid approximation}
\label{appsubsec:fast_sink_formulation}

Our first step is to trade the constrained moving-Dirichlet problem for an equivalent point-sink formulation of the fixed-$X$ saddle. We therefore introduce a sink of strength $\lambda$ at the coherence position. The constraint $\rho(X(t),t)=0$ is recovered in the fast-sink limit $\lambda\to\infty$. In this limit, the gradient tilt in Eq.~\eqref{eq:gradient_tilt} may equivalently be viewed as a tilt on the fluid current into the point sink, since the fluid mobility vanishes at $x=X(t)$.

Indeed, the optimal current in the one-fluid problem is
\begin{equation}
j_\ast(x,t)
=
- D\,\partial_x \rho(x,t)
+
\sigma(\rho(x,t))\,\partial_x \pi(x,t),
\label{eq:jstar_appendix}
\end{equation}
with mobility $\sigma(\rho)=2D\rho(1-\rho)$. At the coherence, $\rho(X(t),t)=0$, so the mobility term vanishes and the current is purely Fickian:
\begin{equation}
j_\ast(X(t),t)
=
- D\,\partial_x \rho\big|_{x=X(t)}.
\label{eq:fickian_sink_current_appendix}
\end{equation}
Thus the current into the sink is determined entirely by the local density gradient at the coherence.

For a local Poisson sink, the instantaneous absorption rate is $\lambda \rho(X(t),t)$. In the Martin--Siggia--Rose/MFT representation, absorption of one unit of density carries the usual response-field factor $e^{-\pi(X(t),t)}$~\cite{M_Doi_1976,Peliti85,Jona-Lasinio1993,Tailleur_2008,Bertini_2015}. Applying a bias to the integrated absorbed current by a factor $e^{-s}$ per absorbed unit biases this same local Poisson process, and therefore enters in exactly the same way as the conjugate field: it shifts $e^{-\pi}\to e^{-(\pi+s)}$. The corresponding local tilted sink term is
\begin{align}
S_{\rm sink}^{(\lambda)}[\rho,\pi;X]
=
\int_0^T dt&\int dx\,
\lambda\,\delta(x-X(t)) \nonumber\\
&\times \rho(x,t)\Bigl(1-e^{-\pi(x,t)-s}\Bigr).
\label{eq:tilted_sink_action_appendix}
\end{align}

The fixed-$X$ fluid partition sum \eqref{eq:Z_X_fixed_path_appendix} may therefore be written equivalently as
\begin{equation}
\mathcal Z_T[X]
=
\lim_{\lambda\to\infty}
\int \mathcal D\rho\,\mathcal D\pi\;
e^{-S^{(\lambda)}_{\rm fluid}[\rho,\pi;X]-\Phi_0[\rho(\cdot,0)]},
\label{eq:Z_X_fast_sink_appendix}
\end{equation}
with terminal condition $\pi(x,T)=0$, and where
\begin{equation}
S^{(\lambda)}_{\rm fluid}[\rho,\pi;X]
=
S_{\rm bulk}[\rho,\pi]
+
S_{\rm sink}^{(\lambda)}[\rho,\pi;X].
\label{eq:onefluid_tilted_sink_action_appendix}
\end{equation}
Here $S_{\rm bulk}$ denotes the bulk MFT fluid action. 

We now replace the full fluid fixed-$X$ theory by a dilute effective version, appropriate inside the conditioned void core. This approximation is not merely the replacement of the mobility by its low-density form. Rather, once exclusion is neglected in the bulk, the corresponding bath should also be replaced by the ideal-gas birth--death bath for a non-excluding diffusive fluid. Concretely, we approximate the mobility by
\begin{equation}
\sigma(\rho)\simeq 2D\rho,
\label{eq:dilute_sigma_appendix}
\end{equation}
and replace the full exclusion bath Hamiltonian by the dilute ideal-gas form
\begin{equation}
H_{\rm bath}^{\rm dil}(\rho,\pi)
=
\rho(e^{-\pi}-1)+\frac12(e^\pi-1).
\label{eq:dilute_bath_hamiltonian_appendix}
\end{equation}
This corresponds to local particle removal at rate $\gamma\rho$ together with injection from the half-filled environment at rate $\gamma/2$. In the unforced far field, where the sink is absent and $\pi\to 0$, the density equation therefore reduces to
\begin{equation}
\partial_t\rho
=
D\,\partial_x^2\rho-\gamma\Big(\rho-\tfrac12\Big),
\label{eq:dilute_farfield_rho_appendix}
\end{equation}
so the dilute bath preserves the same half-filled fixed point as the full theory.

Under dilute approximation, we also replace the Bernoulli initial functional Eq.~\eqref{eq:Phi0_SSEP_appendix} by the corresponding ideal-gas initial functional at mean density $1/2$,
\begin{equation}
\Phi_0^{\rm dil}[\rho_0]
=
\int dx\,
\Big[
\rho_0(x)\log\!\big[2\rho_0(x)\big]-\rho_0(x)+\tfrac12
\Big].
\label{eq:Phi0_dil_appendix}
\end{equation}
This preserves the correct mean filling and the exponential large-deviation cost for creating an empty interval, while remaining compatible with the non-excluding dilute bulk theory.

Thus, under the dilute approximation, we take the initial state functional $\Phi_0[\rho_0]\equiv \Phi_0^{\rm dil}[\rho_0]$, and replaced the fluid action $S_{\rm fluid}^{(\lambda)}$ by
\begin{align}
S^{(\lambda)}_{\rm dil}[\rho,\pi;X]
=
\int \! dt \!\int \! dx
&\Big[
\pi\partial_t \rho
+D\partial_x\rho\partial_x\pi
-D\rho(\partial_x\pi)^2
\nonumber\\
&-\lambda\,\delta(x-X(t))\,\rho\,(e^{-\pi-s}-1)\nonumber\\
&\qquad -\gamma H_{\rm bath}^{\rm dil}(\rho,\pi)
\Big].
\label{eq:dilute_rho_pi_action}
\end{align}
The associated bulk saddle equations, obtained by varying
$S^{(\lambda)}_{\rm dil}[\rho,\pi;X]$ for $0<t<T$, are
\begin{align}
\partial_t\rho
&=
D\,\partial_x\!\Big(\partial_x\rho-2\rho\,\partial_x\pi\Big)
+\gamma\Big(\tfrac12 e^\pi-\rho e^{-\pi}\Big)\nonumber\\
&\qquad -\lambda\,\delta(x-X(t))\,\rho\,e^{-\pi-s},
\label{eq:dilute_rho_eq_appendix}
\\
\partial_t\pi
&=
-\,D\,\partial_x^2\pi
-D(\partial_x\pi)^2
+\gamma\bigl(1-e^{-\pi}\bigr)\nonumber\\
&\qquad +\lambda\,\delta(x-X(t))\bigl(1-e^{-\pi-s}\bigr).
\label{eq:dilute_pi_eq_appendix}
\end{align}
The endpoint data are supplied separately by $\pi(x,T)=0$ and $\Phi_0[\rho(\cdot,0)]$. Variation with respect to the initial fluid profile gives
\begin{equation}
\pi(x,0)
=
\frac{\delta \Phi_0}{\delta \rho_0(x)}
=
\log\! \big[2\rho(x,0)\big].
\label{eq:initial_boundary_condition_appendix}
\end{equation}

The main benefit of this dilute fluid replacement is that the resulting bulk action admits an exact linearization under a Cole--Hopf transformation. 

\subsection{Cole--Hopf linearization}
\label{appsubsec:dilute_core_volterra}

We introduce the Cole-Hopf variables
\begin{equation}
P(x,t)=e^{\pi(x,t)},
\quad
Q(x,t)=\rho(x,t)e^{-\pi(x,t)}.
\label{eq:cole_hopf_appendix}
\end{equation}
This transformation has unit Jacobian, so the functional measure is unchanged. The canonical time-derivative term transforms as
\begin{equation}
\pi\,\partial_t\rho
=
\partial_t(\rho\pi-\rho)+P\,\partial_t Q,
\label{eq:canonical_CH_identity_appendix}
\end{equation}
which yields both a finite-time boundary contribution and a bilinear bulk term. The remaining terms transform directly:
\begin{align}
D\,\partial_x\rho\,\partial_x\pi
-D\rho(\partial_x\pi)^2
&=
D\,\partial_xP\,\partial_xQ,
\nonumber
\\
-\gamma H_{\rm bath}^{\rm dil}(\rho,\pi)
&=
\gamma(P-1)\Big(Q-\tfrac12\Big),
\nonumber
\\
-\rho\,(e^{-\pi-s}-1)
&=
(P-e^{-s})Q.
\nonumber
\end{align}
Thus the bulk action becomes
\begin{align}
S^{(\lambda)}_{\rm dil}[Q,P;X]
=
&\ \bigg[\int dx\,(\rho \pi-\rho)\bigg]_0^T
\nonumber\\
&+
\int \! dt \!\int \! dx
\Big[
P\partial_tQ
+D\partial_xP\partial_xQ
\nonumber\\
&\hspace{1.2cm}+\lambda\delta(x-X(t))(P-e^{-s})Q\nonumber\\
&\hspace{1.2cm}+\gamma(P-1)\Big(Q-\tfrac12\Big)
\Big].
\label{eq:bilinear_QP_appendix}
\end{align}
The full fixed-$X$ partition sum in the dilute theory is therefore
\begin{equation}
\mathcal Z_T[X]
=
\lim_{\lambda\to\infty}
\int \mathcal DQ\,\mathcal DP\;
e^{-S^{(\lambda)}_{\rm dil}[Q,P;X]-\Phi_0[\rho(\cdot,0)]},
\label{eq:Z_X_QP_appendix}
\end{equation}
with the final-time condition $P(x,T)=1$ ($\pi(x,T)=0$).

Varying Eq.~\eqref{eq:bilinear_QP_appendix} with respect to $P$ and $Q$ gives the
linear bulk saddle equations
\begin{align}
\partial_t Q
&=
D\,\partial_x^2 Q
-
\gamma\Big(Q-\tfrac12\Big)
-
\lambda\,\delta(x-X(t))\,Q,
\label{eq:Q_forward_appendix}
\\
-\partial_t P
&=
D\,\partial_x^2 P
-
\gamma(P-1)
-
\lambda\,\delta(x-X(t))(P-e^{-s}),
\label{eq:P_backward_appendix}
\end{align}
so that $Q$ evolves forward in time while $P$ obeys the corresponding backward equation.

From Eq.~\eqref{eq:initial_boundary_condition_appendix}, using $P = e^\pi$ we find $P(x,0)=2\rho(x,0)$ and thus the initial condition for $Q(x,0)$,
\begin{equation}
Q(x,0)=\rho(x,0)e^{-\pi(x,0)}=\frac12.
\label{eq:Q_initial_condition_appendix}
\end{equation}

In terms of the Cole--Hopf variables, the optimal current is
\begin{equation}
j_\ast(x,t)
=
D\bigl(Q\,\partial_x P-P\,\partial_x Q\bigr).
\label{eq:jstar_QP_appendix}
\end{equation}
We define the finite-$\lambda$ sink rate by
\begin{equation}
r_X^\lambda(t)\equiv -j_\ast(X^+(t),t)+j_\ast(X^-(t),t),
\label{eq:rx_def_appendix}
\end{equation}
so that $r_X^\lambda(t)$ is the current into the sink. Integrating Eq.~\eqref{eq:Q_forward_appendix} and Eq.~\eqref{eq:P_backward_appendix} across the sink gives the jump conditions
\begin{align}
D\Big[\partial_xQ\Big]_{X(t)^-}^{X(t)^+}
&=
\lambda\,Q(X(t),t),
\label{eq:Q_jump_appendix}
\\
D\Big[\partial_xP\Big]_{X(t)^-}^{X(t)^+}
&=
\lambda\,\bigl(P(X(t),t)-e^{-s}\bigr).
\label{eq:P_jump_appendix}
\end{align}
Using these in Eq.~\eqref{eq:jstar_QP_appendix}, the sink rate may be written exactly as
\begin{equation}
r_X^\lambda(t)=e^{-s}\lambda\,Q(X(t),t).
\label{eq:rx_from_Q_appendix}
\end{equation}
In the fast-sink limit $\lambda\to\infty$ one has
\begin{equation}
Q(X(t),t)\to 0,
\qquad
P(X(t),t)\to e^{-s},
\label{eq:fast_sink_boundary_appendix}
\end{equation}
and the physical fluid filtering rate is
\begin{equation}
r_X(t)\equiv \lim_{\lambda\to\infty}r_X^\lambda(t).
\label{eq:rx_fast_sink_def_appendix}
\end{equation}
Since $\rho=PQ$, this reduces precisely to the gradient-jump quantity penalized by the tilted measure in Eq.~\eqref{eq:gradient_tilt} of the main text,
\begin{equation}
r_X(t)
=
-\,D\Big[\partial_x\rho_X(x,t)\Big]_{x=X(t)^-}^{x=X(t)^+},
\label{eq:rx_gradientjump_appendix}
\end{equation}
so that $\lambda \to \infty$ recovers the physical conditioned fluid functional.

Because the forward equation EQ.~\eqref{eq:Q_forward_appendix} is linear, it admits
an explicit Duhamel solution. Using the screened heat kernel
\begin{equation}
G_\gamma(\Delta x,\tau)
=
\frac{e^{-\gamma\tau}}{\sqrt{4\pi D\tau}}
\exp\!\left[-\frac{(\Delta x)^2}{4D\tau}\right],
\label{eq:screened_kernel_appendix}
\end{equation}
one finds
\begin{align}
Q(x,t)
&=
\frac12
+
\int dx'\,G_\gamma(x-x',t)\Bigl(Q(x',0)-\frac12\Bigr)
\nonumber\\
&\qquad
-
\int_0^t d\tau\,
G_\gamma\!\left(x-X(\tau),t-\tau\right)e^{s}r_X^\lambda(\tau).
\nonumber
\end{align}
Using Eq.~\eqref{eq:Q_initial_condition_appendix}, the initial-data term vanishes identically, so that $Q(x,t)$ is determined purely by the sink-rate history:
\begin{equation}
Q(x,t)
=
\frac12
-
\int_0^t d\tau\,
G_\gamma\!\left(x-X(\tau),t-\tau\right)e^{s}r_X^\lambda(\tau).
\label{eq:Q_duhamel_appendix}
\end{equation}
%We will use this retarded representation in the next subsection to derive the corresponding closure for $r_X(t)$ and to evaluate the on-shell action.

\subsection{On-shell action}
\label{appsubsec:onshell_action}

We now evaluate the dilute fixed-$X$ action on the saddle. It is convenient to begin with the \emph{time-bulk} part of the bilinear action Eq.~\eqref{eq:bilinear_QP_appendix}, which we call $\mathcal I_{{\rm bulk,cl}}^{(\lambda)}[X]$, and only afterwards restore the finite-time boundary term \( \big[\int dx\,(\rho\pi-\rho)\big]_0^T \) and the separate initial functional $\Phi_0[\rho(\cdot,0)]$. On the saddle, and away from the sink position $x=X(t)$, the forward field obeys
\begin{equation}
\partial_t Q
=
D\,\partial_x^2 Q
-
\gamma\Big(Q-\tfrac12\Big).
\label{eq:Q_forward_away_from_sink_appendix}
\end{equation}
Using this equation, the time-bulk integrand simplifies as
\begin{align}
P\,\partial_tQ+D\,\partial_xP\,\partial_xQ
&=
P\Big[D\,\partial_x^2Q-\gamma\Big(Q-\tfrac12\Big)\Big]\nonumber\\
&\qquad +D\,\partial_xP\,\partial_xQ
\nonumber\\
&=
D\,\partial_x\!\big(P\,\partial_xQ\big)
-\gamma P\Big(Q-\tfrac12\Big),
\end{align}
and therefore
\begin{align}
&P\partial_tQ
+D\partial_xP\partial_xQ
+\gamma(P-1)\Big(Q-\tfrac12\Big)
\nonumber\\
&\qquad=
D\,\partial_x\!\big(P\,\partial_xQ\big)
+\gamma\Big(\tfrac12-Q\Big).
\label{eq:bulk_integrand_onshell_appendix}
\end{align}
Thus, on shell, the bilinear bulk action reduces to a spatial total derivative plus a term proportional to the instantaneous depletion from the half-filled background.

Integrating Eq.~\eqref{eq:bulk_integrand_onshell_appendix} over the punctured line (i.e., excluding the sink location $x=X(t)$), the total derivative contributes only at the sink. Since $P(x,t)$ is continuous across $x=X(t)$, this gives
\begin{equation}
\int_{\mathbb R\setminus\{X(t)\}} dx\,
D\,\partial_x\!\big(P\,\partial_xQ\big)
=
-\,D\,P(X(t),t)\Big[\partial_xQ\Big]_{X(t)^-}^{X(t)^+}.
\end{equation}
Introducing the depletion mass of $Q$
\begin{equation}
\mathcal M(t)\equiv \int dx\,\Big(\tfrac12-Q(x,t)\Big),
\label{eq:MX_def_appendix_rewrite}
\end{equation}
we obtain the time-bulk part of the fluid action
\begin{align}
\mathcal I_{{\rm bulk,cl}}^{(\lambda)}[X]
=
\int_0^T dt
\Big[&
-DP(X(t),t)\Big[\partial_xQ\Big]_{X^-}^{X^+}
+\gamma\mathcal M(t)
\nonumber\\
&
+\lambda\big(P(X(t),t)-e^{-s}\big)Q(X(t),t)
\Big].
\label{eq:Ibulk_onshell_appendix}
\end{align}

At this stage we restore the endpoint piece and the initial functional. The full on-shell exponent is
\begin{align}
\mathcal I_{\rm cl}^{(\lambda)}[X]
&\equiv
\left[
\Phi_0[\rho(\cdot,0)]
+
S_{\rm dil}^{(\lambda)}[Q,P;X]
\right]_{\rm saddle}\nonumber\\
&=
\Phi_0[\rho(\cdot,0)]
+
\bigg[\int dx\,(\rho\pi-\rho)\bigg]_0^T
+
\mathcal I_{{\rm bulk,cl}}^{(\lambda)}[X].
\label{eq:Icl_def_appendix_rewrite}
\end{align}

We now simplify the sink contribution. Using the jump condition
\eqref{eq:Q_jump_appendix},
\begin{equation}
D\Big[\partial_xQ\Big]_{X(t)^-}^{X(t)^+}
=
\lambda Q(X(t),t),
\end{equation}
the two sink terms in \eqref{eq:Ibulk_onshell_appendix} combine to
\begin{align}
&-\,D\,P(X,t)\Big[\partial_xQ\Big]_{X^-}^{X^+}
+\lambda\big(P(X,t)-e^{-s}\big)Q(X,t)
\nonumber\\
&\qquad=
-\,P(X,t)\lambda Q(X,t)
+\lambda\big(P(X,t)-e^{-s}\big)Q(X,t)
\nonumber\\
&\qquad=
-\,e^{-s}\lambda Q(X,t)
=
-\,r_X^\lambda(t),
\label{eq:sink_terms_combine_appendix}
\end{align}
where in the last step we used \eqref{eq:rx_from_Q_appendix}. Hence
\begin{align}
\mathcal I_{\rm cl}^{(\lambda)}[X]
=\
&\Phi_0[\rho(\cdot,0)]
+
\bigg[\int dx\,(\rho\pi-\rho)\bigg]_0^T\nonumber\\
&+
\gamma\int_0^T dt\,\mathcal M(t)
-
\int_0^T dt\,r_X^\lambda(t).
\label{eq:onshell_intermediate_appendix_rewrite}
\end{align}

The endpoint terms are now straightforward to evaluate. At the final time, $\pi(x,T)=0$, so $P(x,T)=1$ and therefore
\begin{equation*}
\int dx\,(\rho\pi-\rho)\Big|_{t=T}
=
-\,\int dx\,\rho(x,T)
=
-\,\int dx\,Q(x,T).
\end{equation*}
At the initial time, the dilute ideal-gas functional Eq.~\eqref{eq:Phi0_dil_appendix} together with
\(
\pi_0(x)=\log\!\big[2\rho_0(x)\big]
\)
implies
\begin{equation}
\Phi_0[\rho_0]-\int dx\,(\rho_0\pi_0-\rho_0)=\int dx\,\frac12,
\end{equation}
which is an $X$-independent background constant. Therefore, up to an $X$-independent constant,
\begin{align}
\Phi_0[\rho(\cdot,0)]
+
\bigg[\int dx\,(\rho\pi-\rho)\bigg]_0^T
&=
\int dx\,\Big(\tfrac12-Q(x,T)\Big)\nonumber\\
&=
\mathcal M(T).
\label{eq:boundary_to_MT_appendix_rewrite}
\end{align}
The full on-shell exponent thus becomes
\begin{equation}
\mathcal I_{\rm cl}^{(\lambda)}[X]
=
\mathcal M(T)
+
\gamma\int_0^T dt\,\mathcal M(t)
-
\int_0^T dt\,r_X^\lambda(t).
\label{eq:onshell_mass_form_appendix_rewrite}
\end{equation}

It remains to eliminate $\mathcal M(t)$ in favour of the sink-rate history. Integrating the forward equation \eqref{eq:Q_forward_appendix} over space, and using
\eqref{eq:rx_from_Q_appendix}, gives
\begin{equation}
\dot{\mathcal M}(t)+\gamma\mathcal M(t)
=
\lambda Q(X(t),t)
=
e^s r_X^\lambda(t).
\label{eq:Mdot_identity_appendix_rewrite}
\end{equation}
Since $Q(x,0)=1/2$, the initial depletion vanishes:
\begin{equation}
\mathcal M(0)=0.
\label{eq:M0_zero_appendix_rewrite}
\end{equation}
Integrating \eqref{eq:Mdot_identity_appendix_rewrite} from $0$ to $T$ therefore yields
\begin{equation}
\mathcal M(T)+\gamma\int_0^T dt\,\mathcal M(t)
=
e^s\int_0^T dt\,r_X^\lambda(t).
\label{eq:M_integrated_identity_appendix_rewrite}
\end{equation}
Substituting this into \eqref{eq:onshell_mass_form_appendix_rewrite}, the depletion-mass terms cancel exactly, and the full fixed-$X$ exponent reduces to
\begin{equation}
\mathcal I_{\rm cl}^{(\lambda)}[X]
=
\bigl(e^s-1\bigr)\int_0^T dt\,r_X^\lambda(t),
\label{eq:onshell_exact_finite_time_appendix}
\end{equation}
up to an $X$-independent normalization.

In the fast-sink limit, this gives the exact one-fluid action
\begin{equation}
\mathcal I_{\rm cl}[X]
=
\bigl(e^s-1\bigr)\int_0^T dt\,r_X(t).
\label{eq:onshell_exact_fastsink_appendix}
\end{equation}
Restoring the two identical fluid replicas then yields the coherence-only action used in the main text,
\begin{equation}
S_{\rm eff}[X]
=
S_X[X]
+
2\bigl(e^s-1\bigr)\int_0^T dt\,r_X(t).
\label{eq:Seff_retarded_appendix}
\end{equation}

\subsection{Volterra equation for the sink-rate history}
\label{appsubsec:volterra_equation}

We now derive the closed equation that determines the sink-rate history for a prescribed trajectory $X(t)$. Starting from the retarded Duhamel solution \eqref{eq:Q_duhamel_appendix}, we evaluate the forward field $Q$ at the sink position $x=X(t)$ and use Eq.~\eqref{eq:rx_from_Q_appendix}, which relates the sink rate to $Q(X(t),t)$, to obtain the finite-$\lambda$ self-consistency equation
\begin{equation}
\frac{r_X^\lambda(t)}{\lambda e^{-s}}
=
\frac12
-
\int_0^t d\tau\,
G_\gamma\!\left(X(t)-X(\tau),t-\tau\right)e^{s}r_X^\lambda(\tau).
\label{eq:retarded_volterra_finite_lambda_appendix}
\end{equation}
This is a Volterra equation of the second kind for the sink-rate history in the regularized finite-$\lambda$ problem.

We now take the fast-sink limit $\lambda\to\infty$. Since $r_X^\lambda(t)=e^{-s}\lambda Q(X(t),t)$ remains finite, one has $Q(X(t),t)\sim \lambda^{-1}$ and therefore $Q(X(t),t)\to 0$. The left-hand side of Eq.~\eqref{eq:retarded_volterra_finite_lambda_appendix} then vanishes, and the physical sink-rate history $r_X(t)\equiv \lim_{\lambda\to\infty} r_X^\lambda(t)$ obeys the first-kind Volterra equation
\begin{equation}
\int_0^t d\tau\,
G_\gamma\!\left(X(t)-X(\tau),t-\tau\right)\,r_X(\tau)
=
\frac{e^{-s}}{2}.
\label{eq:retarded_volterra_finite_time_appendix}
\end{equation}

Equation~\eqref{eq:retarded_volterra_finite_time_appendix} is the exact finite-time closure of the fixed-$X$ saddle in the fast-sink limit within the approximate dilute fluid theory. The memory kernel is purely retarded and depends on the trajectory only through the relative displacements $X(t)-X(\tau)$.

\subsection{Jacobian determinant contribution}

Having determined the on-shell action of the Gaussian fixed-$X$ fluid theory. What remains, in principle, is the determinant prefactor produced by Gaussian integration. We now show that this contributes only an $X$-independent normalization.

Discretizing time as $t_n=n\Delta t$, and suppressing the spatial indices, the bilinear action may be written in It\^o form as
\begin{equation}
S^{\lambda=0}_{\rm dil}
=
\sum_{n=1}^{N}
\Big[
-\,Q_n(P_n-P_{n-1})
+\Delta t\,Q_n(A_nP_n-J_n)
\Big],
\label{eq:discrete_bilinear_action_appendix}
\end{equation}
where each $P_n$ and $Q_n$ are spatial vectors, and where
\begin{equation}
A_n = -D\,\partial_x^2+\gamma+\lambda\,\delta(x-X_n),
\quad
J_n=\gamma+\lambda e^{-s}\delta(x-X_n),
\label{eq:An_Jn_appendix}
\end{equation}
and the final value $P_N$ is fixed by the terminal condition on $P$. Integrating over all $Q_n$ therefore imposes the sequence of linear constraints, i.e., $\delta(F_n)$, $n=1,\cdots,N$ (where $\delta$ is the Dirac delta function), with
\begin{equation}
F_n\equiv -P_n+P_{n-1}+\Delta t\,(A_nP_n-J_n),
\label{eq:Fn_constraint_appendix}
\end{equation}
so that, starting from the fixed terminal value $P_N$, the fields on
earlier time slices are determined recursively by
\begin{equation}
P_{n-1}=P_n-\Delta t\,(A_nP_n-J_n).
\label{eq:backward_recursion_appendix}
\end{equation}
Thus, at each step, we have $\frac{\partial F_n}{\partial P_{n-1}} = \mathbb{1}$. The Jacobian of the full sequence of constraints is therefore just a product of identity determinants, $\prod_{n=1}^{N} \det \mathbb{1} = 1$, and is independent of the position $X_n$ at every time slice.

Thus the exact Gaussian integration differs from the classical result only by an $X$-independent normalization. Within the dilute-core theory, the full $X$-dependence of the fixed-$X$ fluid functional is therefore exactly given by the on-shell action derived above.

\section{Weak-noise coherence--void polaron diffusivity}
\label{app:weak_noise_polaron_diffusivity}

In the main text, we extracted the weak-noise dressed diffusion constant \(D_{\rm eff}\) from the ballistic large-deviation cost of translating the coherence--void polaron. In this appendix we give two complementary derivations of the same scaling. First, we show that the retarded \(X\)-only action reduces at low frequency to a local Gaussian translation mode with \(D_{\rm eff}\sim \sqrt{\gamma}\). Second, we verify the same scaling directly from the full nonlinear moving-coherence MFT saddle, without making the dilute-fluid approximation. Finally, we give a direct TEBD comparison showing that the full two-replica statistical mechanics model and the dilute approximation give the same time-dependent coherence diffusivity at small $\gamma$ over the accessible late-time window.

\subsection{Low-frequency limit of the retarded coherence action}
\label{appsubsec:low_frequency_retarded_action}

We start from the effective coherence action
\begin{equation}
S_{\rm eff}[X] = S_X[X] + 2(e^s-1) \int_0^T dt \, r_X(t),
\label{eq:app_lowfreq_Seff}
\end{equation}
with bare kinetic term
\begin{equation}
S_X[X] = \frac{1}{4D}\int_0^T dt \, \dot X(t)^2,
\label{eq:app_lowfreq_SX}
\end{equation}
and where the filtering rate \(r_X(t)\) is determined implicitly by the Volterra equation
\begin{equation}
\int_0^t d\tau \,
G_\gamma\bigl(X(t)-X(\tau),t-\tau\bigr) \, r_X(\tau)
= \frac{e^{-s}}{2},
\label{eq:app_lowfreq_volterra0}
\end{equation}
with screened kernel
\begin{equation}
G_\gamma(\Delta x,u)
= \frac{e^{-\gamma u}}{\sqrt{4\pi D u}}
\exp\!\left[-\frac{(\Delta x)^2}{4Du}\right].
\label{eq:app_lowfreq_kernel}
\end{equation}
Introducing \(u=t-\tau\), Eq.~\eqref{eq:app_lowfreq_volterra0} may be written as
\begin{equation}
\int_0^t du\, K_\gamma(u)
\exp\!\left[-\frac{\bigl(X(t)\!-\!X(t\!-\!u)\bigr)^2}{4Du}\right]
r_X(t-u)=\frac{e^{-s}}{2},
\label{eq:app_lowfreq_volterra1}
\end{equation}
where
\begin{equation}
K_\gamma(u) \equiv \frac{e^{-\gamma u}}{\sqrt{4\pi D u}}.
\label{eq:app_lowfreq_K}
\end{equation}
For the long-time, low-frequency translation mode, \(T\gg \gamma^{-1}\), we may extend the upper limit to infinity, since the memory kernel is exponentially cut off on the screening timescale \(\gamma^{-1}\).

For the static trajectory \(X(t)=\mathrm{const}\), the filtering rate is time-independent, \(r_X(t)=r_0\), and Eq.~\eqref{eq:app_lowfreq_volterra1} gives
\begin{equation}
r_0 \int_0^\infty du\, K_\gamma(u)=\frac{e^{-s}}{2}.
\label{eq:app_lowfreq_static1}
\end{equation}
Therefore
\begin{equation}
r_0=e^{-s}\sqrt{D\gamma}.
\label{eq:app_lowfreq_r0}
\end{equation}

We now expand about this pinned solution for a slowly varying trajectory, writing
\begin{equation}
r_X(t)=r_0+\delta r(t).
\label{eq:app_lowfreq_rsplit}
\end{equation}
If \(X(t)\) varies only weakly over one memory time \(u\sim \gamma^{-1}\), then
\begin{equation}
X(t)-X(t-u)=u\dot X(t)+O\!\left(u^2\ddot X\right),
\label{eq:app_lowfreq_Xexpand}
\end{equation}
so that
\begin{equation}
\exp\!\left[-\frac{\bigl(X(t)-X(t-u)\bigr)^2}{4Du}\right]
=1-\frac{u\dot X(t)^2}{4D}+\cdots.
\label{eq:app_lowfreq_expansion}
\end{equation}
At the same order it is sufficient to replace \(r_X(t-u)\) by \(r_0+\delta r(t)\), since the omitted retardation corrections are higher order in temporal gradients.

Substituting Eqs.~\eqref{eq:app_lowfreq_rsplit} and \eqref{eq:app_lowfreq_expansion} into Eq.~\eqref{eq:app_lowfreq_volterra1}, and using Eq.~\eqref{eq:app_lowfreq_static1} to cancel the static part, gives
\begin{equation}
\delta r(t)\int_0^\infty du\,K_\gamma(u)
= r_0\frac{\dot X(t)^2}{4D}\int_0^\infty du\,uK_\gamma(u).
\label{eq:app_lowfreq_deltar1}
\end{equation}
The required moments of the kernel are
\begin{equation}
\int_0^\infty du\,K_\gamma(u)=\frac{1}{2\sqrt{D\gamma}},
\ \
\int_0^\infty du\,uK_\gamma(u)=\frac{1}{4\sqrt{D}\,\gamma^{3/2}}.
\label{eq:app_lowfreq_kernelmoments}
\end{equation}
Using Eq.~\eqref{eq:app_lowfreq_r0}, we obtain
\begin{equation}
\delta r(t)=\frac{e^{-s}}{8\sqrt{D\gamma}}\,\dot X(t)^2.
\label{eq:app_lowfreq_deltar2}
\end{equation}

Substituting this back into Eq.~\eqref{eq:app_lowfreq_Seff}, and dropping the path-independent baseline \(2(e^s-1)r_0T\), gives the local low-frequency form
\begin{align}
S_{\rm eff}[X]
&= \mathrm{const}
+ \int_0^T dt\left[
\frac{1}{4D}
+\frac{1-e^{-s}}{4\sqrt{D\gamma}}
\right]\dot X(t)^2
+\cdots\nonumber\\
&\equiv
\mathrm{const}
+\frac{1}{4D_{\rm eff}}\int_0^T dt\,\dot X(t)^2+\cdots.
\label{eq:app_lowfreq_localaction}
\end{align}
Therefore,
\begin{equation}
D_{\rm eff}^{-1}=D^{-1}+\frac{1-e^{-s}}{\sqrt{D\gamma}},
\qquad
D_{\rm eff}(\gamma)\sim \sqrt{\gamma}.
\label{eq:app_lowfreq_Deff}
\end{equation}
Thus the retarded coherence action reduces, at asymptotically low frequency, to a local Gaussian translation mode with parametrically small dressed diffusion constant \(D_{\rm eff}\sim\sqrt{\gamma}\). The omitted terms are controlled by the small parameter \(\omega/\gamma\), where \(\omega\) is the characteristic frequency of the translation mode, and therefore do not affect the leading weak-noise scaling.

\subsection{Large-deviation rate function of the full nonlinear fluid saddle}
\label{appsubsec:full_fluid_Deff_check}

We now verify the same scaling directly within the full nonlinear hydrodynamic large-deviation problem. Unlike the derivation above, this calculation does not use the dilute-fluid approximation. Instead, we work directly with the full replica-symmetric moving-coherence saddle and extract the velocity dependence of the large-deviation rate function.

We start from the phase-averaged two-replica moving-coherence action, Eq.~\eqref{eq:two_rep_moving_action}, and use the replica symmetry of the saddle equations and boundary conditions to restrict to symmetric saddle solutions \(\rho_1=\rho_2\equiv \rho\) and \(\pi_1=\pi_2\equiv \pi\). Working in the branch-projection limit established in the main text, Eq.~\eqref{eq:hard-repair}, the replica-symmetric action reduces to
\begin{equation}
S[\rho,\pi,X]
=
S_X[X]
+
2S_{\rm bulk}[\rho,\pi]
+
2S_{\rm bdy}[\rho,X],
\label{eq:app_fullfluid_action}
\end{equation}
with \(S_{\rm bulk}\) and \(S_{\rm bdy}\) given by Eqs.~\eqref{eq:Sbulk_1rep_general} and \eqref{eq:gradient_tilt}, respectively, and \(S_X[X]=\frac{1}{4D}\int dt \, \dot{X}^2\), and where the factor of \(2\) multiplying the fluid contribution reflects the two replicas. As in the main text, varying the boundary gradients at the coherence gives the equivalent Dirichlet conditions
\begin{align}
\rho(X(t),t)&=0,
&
\pi(X(t),t)&=-s,
\nonumber\\
\rho(|x|\to\infty,t)&=\tfrac12,
&
\pi(|x|\to\infty,t)&=0.
\label{eq:app_fullfluid_bcs}
\end{align}

We define the velocity-dependent rate function by
\begin{equation}
I(v)
\equiv
\lim_{T\to\infty}
\frac{1}{T}
\inf_{\substack{X(T)=vT\\ \rho,\pi}}
\Bigl\{
S[\rho,\pi,X]-S[\rho,\pi,X]\big|_{v=0}
\Bigr\}.
\label{eq:app_Iv_def}
\end{equation}
For small \(v\),
\begin{equation}
I(v)=\frac{v^2}{4D_{\rm eff}}+O(v^4).
\label{eq:app_Iv_smallv}
\end{equation}
To evaluate \(I(v)\), we assume that at large \(T\) the optimal trajectory compatible with \(X(T)=vT\) is the uniform-drift saddle
\begin{equation}
X(t)=vt,
\label{eq:app_Xeqvt}
\end{equation}
with stationary fields in the comoving coordinate
\begin{equation}
y=x-vt,
\quad
\rho(x,t)=\rho_v(y),
\quad
\pi(x,t)=\pi_v(y).
\label{eq:app_traveling_wave}
\end{equation}
This retains the full nonlinear fluid problem, while restricting to the natural time-translation-invariant saddle in the moving frame.

Substituting Eq.~\eqref{eq:app_traveling_wave} into the full fluid bulk action yields the stationary comoving functional on the whole line. Since the coherence sits at \(y=0\), the problem splits into two half-lines, with the left half-line obtained from the right by \(v\to -v\). It is therefore enough to work on \(y\ge 0\).

For the SSEP diffusivity and mobility,
\begin{equation}
D(\rho)=D,
\qquad
\sigma(\rho)=2D\rho(1-\rho),
\label{eq:app_ssep_constitutive}
\end{equation}
and the weak depolarizing bath term
\begin{equation}
H_{\rm bath}(\rho,\pi)
=
\frac12\Big[\rho(e^{-\pi}-1)+(1-\rho)(e^\pi-1)\Big],
\label{eq:app_Hbath}
\end{equation}
the right-half-line stationary comoving action per unit time is
\begin{equation}
\mathcal{F}_v[\rho,\pi]
=
\!\int_0^\infty\!\!\!\! dy
\Big[
D\pi'\rho'
-\frac{\sigma(\rho)}{2}(\pi')^2
-\gamma H_{\rm bath}(\rho,\pi)
-v\pi\rho'
\Big],
\label{eq:app_Fv_dimful}
\end{equation}
where primes denote derivatives with respect to \(y\). The corresponding Euler--Lagrange equations are
\begin{align}
D\,\rho'' &=
\bigl(\sigma(\rho)\pi'\bigr)'
-\frac{\gamma}{2}\Big[(1-\rho)e^\pi-\rho e^{-\pi}\Big]
-v\,\rho',
\label{eq:app_dimful_rho}
\\
D\,\pi'' &=
\gamma \sinh\pi
-\frac{\sigma'(\rho)}{2}(\pi')^2
+v\,\pi'.
\label{eq:app_dimful_pi}
\end{align}
The boundary conditions on the right half-line are
\begin{equation}
\rho(0)=0,
\quad
\pi(0)=-s,
\quad
\rho(\infty)=\tfrac12,
\quad
\pi(\infty)=0.
\label{eq:app_dimful_bcs}
\end{equation}

Introduce the screened coordinate \(z\) and rescaled velocity \(u\),
\begin{equation}
z\equiv \sqrt{\frac{\gamma}{D}}\,y,
\qquad
u\equiv \frac{v}{\sqrt{D\gamma}}.
\label{eq:app_z_def}
\end{equation}
Writing $\hat\rho(z)\equiv \rho_v(y)$, and $\hat\pi(z)\equiv \pi_v(y)$, the half-line action becomes
\begin{equation}
\mathcal{F}_v[\rho,\pi]
=
\sqrt{D\gamma}\,
\mathcal{F}[u;\hat\rho,\hat\pi],
\label{eq:app_scaling_of_action}
\end{equation}
with dimensionless functional
\begin{align}
\mathcal{F}[u;\hat\rho,\hat\pi]
=
\int_0^\infty dz\,
\Big[&
\hat\pi'\hat\rho'
-\frac12\sigma_0(\hat\rho)(\hat\pi')^2\nonumber\\
&\quad -H_{\rm bath}(\hat\rho,\hat\pi)
-u\,\hat\pi\,\hat\rho'
\Big],
\label{eq:app_dimensionless_functional}
\end{align}
where \(\sigma_0(\hat\rho)=2\hat\rho(1-\hat\rho)\), and primes now denote derivatives with respect to \(z\).

Extremizing Eq.~\eqref{eq:app_dimensionless_functional} gives a dimensionless boundary-value problem
\begin{align}
\hat\rho''
&=
\bigl(\sigma_0(\hat\rho)\hat\pi'\bigr)'
-\frac12\Big[(1-\hat\rho)e^{\hat\pi}-\hat\rho e^{-\hat\pi}\Big]
-u\,\hat\rho',
\label{eq:app_bvp_rho}
\\
\hat\pi''
&=
\sinh(\hat\pi)
-\frac12\sigma_0'(\hat\rho)(\hat\pi')^2
+u\,\hat\pi',
\label{eq:app_bvp_pi}
\end{align}
with boundary conditions
\begin{equation}
\hat\rho(0)=0,
\quad
\hat\pi(0)=-s,
\quad
\hat\rho(\infty)=\tfrac12,
\quad
\hat\pi(\infty)=0.
\label{eq:app_bvp_bcs}
\end{equation}
We denote the extremal value of the right-half-line action by
\begin{equation}
\Phi(u)\equiv \mathcal{F}[u;\hat\rho_u,\hat\pi_u],
\label{eq:app_Phi_def}
\end{equation}
where \((\hat\rho_u,\hat\pi_u)\) solves Eqs.~\eqref{eq:app_bvp_rho}--\eqref{eq:app_bvp_bcs}.

The full action for the constant-velocity saddle contains the bare coherence kinetic contribution plus the two fluid half-lines. Subtracting the static \(v=0\) saddle, the resulting rate function is
\begin{equation}
I(v)
=
\frac{v^2}{4D}
+
\sqrt{D\gamma}\,
\Big(
\Phi(u)+\Phi(-u)-2\Phi(0)
\Big),
\label{eq:app_Iv_final}
\end{equation}
with \(u=v/\sqrt{D\gamma}\). This expression already determines the weak-noise scaling. All dependence on \(\gamma\) enters through the overall factor \(\sqrt{D\gamma}\) and through the scaled velocity \(u\).

Since the right- and left-half-line contributions are related by \(u\to -u\), the odd terms cancel. Expanding around \(u=0\),
\begin{equation}
\Phi(u)+\Phi(-u)-2\Phi(0)
=
2a(s)\,u^2+O(u^4),
\label{eq:app_Phi_expand}
\end{equation}
for some dimensionless coefficient \(a(s)\) determined by the full nonlinear dimensionless boundary-value problem. Substituting Eq.~\eqref{eq:app_Phi_expand} into Eq.~\eqref{eq:app_Iv_final} gives
\begin{equation}
I(v)
=
\left(
\frac{1}{4D}
+
\frac{2a(s)}{\sqrt{D\gamma}}
\right)v^2
+
O(v^4).
\label{eq:app_Iv_smallv_final}
\end{equation}
Comparing with Eq.~\eqref{eq:app_Iv_smallv}, we obtain
\begin{equation}
D_{\rm eff}(\gamma,s)
=
\frac{\sqrt{D\gamma}\,D}{\sqrt{D\gamma}+8a(s)D}
\sim \sqrt{D\gamma}.
\label{eq:app_Deff_scaling}
\end{equation}
Thus the dressed translation mode remains diffusive, but its diffusion constant is parametrically smaller than the bare coherence diffusion constant:
\begin{equation}
D_{\rm eff}(\gamma)\propto \sqrt{\gamma}.
\label{eq:app_Deff_final_scaling}
\end{equation}
This result is obtained directly from the full nonlinear moving-coherence fluid saddle, without using the dilute-fluid approximation.

\subsection{Numerical comparison of the exact two-replica dynamics and the dilute approximation}
\label{appsubsec:tebd_full_dilute_diffusivity}

We finally give an additional numerical check of the dilute-coherence approximation for the weak-noise translation dynamics in $U(1)$-conserving random unitary circuits subject to weak depolarizing noise. In the main text, the long-time diffusivity of the coherence--void polaron was extracted most cleanly from population-dynamics sampling of the sub-stochastic process obtained under the dilute-coherence approximation (Sec.~\ref{subsec:screened_Deff_numerics}), using the void center of mass as a proxy for the slow polaron coordinate. Here we instead compare the physical coherence spreading directly in the full two-replica statistical-mechanics transfer matrix (Sec.~\ref{RUCsection}) and in the dilute-coherence approximation.

Specifically, we compute the normalized coherence-position distribution \(P(X,t)\) (obtained by normalizing $\overline{|\mathcal{G}(x,t)|^2}$) by TEBD simulation and extract the mean-squared displacement and time-dependent diffusion constant $D_X(t)$.
\begin{equation}
    \langle X(t)^2\rangle
    =
    \sum_X X^2 P(X,t),
    \qquad
    D_X(t)
    =
    {1\over 2}\partial_t \langle X(t)^2\rangle .
\label{eq:app_tebd_Dt_def}
\end{equation}
Because this calculation uses TEBD rather than population dynamics sampling, the accessible times are shorter than those reached in the main-text population-dynamics simulations. Nevertheless, it provides a direct test of whether the full microscopic statistical mechanics model and the substochastic model obtained under the dilute-coherence approximation give the same time-dependent coherence diffusivity at weak noise $\gamma$.

\begin{figure}
    \centering
    \includegraphics[width=\linewidth]{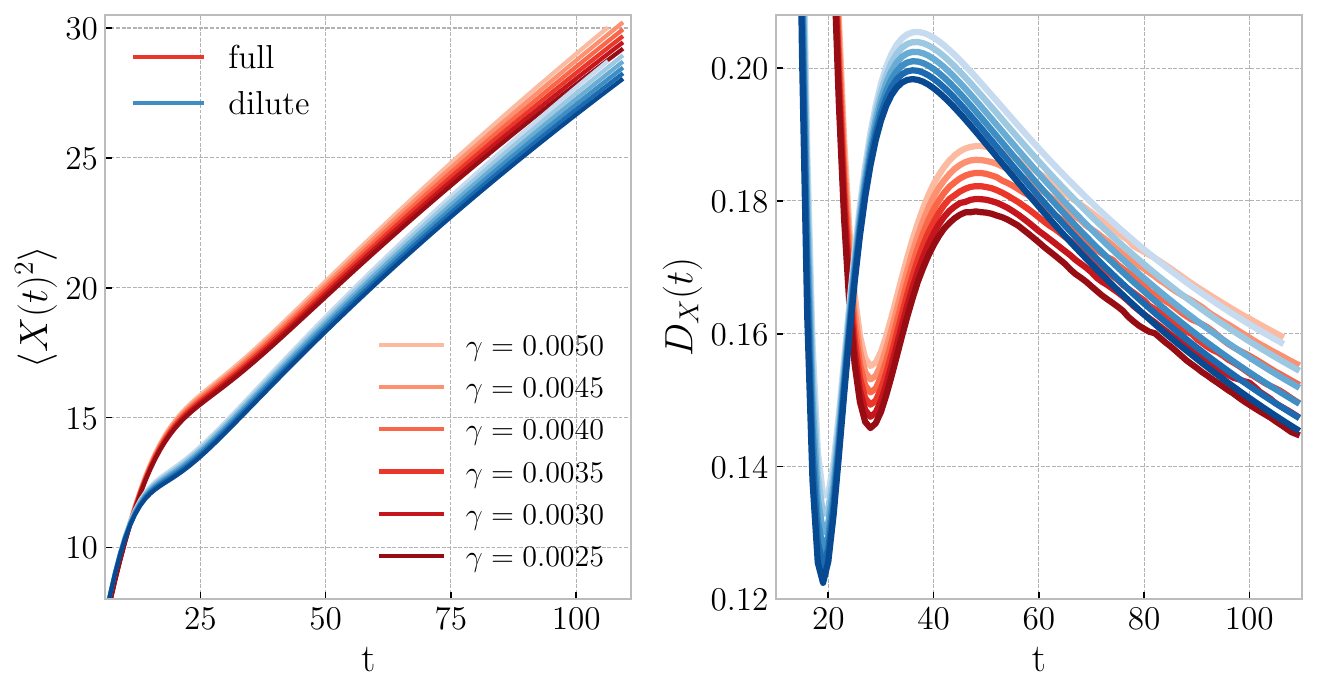}
    \caption{\textbf{Weak noise coherence--void polaron diffusion}. A comparison of the time-dependent coherence diffusion in the full random-circuit dynamics (using the exact two-replica statistical mechanics model in Sec.~\ref{RUCsection}) and within the dilute-coherence approximation. Left: coherence-position variance \(\langle X(t)^2\rangle\) for    \(\gamma=0.0025,\ldots,0.0050\). Right: corresponding time-dependent diffusion constant \(D_X(t)\equiv\frac{1}{2}\partial_t\langle X(t)^2\rangle\). Over the accessible late-time window the time-dependent coherence diffusion constant in the full and dilute-coherence models agree well. These simulations use bond dimension $\chi = 650$ and system size $L=84$.
    }
    \label{fig:full_dilute_Dt}
\end{figure}

As shown in Fig.~\ref{fig:full_dilute_Dt}, the corresponding time-dependent diffusivities \(D_X(t)\), initially far apart, agree at late times between the full and approximate statistical mechanics transfer matrix evolution. This supports the use of the dilute approximation in the population-dynamics simulations of the main text, where it allows access to later times and to the void center-of-mass motion. At earlier and intermediate times, the dilute-coherence approximation and full transfer-matrix evolution show a sizable discrepancy, suggesting that the coherence-pair creation and recombination processes neglected in the dilute approximation play a role in the short-time dynamics of the coherence within the void, even though they do not appear to change the late-time translation dynamics. We leave the exploration of such creation/recombination processes to future work.

\section{Annealed evaluation of the nonlocal coherence action under the Feynman trial measure}
\label{app:annealed_trial_expectation}

In the Feynman--Jensen estimates used in the main text, the only nontrivial quantity is the trial expectation of the fluid conditioning term appearing in Eq.~\eqref{eq:FJ-functional} (through the non-local coherence action Eq.~\eqref{eq:Seff_maintext_corrected}),
\begin{equation}
\Big\langle \int_0^T dt\, r_X(t)\Big\rangle_{\rm tr},
\label{eq:app_target_quantity}
\end{equation}
where \(r_X(t)\) is the outward fluid-gradient jump at the coherence, equivalently the instantaneous fluid filtering rate, introduced in Eq.~\eqref{eq:rX_maintext}. For a prescribed trajectory \(X(t)\), this rate is determined implicitly by the retarded Volterra closure. In the weak-noise problem,
\begin{equation}
\int_0^t d\tau\,
G_\gamma\!\left(X(t)-X(\tau),t-\tau\right)\,r_X(\tau)
=
\frac{e^{-s}}{2},
\label{eq:app_volterra_screened}
\end{equation}
with screened kernel
\begin{equation}
G_\gamma(\Delta x,\tau)
=
\frac{e^{-\gamma \tau}}{\sqrt{4\pi D \tau}}
\exp\!\left[-\frac{(\Delta x)^2}{4D\tau}\right],
\label{eq:app_screened_kernel}
\end{equation}
while in the noiseless problem the same relation holds with \(\gamma=0\).

The dependence of \(r_X(t)\) on the coherence trajectory is therefore non-Gaussian and not available in closed form under the trial measure. In this appendix we estimate Eq.~\eqref{eq:app_target_quantity} in the annealed approximation, by averaging the Volterra kernel under the Gaussian trial process used in the main text. We first treat the weak-noise stationary problem, and then the noiseless aging problem.

\subsection{Weak noise}
\label{appsubsec:annealed_rX_weak}

In the weak-noise regime we use the stationary Gaussian trial measure of Sec.~\ref{subsec:feynman-polaron},
\begin{equation}
S_{\rm tr}[Y,\zeta]
=
\int_0^T dt\,
\left[
\frac{\dot Y(t)^2}{4D_{\rm eff}}
+
\frac{\dot\zeta(t)^2}{4D}
+
\frac{D}{4\ell_\zeta^4}\,\zeta(t)^2
\right],
\label{eq:app_trial_action_weak}
\end{equation}
with \(X(t)=Y(t)+\zeta(t)\). Since this trial process is stationary, the annealed filtering rate is time independent, and Eq.~\eqref{eq:app_target_quantity} reduces to
\begin{equation}
\Big\langle \int_0^T dt\, r_X(t)\Big\rangle_{\rm tr}
\;\approx\;
T\,\overline r_{\ell_\zeta},
\label{eq:app_target_quantity_weak}
\end{equation}
where \(\overline r_{\ell_\zeta}\) is obtained by annealing the screened kernel in Eq.~\eqref{eq:app_volterra_screened}.

Using \(\langle e^{-aZ^2}\rangle=(1+2a\langle Z^2\rangle)^{-1/2}\) for a centered Gaussian variable \(Z\), the annealed kernel is
\begin{align}
\overline G_{\gamma,\ell_\zeta}(u)
&\equiv
\left\langle
G_\gamma(X(t)-X(t-u),u)
\right\rangle_{\rm tr}\nonumber\\
&=
\frac{e^{-\gamma u}}{\sqrt{4\pi D u}}
\left(
1+\frac{\Sigma_{\ell_\zeta}(u)}{2Du}
\right)^{-1/2},
\label{eq:app_annealed_kernel_weak}
\end{align}
with increment variance
\begin{equation}
\Sigma_{\ell_\zeta}(u)
=
2D_{\rm eff}u
+
2\ell_\zeta^2\!\left(1-e^{-Du/\ell_\zeta^2}\right).
\label{eq:app_sigma_weak}
\end{equation}
The annealed rate is then determined by
\begin{equation}
\overline r_{\ell_\zeta}
\int_0^\infty du\,\overline G_{\gamma,\ell_\zeta}(u)
=
\frac{e^{-s}}{2}.
\label{eq:app_annealed_volterra_weak}
\end{equation}

To extract the \(\ell_\zeta\)-dependence, we compare with the case with a stationary coherence
\(X(t)=\mathrm{const}\), for which
\begin{equation}
I_{\rm pin}
\equiv
\int_0^\infty du\,G_\gamma(0,u)
=
\frac{1}{2\sqrt{D\gamma}}.
\label{eq:app_Ipin_weak}
\end{equation}
Writing
\begin{equation}
\Delta I_{\ell_\zeta}
\equiv
\int_0^\infty du\,
\Big[
G_\gamma(0,u)-\overline G_{\gamma,\ell_\zeta}(u)
\Big],
\end{equation}
we have
\begin{equation}
\int_0^\infty du\,\overline G_{\gamma,\ell_\zeta}(u)
=
I_{\rm pin}-\Delta I_{\ell_\zeta}.
\label{eq:app_Ibar_split_weak}
\end{equation}

The relevant crossover time is
\begin{equation}
\tau_\zeta\sim \omega_\zeta^{-1}\sim \frac{\ell_\zeta^2}{D}.
\end{equation}
For \(u\ll \tau_\zeta\), the internal displacement has not yet saturated, so
\(\Sigma_{\ell_\zeta}(u)\simeq 2D_{\rm eff}u+2Du\), and the annealed kernel is reduced by an \(O(1)\) fraction relative to the pinned kernel. This gives
\begin{equation}
\Delta I_{\ell_\zeta}^{\rm short}
\sim
\int_0^{\tau_\zeta} du\,\frac{1}{\sqrt{D u}}
\sim
\frac{\ell_\zeta}{D}.
\label{eq:app_short_window_weak}
\end{equation}
For \(u\gg \tau_\zeta\), the internal motion has saturated,
\(\Sigma_{\ell_\zeta}(u)\simeq 2D_{\rm eff}u+2\ell_\zeta^2\), and the correction is perturbative:
\begin{equation}
G_\gamma(0,u)-\overline G_{\gamma,\ell_\zeta}(u)
\sim
\frac{e^{-\gamma u}}{\sqrt{D u}}\,
\frac{\ell_\zeta^2}{Du}.
\end{equation}
Since \(\ell_\zeta\ll \xi\) implies \(\tau_\zeta\ll \gamma^{-1}\), the long-lag integral contains an intermediate window \(\tau_\zeta\lesssim u\lesssim \gamma^{-1}\) in which the exponential cutoff is ineffective. This window contributes
\begin{equation}
\Delta I_{\ell_\zeta}^{\rm long}
\sim
\frac{\ell_\zeta^2}{D^{3/2}}
\int_{\tau_\zeta}^{\gamma^{-1}} du\,u^{-3/2}
\sim
\frac{\ell_\zeta}{D},
\end{equation}
the same scaling as the short-lag contribution. Only the far screened tail \(u\gtrsim \gamma^{-1}\) is parametrically smaller. Thus
\begin{equation}
\Delta I_{\ell_\zeta}\sim \frac{\ell_\zeta}{D},
\end{equation}
and therefore
\begin{equation}
\int_0^\infty du\,\overline G_{\gamma,\ell_\zeta}(u)
=
\frac{1}{2\sqrt{D\gamma}}
-
c_1\,\frac{\ell_\zeta}{D}
+\cdots,
\label{eq:app_integral_annealed_weak}
\end{equation}
with \(c_1>0\) an \(O(1)\) constant.

Substituting Eq.~\eqref{eq:app_integral_annealed_weak} into
Eq.~\eqref{eq:app_annealed_volterra_weak} and expanding for
\(\ell_\zeta/\xi\ll 1\) gives
\begin{equation}
\overline r_{\ell_\zeta}
=
e^{-s}\sqrt{D\gamma}
+
c_2\,\gamma \ell_\zeta
+\cdots,
\label{eq:app_rbar_weak}
\end{equation}
with \(c_2>0\) also \(O(1)\). Hence
\begin{equation}
\Big\langle \int_0^T dt\,r_X(t)\Big\rangle_{\rm tr}
\approx
T\left[
e^{-s}\sqrt{D\gamma}
+
c_2\,\gamma \ell_\zeta
+\cdots
\right].
\label{eq:app_rint_weak_final}
\end{equation}
The first term is independent of the variational parameter and may be dropped in the Feynman--Jensen optimization. Thus the \(\ell_\zeta\)-dependent part of the fluid-conditioning cost scales as $\mathcal{O}(\gamma \ell_\zeta T)$ as stated in the main text.

\subsection{Noiseless regime}
\label{appsubsec:annealed_rX_noiseless}

In the noiseless problem we use the one-field Gaussian trial measure introduced in the main text,
\begin{equation}
S_{\rm tr}[X]
=
\int_0^T dt\,
\left[
\frac{\dot X(t)^2}{4D}
+
\frac{D}{4\ell_X(t)^4}\,X(t)^2
\right].
\label{eq:app_trial_action_noiseless}
\end{equation}
Unlike in the weak-noise case, the trial process is now nonstationary, so the annealed filtering rate remains time dependent. Accordingly, we estimate
\begin{equation}
\Big\langle \int_0^T dt\,r_X(t)\Big\rangle_{\rm tr}
\;\approx\;
\int_0^T dt\,\overline r_{\ell_X}(t),
\label{eq:app_target_quantity_noiseless}
\end{equation}
where \(\overline r_{\ell_X}(t)\) is determined by annealing the noiseless Volterra kernel.

For the Gaussian trial process, the annealed kernel is
\begin{align}
\overline G_t(u)
&\equiv
\left\langle
G\!\left(X(t)-X(t-u),u\right)
\right\rangle_{\rm tr}\nonumber\\
&=
\frac{1}{\sqrt{4\pi D u}}
\left(
1+\frac{\Sigma_X(t,u)}{2Du}
\right)^{-1/2},
\label{eq:app_annealed_kernel_noiseless}
\end{align}
with
\begin{equation}
\Sigma_X(t,u)
\equiv
\left\langle [X(t)-X(t-u)]^2\right\rangle_{\rm tr}.
\end{equation}
The annealed rate then satisfies
\begin{equation}
\int_0^t du\,\overline G_t(u)\,\overline r_{\ell_X}(t-u)
=
\frac{e^{-s}}{2}.
\label{eq:app_annealed_volterra_noiseless}
\end{equation}

To extract the \(\ell_X\)-dependence, we compare with the stationary reference problem \(X(t)\equiv 0\), for which
\begin{equation}
\int_0^t d\tau\,G(0,t-\tau)\,r_0(\tau)=\frac{e^{-s}}{2}
\implies
r_0(t)=e^{-s}\sqrt{\frac{D}{\pi t}}.
\label{eq:app_r0_noiseless}
\end{equation}
Writing
\begin{equation}
\delta G_t(u)\equiv \overline G_t(u)-G(0,u),
\qquad
\delta r_{\ell_X}(t)\equiv \overline r_{\ell_X}(t)-r_0(t),
\end{equation}
and subtracting Eq.~\eqref{eq:app_r0_noiseless} from
Eq.~\eqref{eq:app_annealed_volterra_noiseless}, we obtain the exact relation
\begin{equation}
\int_0^t du\,G(0,u)\,\delta r_{\ell_X}(t-u)
=
-\int_0^t du\,\delta G_t(u)\,\overline r_{\ell_X}(t-u).
\label{eq:app_exact_delta_r_noiseless}
\end{equation}

We now assume an algebraic exploration law
\begin{equation}
\ell_X(t)\sim t^\alpha,
\qquad
\alpha<\tfrac12,
\label{eq:app_alpha_assumption_noiseless}
\end{equation}
so that the trial exploration scale remains parametrically smaller than the diffusive void scale, and the exploration time remains parametrically smaller than $t$,
\begin{equation}
\ell_X(t)\ll \sqrt{Dt}, \qquad \tau_X(t)\sim \frac{\ell_X(t)^2}{D}\ll t.
\label{eq:app_tauX_noiseless}
\end{equation}
Under this assumption, $\Sigma_X(t,u)$ has two regimes
\begin{equation}
\Sigma_X(t,u)
\sim
\begin{cases}
D\,u, & u\ll \tau_X(t),\\[0.5ex]
\ell_X(t)^2, & u\gg \tau_X(t).
\end{cases}
\label{eq:app_sigma_noiseless_scaling}
\end{equation}
Hence the kernel deformation obeys
\begin{equation}
\delta G_t(u)\sim
\begin{cases}
-\,u^{-1/2}, & u\ll \tau_X(t),\\[0.5ex]
-\,\dfrac{\ell_X(t)^2}{D}\,u^{-3/2}, & u\gg \tau_X(t).
\end{cases}
\label{eq:app_deltaG_noiseless_scaling}
\end{equation}
The short-lag window is not perturbative, but is confined to the parametrically small interval \(u\lesssim \tau_X(t)\); the long-lag window is perturbative in the small ratio \(\ell_X(t)^2/(Du)\).

At the scaling level, the correction \(\delta r_{\ell_X}\) is subleading, so we may evaluate the right-hand side of Eq.~\eqref{eq:app_exact_delta_r_noiseless} on the pinned problem ($X(t)=0$) reference rate \(r_0\). Since \(r_0(t-u)\sim t^{-1/2}\), the short-lag contribution scales as
\begin{equation}
t^{-1/2}\int_0^{\tau_X(t)} du\,u^{-1/2}
\sim
\frac{\ell_X(t)}{t^{1/2}},
\label{eq:app_short_lag_noiseless}
\end{equation}
while the long-lag contribution scales as
\begin{equation}
\frac{\ell_X(t)^2}{D}\,
t^{-1/2}
\int_{\tau_X(t)}^t du\,u^{-3/2}
\sim
\frac{\ell_X(t)}{t^{1/2}}.
\label{eq:app_long_lag_noiseless}
\end{equation}
Thus the source term in Eq.~\eqref{eq:app_exact_delta_r_noiseless} has the scaling
\begin{equation}
\int_0^t du\,\delta G_t(u)\,\overline r_{\ell_X}(t-u)
\sim
\frac{\ell_X(t)}{t^{1/2}}.
\label{eq:app_source_scaling_noiseless}
\end{equation}

Since the left-hand side of Eq.~\eqref{eq:app_exact_delta_r_noiseless} is an Abel transform, Abel inversion gives
\begin{equation}
\delta r_{\ell_X}(t)\sim \frac{\ell_X(t)}{t}.
\label{eq:app_delta_r_noiseless_final}
\end{equation}
The same result also follows directly by scaling: for \(\ell_X(t)\sim t^\alpha\), the term on the right scales as \(t^{\alpha-1/2}\), while the left-hand side behaves as the convolution of \(u^{-1/2}\) with \(\delta r_{\ell_X}(t)\sim t^\beta\), hence scales as \(t^{\beta+1/2}\); matching powers yields \(\beta=\alpha-1\), so \(\delta r_{\ell_X}(t)\sim \ell_X(t)/t\).

We therefore obtain
\begin{equation}
\overline r_{\ell_X}(t)
=
r_0(t)+c_r\,\frac{\ell_X(t)}{t}+\cdots,
\label{eq:app_rbar_noiseless_final}
\end{equation}
with \(c_r=\mathcal O(1)\). Hence
\begin{equation}
\Big\langle \int_0^T dt\,r_X(t)\Big\rangle_{\rm tr}
\approx
\int_0^T dt\,r_0(t)
+
c_r\int_0^T dt\,\frac{\ell_X(t)}{t}
+\cdots.
\label{eq:app_rint_noiseless_final}
\end{equation}
The first term is independent of the variational parameter and may be dropped in the Feynman--Jensen optimization. Thus the \(\ell_X\)-dependent part of the fluid-conditioning cost scales as $\int_0^T dt\,\frac{\ell_X(t)}{t}$, as stated in the main text.

\section{Subdiffusion of the void center-of-mass in the aging polaron}
\label{appsec:aging_com}

In Sec.~VII we estimated the subdiffusive wandering of the coherence in the noiseless aging-polaron regime. Here we briefly estimate the motion of the void center of mass and of the relative coordinate between the coherence and the void center. The argument is only intended to establish the scaling of these observables, and is not needed for the main scaling form of the coherence Green's function.

We use a simple passive-source approximation: each filtering event at the coherence, which converts an incoming depolarized spin into a polarized spin, is viewed as the emission of a polarized spin from the coherence worldline. The emitted polarized spins are then treated as independent diffusers, so that the aging void is approximated as the cloud of such diffusers generated by a history-dependent source with emission rate \(r(t)\sim t^{-1/2}\).

Writing the position of an emitted polarized spin as
\begin{equation}
x_i(t)=X(\tau_i)+\eta_i(t),
\end{equation}
with \(\langle \eta_i(t)\rangle=0\) and \(\mathrm{Var}[\eta_i(t)]=2D(t-\tau_i)\), the void center of mass may be decomposed as
\begin{equation}
x_{\rm COM}(t)
=
\frac{\int_0^t d\tau\,r(\tau)\,X(\tau)}{\int_0^t d\tau\,r(\tau)}
+
\frac{1}{N(t)}\sum_i \eta_i(t),
\label{eq:Xcom_exact}
\end{equation}
with \(N(t)\sim \int_0^t d\tau\,r(\tau)\sim t^{1/2}\).

The first term is a weighted time average of the coherence trajectory and therefore inherits the same scaling as \(X(t)\). The second is an internal diffusive contribution, whose variance scales as
\begin{equation}
\mathrm{Var}_{\rm int}[x_{\rm COM}(t)]
\sim
N(t)^{-2}\int_0^t d\tau\,r(\tau)\,(t-\tau)
\sim t^{1/2}.
\label{eq:var_xcom_internal_final}
\end{equation}
For the coherence exponent \(z=3\) found in Sec.~\ref{sec:aging-polaron}, the source-driven term therefore dominates, and the void center of mass inherits the same asymptotic wandering scale as the coherence,
\begin{equation}
\mathrm{Var}[x_{\rm COM}(t)]\sim t^{2/3}.
\end{equation}

The same conclusion holds for the relative coordinate
\begin{equation}
\Delta(t)\equiv X(t)-x_{\rm COM}(t).
\end{equation}
Using Eq.~\eqref{eq:Xcom_exact}, we may write
\begin{equation}
\Delta(t)
=
\frac{\int_0^t d\tau\,r(\tau)\,[X(t)-X(\tau)]}{\int_0^t d\tau\,r(\tau)}
-
\frac{1}{N(t)}\sum_i \eta_i(t).
\label{eq:Delta_exact}
\end{equation}
The second term is the same internal diffusive contribution as above, and therefore contributes a variance of order \(t^{1/2}\).

To estimate the first term, assume that the coherence trajectory is self-affine with dynamical exponent \(z\), so that its increment covariance takes the scaling form
\begin{equation}
\mathrm{Cov}\!\left[X(t)-X(\tau),\,X(t)-X(\tau')\right]
=
t^{2/z}
K\!\left(\frac{\tau}{t},\frac{\tau'}{t}\right),
\label{eq:increment_scaling_kernel}
\end{equation}
for some dimensionless kernel \(K\). Since \(r(\tau)\sim \tau^{-1/2}\) and \(\int_0^t d\tau\,r(\tau)\sim t^{1/2}\), rescaling \(\tau=tu\) and \(\tau'=tv\) in Eq.~\eqref{eq:Delta_exact} gives
\begin{equation}
\mathrm{Var}_{\rm src}[\Delta(t)]\sim t^{2/z},
\end{equation}
up to a dimensionless integral over \(u,v\in[0,1]\). Thus the source-history contribution to the relative coordinate has the same scaling as that of the coherence itself.

Therefore,
\begin{equation}
\mathrm{Var}[\Delta(t)]
\sim
t^{2/z}+t^{1/2}.
\end{equation}
For the aging-polaron exponent \(z=3\) found above, the source-history term dominates, so
\begin{equation}
\mathrm{Var}[X(t)-x_{\rm COM}(t)]\sim t^{2/3}.
\end{equation}
Thus, within this passive-source approximation, both the void center and the relative coordinate inherit the same asymptotic subdiffusive scale as the coherence itself.

\begin{figure*}[t!]
    \centering
    \includegraphics[width=\linewidth]{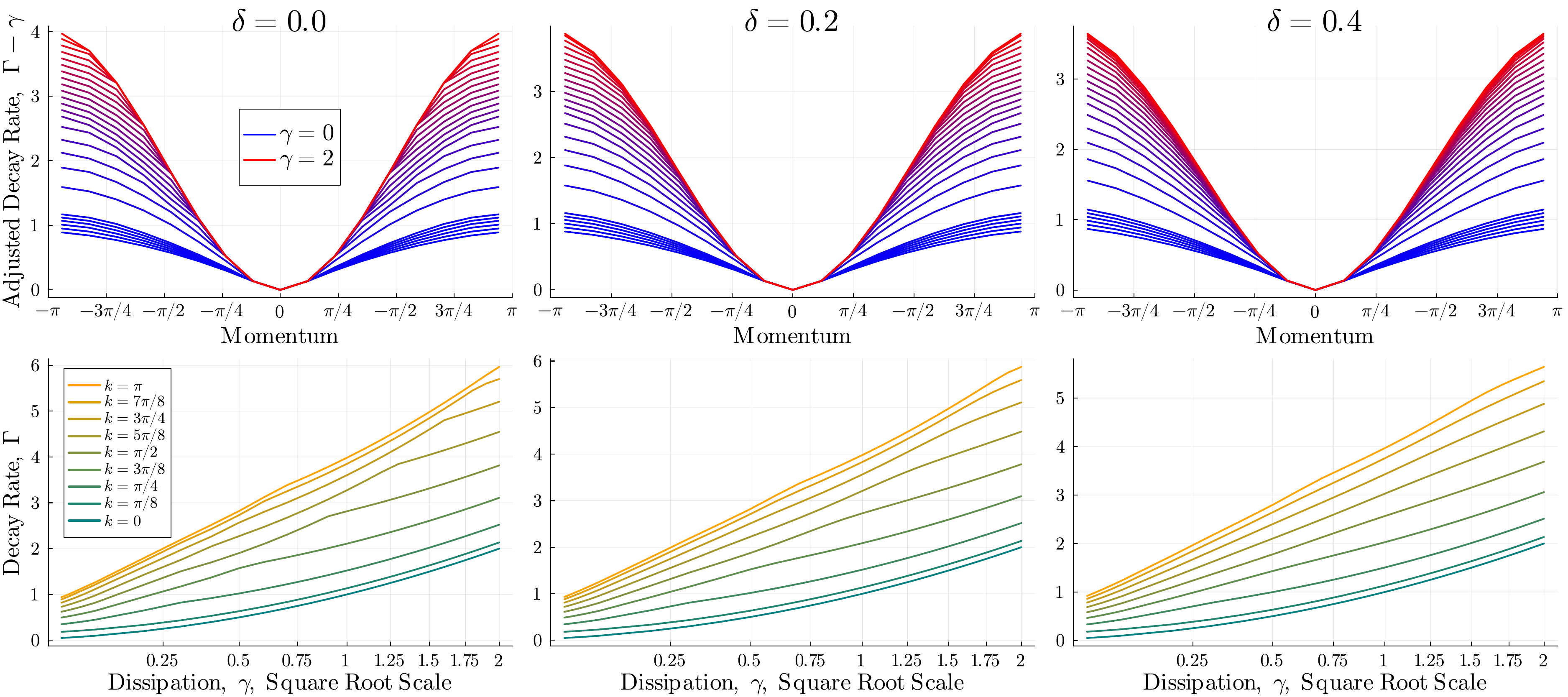}
    \caption{\textbf{(Top Row): Leading eigenvalues of KLS model.} The leading eigenvalues in each momentum sector are shown at three different values of the parameter $\delta$, which sets the background density dependence of the diffusion constant. For weak enough $\gamma$ the dispersion relation becomes linear at intermediate momentum, with a quadratic band minimum gapped by exactly $\gamma$. Eigenvalues were obtained by momentum-resolved exact diagonalization. \textbf{(Bottom Row): Dependence on $\mathbf{\gamma}$.} For each momentum, the scaling with $\gamma$ is shown on a square-root scale, which demonstrates the linear dependence on $\gamma^{1/2}$ for large enough momenta.}
    \label{fig:KLS_dispersion}
\end{figure*}

\section{Details of charge neutral modes}

\label{app:charge_neutral_modes}

In the replicated formalism, $L^2$ norms are represented by overlaps with the cyclic contraction. While the conservation of the cyclic contraction is broken when dissipation is added, the emission of fast operators is largely insensitive to this fact (weak dissipation merely regulates the spread of such operators at large sizes). As such, it is a straightforward exercise to reframe the emission rates in terms of the hydrodynamic modes of the replicated dynamics. For transport dynamics without driving (\textit{e.g.}, SSEP rather than ASEP), we can consider the transport on each replica independently.

We begin by considering $U(1)$ RUC dynamics. On each replica, the transport can be viewed as an imaginary time evolution by $H= -\mc{L}^{\l \alpha \r}$, $\ket{\Phi\l t \r} = e^{- Ht} \ket{\Phi\l 0 \r}$. Dissipation can be added straightforwardly: working in the basis of $P^{\up/\down}\to \ket{\up/\down}$, we get 
\begin{equation}
    H =  \l 1 + \frac{\gamma}{2}\r L - \sum_{i} \l \vec{\sigma}_{i} \cdot \vec{\sigma}_{i+1}-  \frac{\gamma}{2}  \sigma^{x}_{i} \r
\end{equation}
where we have chosen a gate application rate to trivialize the coefficient in front of the $\vec{\sigma} \cdot \vec{\sigma}$ term. Up to an additive constant, this model is simply the ferromagnetic Heisenberg model in a $\hat{x}$-oriented magnetic field proportional to the dissipation, $\gamma$. The elementary excitations of the ferromagnetic Heisenberg model are Goldstone bosons (linear spin waves/magnons), which are gapped by the introduction of the magnetic field, with an energy $\Gamma\l k \r = \sin^{2}\l k/2\r + \gamma$. These magnons are precisely $Z$-operators. Looking in the $n$-magnon sector at some fixed crystalline momentum, $k$, we can approximate the eigenstates as a dilute spin-wave gas (\textit{i.e.}, ignoring Bethe shifts), with total  decay cost
\begin{equation}
    \Gamma\l\lcb k_{i} \rcb\r = \sum_{i=1}^{n} \sin^{2}\l k_{i}/2\r + n\gamma. 
\end{equation}
To find the slowest decaying modes in the regime $1\gg\sin^{2}\l k /2\r \gg \gamma $, we can optimize $\Gamma$ by distributing $k = \sum_{i}k_{i}$ evenly among each magnon, so that $k_{i} = k/n$. Then we find the slowest mode in momentum sector $k$ has $\Gamma\l k\r \sim  k\gamma^{1/2}$ and consists of a gas of $n\sim k \gamma^{-1/2}$ magnons. For small $k$ when $\gamma >  k^{2}$, we cross back over to the standard quadratic dispersion.

However, integrability of the transport model is still preserved: depolarized SSEP retains a $U(1)$ symmetry, expressed as $\sum_{i}\sigma^{x}_{i}$ in the Heisenberg picture in addition to translation invariance. Each $U(1)$ sector experiences a constant ``energy'' shift but with the exact eigenstates of the standard Heisenberg-XXX problem. Therefore, the quasiparticles number remains a good quantum number and different magnon number sectors are dynamically disconnected.

 \begin{figure}[t!]
    \centering
    \includegraphics[width=0.85\linewidth]{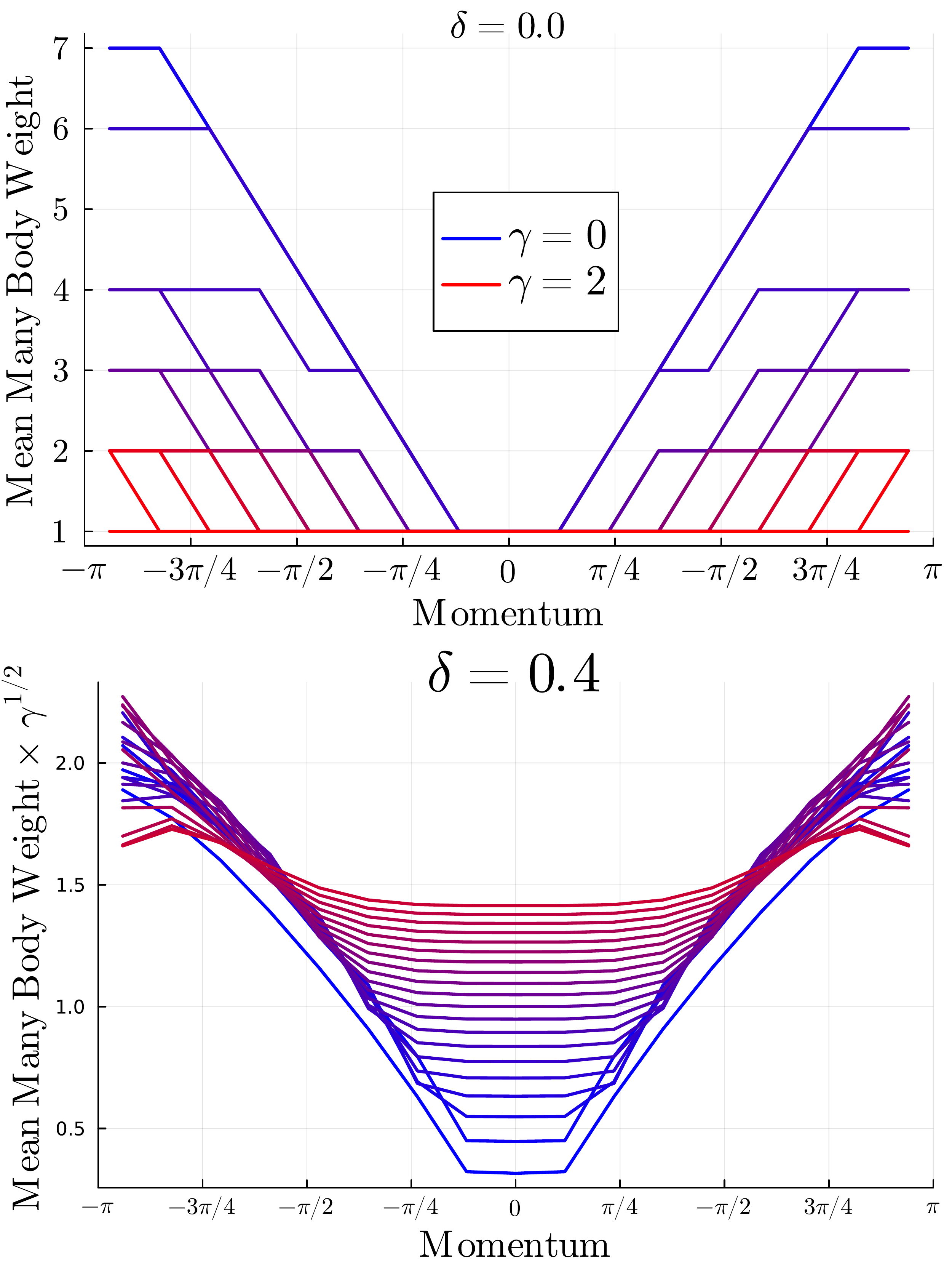}
    \caption{\textbf{Mean magnon number/diagonal weight in KLS model.} (Top) Average magnon number/$Z$-weight of the slowest hydrodynamic mode as a function of crystalline momentum is shown at a variety of dissipation strengths ranging from $\gamma = 0.05$ to $\gamma = 2$ at $\delta =0$. (Bottom) Mean magnon number scaled by $\gamma^{1/2}$ for dissipation strengths ranging from $\gamma = 0.1$ to $\gamma=2$ at $\delta = 0.4$.}
    \label{fig:magnon_number_KLS}
\end{figure}

\subsection{KLS model and the diffuson cascade}

The integrability can be broken by adopting a generalization of SSEP known as the Katz-Lebowitz-Spohn (KLS) model \cite{KLS}. In the KLS model, hopping rates depend on the filling of nearest neighbours. Namely, the two site generator of SSEP is given a rate depending on the occupation of its nearest neighbours. Following the notation of \cite{Michailidis_24}, the KLS model has rules
\begin{eqnarray}
    0100 &\xrightarrow[]{r_{0}\l 1 + \delta\r}& 0010 \nonumber \\
    1101 &\xrightarrow[]{r_{0}\l 1 - \delta\r}& 1011 \nonumber \\
    0101 &\xrightarrow[]{r_{0} }& 0011 \nonumber \\
    1100 &\xrightarrow[]{r_{0}}& 1010, 
\end{eqnarray}
in addition to spatial swaps, where we have set the overall rate $r_{0} =1 $ and the bias (an additional model parameter we will neglect) to $\epsilon =0$. This model has a density dependent diffusion constant $D\l \rho\r = 1 + \delta \l 1 - 2\rho\r$ and does not map to an integrable spin-chain. The slowest hydrodynamic modes, however, exhibit the same $k$ and $\gamma$ dependence in their eigenvalues, modified by a nontrivial diffusion constant: $\Gamma \l k \r \approx 2\sqrt{D\l \rho\r\gamma }k $ for $k \gtrsim \sqrt{\gamma / D\l \rho\r}$ and $D\l \rho\r k ^{2} + \gamma $ for $k \lesssim \sqrt{\gamma / D\l \rho \r}$. Fig.~\ref{fig:KLS_dispersion} shows the decay rates as a function of both $k$ and $\gamma$, and demonstrates consistency with a linear and square-root dependence respectively for intermediate momenta, which both cross over to a quadratic dispersion at low-momenta.

A major distinction between the KLS model away from $\delta=0$ and SSEP is that the magnon quantum number is broken (sectors of different magnon number are dynamically connected). A single $Z$-operator at momentum $k$ is allowed to decay into multiple $Z$-operators with \textit{total} momentum $k$. The magnon number is no longer conserved (see Fig.~\ref{fig:magnon_number_KLS}). Therefore, a so-called ``diffuson cascade'' is expected to occur, in which the magnon number tends towards infinity at late times. It is conjectured that in the thermodynamic limit this results in a long-time decay of density-density correlators as $\exp[- \sqrt{Dk^{2} t}]$ \cite{delacretaz2020heavy}. This conjecture has seen partial support by numerical evidence \cite{Raj_25}, but remains incompletely understood as a dynamical phenomenon. Indeed, while we may characterize the hydrodynamic modes of the KLS model, and believe them to be generic, we are not aware of a circuit model in which these transport dynamics can be derived microscopically.

\subsection{Operator Emission and Many-Body Weight}

The addition of dissipation also has a substantial effect on emitted non-conserved operators. Emission can be understood in simple terms using so-called many-body weight (MBW) distributions. These distributions are defined as the $L^2$ norm partitioned by many body weight, defined in the Pauli basis. Specifically, given a Pauli string, $S\in \mc{P}$, we introduce a function $w(S)$ which counts the number of non-identity elements. Expanding an operator, $O$, in the Pauli basis $O = \sum_{S \in \mc{P}} a_{S} S$, the MBW distribution is defined as
\begin{equation}
    \phi\l w \r = \sum_{\lcb S : \,  w(S) =w\rcb} \lv a_{S}\rv^{2}.
\end{equation}
In terms of the SSEP/KLS models, non-identity elements are always $Z$-operators, and the weight is precisely equivalent to the magnon number as previously defined. Fig.~\ref{fig:magnon_number_KLS}, displays the mean of these distributions for $\delta =0$ at fixed $k$ and $\gamma$ (where the distributions are delta functions at integer values) and the mean scaled by $\gamma^{1/2}$ at $\delta = 0.4$. 

\begin{figure}[b!]
    \centering
    \includegraphics[width=0.9\linewidth]{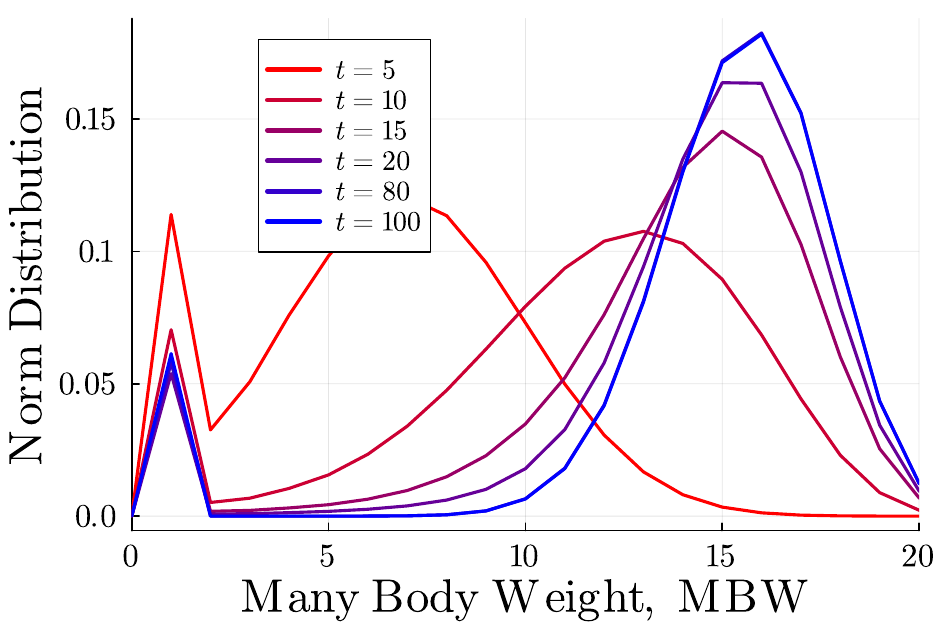}
    \caption{\textbf{Many-body weight distribution.} The relaxation of the many-body weight distribution for a hydrodynamic mode (single magnon) in the $U(1)$-RUC two-copy stat-mech model to its steady state shown at $\gamma =0.0001 $ (matching the very small decay rate). The equilibriation of the decay rates of the single $Z$ (via emission) and the emitted operators as they spread (via the dissipation) leads to the formation of the operator weight ``tail'' shown. These numerics were obtained from TEBD simulations using the library \cite{itensor}.}
    \label{fig:MBWD}
\end{figure}

For the two-copy circuit averaged evolution, the many body weight distribution can detect non-trivial off-diagonal elements as well, namely $\sigma^{+/-}$. %To differentiate from the $Z$-weight, it is helpful to define the off diagonal weight: $w_{\rm od}\l S\r$ counts the number of $\sigma^{\pm}$ in the string with $\phi_{\rm od}\l w\r$ the corresponding distribution. Both $\phi\l w \r$ and $\phi_{\rm od}\l w \r$ are efficiently attainable in the two replica formalism, with detailed constructions given in \autoref{app:mbw_projectors}. The dynamics can then be simulated by TEBD \cite{itensor}, however emission rates for pure linear modes become very small relative to numerically reasonable dissipation rates. As such, we show the relaxation towards the slowest linear mode of the three-magnon sector in Fig.~XXXX.
The dynamics can then be simulated by TEBD \cite{itensor}, and the weight distributions are easily accessible (see the next paragraph). Emission rates for the slowest nontrivial mode (the only one we have access to with open boundary conditions), however, become rather small and require very low dissipation strengths. An example of relaxation to the steady hydrodynamic mode is shown in Fig.~\ref{fig:MBWD}.

To determine many body weight in the two-replica circuit average space, we define the states $\ket{w_{1}} = \ket{\mathbf{zz}} + \ket{\mathbf{p}} + \ket{\mathbf{m}}$ and $\ket{w_{0}} = \ket{\mathbf{11}} $ (the states $\mathbf{1z}$ and $\mbf{z1}$ do not contribute to $L^2$ norms). Then, we can define an MPS with virtual indices $\alpha_{i}$ with $1\leq i \leq L$ and values $0 \leq \alpha_{i}\leq L$, of which the on-site elements are $A^{\l i \r}_{\alpha_{i-1},\alpha_{i}}$ with $\l i \r$ the physical index. This is expressed as
\begin{eqnarray}
    A_{\alpha_{i-1}\alpha_{i}} &=& \delta_{\alpha_{i-1}, \alpha_{i}} \ket{w_{0} } \nonumber \\
    A_{\alpha_{i-1},\alpha_{i} }&=& \delta_{\alpha_{i-1}+1, \alpha_{i} } \ket{w_{1}}
\end{eqnarray}
We then consider the resulting MPS (unnormalized) with $\alpha=0$ and $\alpha_{L}$ unfixed, which we call $\ket{\alpha_{L}}$. The many body weight distribution can then be written
\begin{equation}
    \phi\l \alpha_{L}\r = \braket{\alpha_{L}}{\Psi}
\end{equation}
where $\ket{\Psi}$ is the two-copy stat-mech model wavefunction.
% The off-diagonal many body weight distribution can be determined by redefining $\ket{w_{0}} = \ket{\mbf{11}} + \ket{\mbf{zz}}$ and $ \ket{w_{1}} = \ket{\mbf{p}}+ \ket{\mbf{m}}$.

\end{document}